\documentclass[aps,prd,preprint,groupedaddress,nofootinbib]{revtex4}

\usepackage{amsmath}

\usepackage{hyperref}
\hypersetup{colorlinks=true,citecolor=blue}

\usepackage{slashed}

\usepackage{graphicx}

\usepackage[]{units}

\begin{document}

\title{Anomalous top couplings at hadron colliders revisited}

\author{Fabian Bach}
\email[]{fabian.bach@physik.uni-wuerzburg.de}
\author{Thorsten Ohl}
\email[]{ohl@physik.uni-wuerzburg.de}
\affiliation{Institut f\"ur Theoretische Physik und Astrophysik,
Universit\"at W\"urzburg, Hubland Nord, 97074 W\"urzburg, Germany}

\date{\today}

\begin{abstract}
In an effective operator approach, the full set of leading contributions to
anomalous top couplings comprises various new trilinear as well as higher
interaction vertices, some of which are related to one another by gauge
symmetry or equations of motion. In order to study trilinear top couplings to
SM gauge bosons such as $tt\gamma$, $ttZ$, $tbW$ and $ttg$, the
operator set can be restricted accordingly. However, the complete basis cannot be
mapped onto an on-shell parametrisation of the trilinear
vertices alone. Four-fermion contact terms $qqtt$ and $udtb$ must be
included if the relation to the operator basis is to be retained. In this paper, we point
out how these interactions contribute to the single top search
channels for anomalous trilinear $tbW$ couplings at the LHC and Tevatron,
thus affecting the corresponding bounds.
All results are based on full leading-order partonic matrix elements,
thus automatically
accounting for off-shell and interference effects as well as irreducible
backgrounds.
A discussion of the quantitative effects of going from on-shell tops
to full matrix elements including acceptance cuts
is also provided.
\end{abstract}


\maketitle

\section{Introduction}

The \emph{Standard Model of Particle Physics} (SM) has stood its ground during
the past decades with great success, consistently explaining and predicting
a great variety of high energy experiments with unchallenged
precision. One of the major cornerstones was the discovery of the top
quark at the Tevatron in 1995~\cite{Abe:1995hr,Abachi:1995iq}, confirming
the postulated three-family doublet structure of the SM. While the Tevatron
experiments have continued to collect data and improve
their measurements of top properties, most importantly its
mass~\cite{Lancaster:2011wr,Flyagin:2012zz,Aaltonen:2012va,Aaltonen:2012rz},
most attention is now directed to the LHC up and running at
\mbox{$\sqrt s=\unit[8]{TeV}$}, and the results of its multi-purpose
experiments ATLAS and CMS improving on top statistics by the day. By now,
top pair production has been measured in different channels
with remarkable accuracy~\cite{Aad:2010ey,Chatrchyan:2011ew,Chatrchyan:2011nb,ATLAS:2012aa,Chatrchyan:2012vs,Aad:2012xh}.
Single top production has already been
established for \mbox{$bg\to tW$} associated production despite its small cross
section~\cite{Chatrchyan2012,ATLAS-CONF-2011-104}
and even been definitely observed in the dominant
$t$ channel \mbox{$bq\to tq^\prime$}~\cite{Chatrchyan:2011vp,Aad:2012ux,Chatrchyan1209}.
The ever-growing abundance of top events at the LHC is beginning to allow the determination
of more involved observables such as asymmetries, invariant pair mass
distribution and top couplings to the other SM particles with high
precision (cf.~e.g.,~\cite{Schilling:2012dx,Chiochia:2012vg} for an overview).

On the theoretical side, the top quark takes an outstanding place among the
spectrum of SM particles as a possible
window to new non-SM physics because of its uniquely large mass of the order
of the electroweak symmetry breaking (EWSB) scale
$m_t\sim\upsilon\sim\mathcal O(\unit[100]{GeV})$, with its role
within the dynamics of EWSB still unresolved. Corresponding new physics effects
in the top sector may manifest themselves in the deviation of top properties
from their SM values, where the main attention in this paper is directed
towards the trilinear couplings to SM gauge bosons, especially the charged-current
(CC) interaction $tbW$.
Therefore, even before any experimental analysis, a theoretically robust
parametrisation of these anomalous couplings has to be found, at the same time
reducing the parameter space to
an experimentally manageable
minimum while staying fully general within
the basis of effective operators generating these couplings at Lagrangian
level.  
Indeed, starting from the complete set of effective dimension six
operators as written down by Buchm\"uller and
Wyler already in 1985~\cite{Buchmuller:1985jz},
substantial effort has been put into this task in
the past decades by various authors~\cite{Arzt:1994gp,Gounaris:1996vn,Gounaris:1996yp,
Brzezinski:1997av,Whisnant:1997qu,Yang:1997iv,Grzadkowski:2003tf,
AguilarSaavedra:2008zc,AguilarSaavedra:2009mx,AguilarSaavedra:2010zi,Grzadkowski:2010es}.
The crucial ingredient of most of these analyses is to employ the
theorem~\cite{Arzt:1993gz,Weinberg:1980wa,Gasser:1983yg,Georgi:1991ch,Rujula:1991se}
that the field equations of motion (EOM) can safely be
used at the Lagrangian level at a fixed order in the effective operator
expansion in order to rewrite operators and identify redundant
structures.

As a result of this procedure, it is often argued that these redundancies allow
for a reduction of independent couplings to be incorporated in a
phenomenological analysis of anomalous top couplings. However, as has been
pointed out e.g., in~\cite{Grzadkowski:2003tf,AguilarSaavedra:2008zc} and will
also be reviewed again in more detail later on, the
application of the EOM necessarily generates four-fermion contact interactions,
which are nevertheless often dropped from the analyses for the sake of simplicity. We note
that the latter procedure does not correspond to a \emph{rewriting} but
rather to a \emph{redefinition} of the originally chosen operator basis, thus
also departing from the full generality of the original basis with respect
to the richness of structures in the trilinear couplings. Still, the
operator equalities derived by various
authors~\cite{Grzadkowski:2003tf,AguilarSaavedra:2008gt,AguilarSaavedra:2008zc}
and systematically
presented in~\cite{AguilarSaavedra:2008zc} are very useful to simplify an
implementation of the most general set of trilinear top couplings into a
Monte Carlo (MC) generator in a gauge invariant way, so in our approach,
rather than dropping part of the physics, we make use of these equalities to
implement \emph{all} the trilinear top couplings to SM gauge bosons in the
language of on-shell couplings including the required quartic contact terms into the parton-level
MC~event generator \textsc{Whizard}~\cite{Kilian:2007gr}, also addressing the
interplay of anomalous top and bottom couplings---the latter already heavily
bounded by LEP data---and the repercussions on the top couplings. Finally,
we present phenomenological consequences obtained with our implementation
for the parameter space of the anomalous couplings in the CC sector.

This article is organized as follows: in section~\ref{theo} we review the
procedure described above of defining a complete operator basis to generate
anomalous top couplings to SM gauge bosons and applying the EOM to rewrite
some of these operators, thus arriving at the most suitable form for a
MC~implementation.
In section~\ref{pheno} we discuss the LHC phenomenology with a focus on single
top production, including a comparison of on-shell and full matrix
element approaches
to retrieve the cross sections at detector level as functions of the
anomalous $tbW$ couplings as well as
a presentation of the physical effects and consequences of the newly added
coupling structures.
A discussion and summary of the main statements and results
can be found in section~\ref{sum}.

\section{Theoretical setup}\label{theo}

In order to be self-contained in this article, we start
this section by reviewing in some detail the main steps of the procedure
presented in~\cite{AguilarSaavedra:2008zc} to simplify the most general set of
operators generating the trilinear anomalous top couplings to SM gauge bosons
$tt\gamma$, $ttZ$, $tbW$ and $ttg$. However,
although we emphasize that all of these couplings are implemented in
\textsc{Whizard} in a gauge-invariant way, including all quartic terms generated by
the operator rewriting,
we will restrict ourselves here to the discussion of the CC sector, i.~e.
only those operators generating anomalous contributions to the $tbW$
interaction, for two simple reasons:
\begin{enumerate}

 \item the complexity of the parameter space in the CC sector is increased in
a minimal way compared to previous
studies~\cite{delAguila:2002nf,AguilarSaavedra:2006fy,AguilarSaavedra:2007rs,AguilarSaavedra:2008gt},
because it turns out that
only one additional operator (and hence coupling) has to be considered;

 \item at hadron colliders, the experimental access to the new effects
is most straightforward, combining studies of CC single top production and of top
decay products\footnotemark.

\footnotetext{
The anomalous NC sector, while of course related to the CC sector by gauge
symmetry (cf.~the end of Sec.~\ref{anom_top}), is much harder to access experimentally,
because one would have to identify the final state $ttZ$, which is an even
more complex analysis than the already challenging $tt\gamma$ study due
to the further reduced cross section and the necessity to reconstruct the
decaying $Z$.
In the QCD sector, anomalous $ttg$ (and $ttgg$) chromomagnetic
dipole couplings have been studied
by~\cite{Cheung:1995nt,Hioki:2009hm,Choudhury:2009wd}.
The vector-like $ttqq$ operators which are related to the $ttg$ sector by the EOM
contribute only in the $q\bar{q}\to t\bar{t}$
amplitudes and are therefore suppressed by the pdfs with respect to the dominant
gluon fusion channel.  Non negligible effects of quartic $ttqq$ couplings
have been widely discussed in the
literature as possible explanations of the $t\bar t$ forward-backward
asymmetry observed at the Tevatron,
cf.~e.g.,~\cite{Jung:2009pi,Gabrielli:2011jf}.  However,
for this purpose axial $ttqq$ couplings are required as
well, which are not related to vector-like anomalous
$ttg$ sector by the EOM.
}

\end{enumerate}
In the following two subsections we develop the basic ingredients
of the effective operator analysis, recapitulate the operator rewriting
procedure and finally present our extended parameter space for the
anomalous $tbW$ couplings.

\subsection{Effective operator approach and operator basis}\label{eff_op}

There are basically two ways
to tackle new physics beyond the SM in a systematic and consistent manner:
Either the model building (top-down) approach, i.~e. starting from a
postulated Lagrangian---which incorporates
a sensible UV completion---and deriving from it physical effects to which
present or planned experiments might be sensitive, or the effective
(bottom-up) approach, i.~e. starting from the established SM symmetries
and \emph{a priori} considering \emph{all} possible new physics effects compatible
with these symmetries at the Lagrangian level, postponing the question which
larger theory might generate the relevant parameters at a higher energy scale
$\Lambda$.

Since we follow the second approach, it shall be clarified a little
further. The idea is to confront new physics completely unbiased, that is
without any assumptions about the dynamical degrees of freedom generating it,
and to study the effects that are manifest at a testable energy scale
(considerably smaller than the resonant scale $\Lambda$) where the degrees
of freedom are the well known SM particles. This
corresponds to integrating out the heavy modes, thus generating effective
operators $O^{(d)}_i$ of mass dimension $d>4$ which are normalized
by appropriate powers of $\Lambda$. In the model-independent approach, the
effective Lagrangian can be written as an expansion in
$1/\Lambda$~\cite{Burges:1983zg,Leung:1984ni,Buchmuller:1985jz}:
\begin{equation}
 \mathcal{L}_{\text{eff}} = \mathcal{L}_{\text{SM}} + \sum_{d>4,i} \frac{C_i^{(d)}}{\Lambda^{d-4}} O_i^{(d)} + \text{h.c.}
\end{equation}
with dimensionless operator coefficients $C^{(d)}_i$, comprising \emph{all}
possible effective operators built from SM fields and derivatives only, and
compatible with all local and global SM symmetries. A complete set of
these operators for $d=5,6$ can be found in~\cite{Buchmuller:1985jz}.

The only possible $d=5$ operator in this setup is a neutrino
mass term~\cite{Buchmuller:1985jz}, so the leading contributions to anomalous
trilinear top couplings must be $d=6$.
The complete operator list at this order can
be found in~\cite{AguilarSaavedra:2008zc,AguilarSaavedra:2009mx,AguilarSaavedra:2010zi},
of which we now quote the ones relevant
to trilinear $tbW$ interactions (also adopting the nomenclature
of~\cite{AguilarSaavedra:2008zc}):
\begin{subequations}\label{op_basis}
\begin{align}
 O^{(3,ij)}_{\phi q} &= i\big( \phi^\dagger \tau^I D_\mu \phi \big) \big( \bar{q}_{Li} \gamma^\mu \tau^I q_{Lj} \big)~,\label{Opq} \\
 O^{ij}_{\phi\phi}   &= i\big( \tilde{\phi}^\dagger D_\mu \phi \big) \big( \bar{u}_{Ri} \gamma^\mu d_{Rj} \big)~,\label{Opp} \\
 O^{ij}_{uW}         &=  \big( \bar{q}_{Li} \sigma^{\mu\nu} \tau^I u_{Rj} \big) \tilde{\phi}\, W^I_{\mu\nu}~,\label{OuW} \\
 O^{ij}_{dW}         &=  \big( \bar{q}_{Li} \sigma^{\mu\nu} \tau^I d_{Rj} \big) \phi\, W^I_{\mu\nu}~,\label{OdW} \\
 O^{ij}_{qW}         &=  \big( \bar{q}_{Li} \gamma^\mu \tau^I D^\nu q_{Lj} \big) W^I_{\mu\nu}~,\label{OqW}
\end{align}
\end{subequations}
with generation indices $i,j=1,2,3$ and the non-Abelian $\mathbf{SU}(2)_L$
field strength components
\begin{equation}
 W_{\mu\nu}^I = \partial_\mu W^I_\nu - \partial_\nu W^I_\mu -g\, \varepsilon_{IJK} W^J_\mu W^K_\nu
\end{equation}
to be contracted with the Pauli matrices $\tau^I$ ($I=1,2,3$). The $q_{L(R)i}$
are left(right)-handed quark spinors in the electroweak isodoublet
(isosinglet) representation, and $\phi$ is the isodoublet complex SM scalar
field acquiring a vev $\langle\phi\rangle=\frac{1}{\sqrt 2}(0,\upsilon)^T$,
and $\tilde\phi=i\tau^2\phi^*$.
Of the other electroweak operators listed in~\cite{AguilarSaavedra:2008zc},
$O^{(1,ij)}_{\phi q}$ and $O^{ij}_{\phi u}$ as well as all those containing
the hypercharge field strength $B_{\mu\nu}$ only contribute to NC
interactions, whereas the operators $O^{ij}_{Du}$, $O^{ij}_{\bar{D}u}$,
$O^{ij}_{Dd}$ and $O^{ij}_{\bar{D}d}$
appear to contribute to the $tbW$ vertex.
However, the differences
$O^{ij}_{Du(d)} - O^{ij}_{\bar{D}u(d)}$
are entirely redundant as is shown in~\cite{AguilarSaavedra:2008zc}, and the sums
$O^{ij}_{Du(d)} + O^{ij}_{\bar{D}u(d)}$
are proportional to the gauge
boson momentum $q^\mu=(p_i-p_j)^\mu$ so that amplitudes containing these
vertices vanish either for physical on-shell $W$ or for on-shell light fermions
coupling to the $W$, which is always
the case at parton level for all processes to be considered
for single top and top decay studies discussed here. So Eq.~\eqref{op_basis}
represents the most general $d=6$ operator basis generating anomalous $tbW$
couplings, in which we shall therefore be complete in our
phenomenological studies.

We could now straightforwardly start off to find and implement all
interactions generated by the operator basis which could appear in the relevant
amplitudes. However, this can become a rather involved business
particularly for $O^{ij}_{qW}$ which contains, apart from the trilinear
coupling, also some relevant quartic terms such as e.g.,~$tbW\!g$ with a
complicated Dirac and momentum structure. Therefore, in order to facilitate the
implementation work we rather follow the operator rewriting procedure
of~\cite{AguilarSaavedra:2008zc}, illustrating the main steps here for
$O^{ij}_{qW}$: starting from its decomposition into hermitian and
anti-hermitian parts,
the hermitian part becomes
\begin{equation}\label{OqW_herm}
 \frac{1}{2} \left[ O^{ij}_{qW} + \big( O^{ji}_{qW} \big)^\dagger \right] = \frac{1}{2} \big( \bar{q}_{Li} \gamma^\mu \tau^I q_{Lj} \big)
\big( D^\nu  W_{\nu\mu} \big)^I
\end{equation}
(dropping the total derivative), where the EOM of the $W$ field
\begin{equation}\label{eom_W}
 \big( D^\nu  W_{\nu\mu} \big)^I = g \left\{ \bar{\ell}_{Li} \gamma^\mu \frac{\tau^I}{2} \ell_{Li}
                                        + \bar{q}_{Li} \gamma^\mu \frac{\tau^I}{2} q_{Li}
                                        + i\left[ \phi^\dagger \frac{\tau^I}{2} D^\mu \phi
                                        - \bigl( D^\mu \phi^\dagger \bigr) \frac{\tau^I}{2} \phi \right] \right\}
\end{equation}
can be applied to replace the derivative. On the other hand,
with some algebra~\cite{Buchmuller:1985jz,AguilarSaavedra:2008zc}
the anti-hermitian part can be brought in the form
\begin{equation}
 \frac{1}{2} \left[ O^{ij}_{qW} - \big( O^{ji}_{qW} \big)^\dagger \right] = - \frac{1}{4} \left( \bar{q}_{Li} \sigma^{\mu\nu} \tau^I i\slashed D q_{Lj} W^I_{\mu\nu} - \text{h.c.} \right)
\end{equation}
up to total derivatives, where the EOM of the quark field
\begin{equation}\label{eom_q}
 i\slashed D q_{Li} = Y^u_{ij} u_{Rj} \tilde\phi +Y^d_{ij} d_{Rj} \phi
\end{equation}
(introducing Yukawa matrices $Y^{u/d}$) can be inserted.
Joining it all together, one arrives at the operator equality
\begin{subequations}\label{rewrite}
\begin{align}
 O^{ij}_{qW} =& +\frac{g}{4} \bigl( \bar{q}_{Li} \gamma^\mu \tau^I q_{Lj} \bigr) \Bigl[ \bigl( \bar{\ell}_{Lk} \gamma_\mu \tau^I \ell_{Lk} \bigr)
                  +\bigl( \bar{q}_{Lk} \gamma_\mu \tau^I q_{Lk} \bigr) \Bigr] \label{rewrite_contact} \\
              & +\frac{g}{4} \left[ O^{(3,ij)}_{\phi q} + \bigl( O^{(3,ji)}_{\phi q} \bigr)^\dagger \right] \label{rewrite_Opq} \\
              & -\frac{1}{4} \left[ Y^u_{jk}\, O^{ik}_{uW} + Y^d_{jk}\, O^{ik}_{dW} - Y^{u\dagger}_{ki} \bigl( O^{jk}_{uW} \bigr)^\dagger - Y^{d\dagger}_{ki} \bigl( O^{jk}_{dW} \bigr)^\dagger \right]~.\label{rewrite_OuWdW}
\end{align}
\end{subequations}
Obviously, the terms in~\eqref{rewrite_Opq} and~\eqref{rewrite_OuWdW}
are redundant and can be absorbed into the operators~\eqref{Opq}--\eqref{OdW},
whereas~\eqref{rewrite_contact} generates four-fermion contact interactions.

Although it is clear that without further restrictions there is enough
freedom within the operator basis to independently vary all the couplings
emerging from~\eqref{Opq}--\eqref{OdW} and the associated
contact terms coming from the rewriting~\eqref{rewrite}, the
rewriting procedure corresponds to a shift of the original operator
coefficients. Setting $i=j=3$ and dropping all generation superscripts from
now on, these shifts are:
\begin{align}\label{coeff_shift}
 \delta \text{Re}\,C_{\phi q}^{(3)} &=  \frac{g}{2} \text{Re}\,C_{qW}~, \nonumber \\
 \delta \text{Im}\,C_{uW}     &= -\frac{m_t}{\sqrt 2\upsilon} \text{Im}\,C_{qW}~, \nonumber \\
 \delta \text{Im}\,C_{dW}     &= -\frac{m_b}{\sqrt 2\upsilon} \text{Im}\,C_{qW} \simeq 0~,
\end{align}
assuming an approximate decoupling of the third generation in the Yukawa
matrices.
With this setup, we can now go on to physical states of the gauge and matter
fields and write down the interaction terms generated by our operator basis.

\subsection{Parametrisation of anomalous charged-current couplings}\label{anom_top}

Inserting the scalar vev and physical states of the gauge fields into the
operators~\eqref{op_basis} and forming hermitian combinations
$C_xO_x+C_x^*O_x^\dagger$, one finds various trilinear interaction terms
$tbW$, $ttZ$, $ttA$ but also $bbZ$ and $bbA$, as well as associated quartic
interactions $ttWW$, $bbWW$, $tbW\!Z$ and $tbW\!A$ which are all necessary to
maintain gauge invariance in the resulting amplitudes, and have therefore
been included in our implementation.
The resulting effective $tbW$ interaction Lagrangian can be written as
\begin{subequations}\label{L_tbW}
\begin{align}
 \mathcal{L}_{tbW} =& -\frac{g}{\sqrt 2} \bar{b}\,\gamma^\mu \bigl( V_L P_L + V_R P_R \bigr)\,t\,W^-_\mu  + \text{h.c.} \label{V_LR} \\
                    & -\frac{g}{\sqrt 2} \bar{b}\,\frac{i\sigma^{\mu\nu}q_\nu}{m_W} \bigl( g_L P_L + g_R P_R \bigr)\,t\,W^-_\mu  + \text{h.c.} \label{g_LR} \\
                    & -\frac{g}{\sqrt 2} \bar{b}\,\gamma^\mu \frac{q^2-m_W^2}{m_W^2} \bigl( V_L^\text{off} P_L \bigr)\,t\,W^-_\mu + \text{h.c.}~, \label{V_off}
\end{align}
\end{subequations}
where all couplings except for $V_L\equiv V_{tb}\simeq1$ vanish in the SM
at tree level, and get the following anomalous contributions from operator
coefficients\footnotemark:
\begin{align}\label{cpl}
 \delta V_L &= \left( C_{\phi q}^{(3)*} + \frac{g}{2} \text{Re}\,C_{qW} \right) \frac{\upsilon^2}{\Lambda^2}~, &
 \delta g_L &= \sqrt 2 C_{dW}^* \frac{\upsilon^2}{\Lambda^2}~, \nonumber \\
 \delta V_R &= \frac{1}{2} C_{\phi\phi}^* \frac{\upsilon^2}{\Lambda^2}~, &
 \delta g_R &= \sqrt 2 C_{uW} \frac{\upsilon^2}{\Lambda^2}~, \nonumber \\
 \delta V_L^\text{off} &= \frac{g}{2} \text{Re}\,C_{qW} \frac{\upsilon^2}{\Lambda^2}~.
\end{align}
\footnotetext{Note that in Eq.~(37) of~\cite{AguilarSaavedra:2008zc}
the operator coefficient $C_{\phi\phi}^{33}$ appearing in $\delta V_R$
should also be complex-conjugated.}
The interaction terms~\eqref{V_LR} and~\eqref{g_LR} represent the on-shell
parametrisation widely used in various phenomenological studies
(normalization convention taken from~\cite{AguilarSaavedra:2008zc}),
which is retrieved from the operators~\eqref{Opq}--\eqref{OdW}.
The interaction~\eqref{V_off} emerges from the hermitian part of $O_{qW}$,
\begin{equation}
 O_{qW} + O_{qW}^\dagger = \big( \bar{q}_{L3} \gamma^\mu \tau^I q_{L3} \big)
\big( \partial^2  W_\mu^I \big) + \text{higher contact interactions,}
\end{equation}
cf. Eq.~\eqref{OqW_herm}, which---unlike the anti-hermitian part---cannot
be completely recast into a combination of the other four operators.
However, the \emph{partial} redundance of $O_{qW}$ has been made explicit
in the parametrisation~\eqref{L_tbW} by defining its on-shell part into
$V_L$ so that any contribution $\sim V_L^\text{off}$ vanishes when the
$W$ goes on the mass shell. Hence it is no surprise that in $\delta V_L$
of Eq.~\eqref{cpl} we find again the shift of the coefficient
$C_{\phi q}^{(3)}$ already stated in Eq.~\eqref{coeff_shift} after the
operator rewriting. Furthermore, by comparison to Eq.~\eqref{rewrite},
one finds that all contributions $\sim V_L^\text{off}$ must be in one-to-one
correspondence to the four-fermion contact interactions given
in~\eqref{rewrite_contact}, which is also highlighted by the fact that in
physical amplitudes the kinematic structure of the $W$ propagator is exactly
cancelled by the $q$-dependent vertex.

We have now isolated the non-redundant
contribution of $O_{qW}$ to the $tbW$ interaction Lagrangian, and also
identified the most convenient way to implement it in a gauge-invariant way,
namely by adding the quartic fermion vertices
\begin{align}
 \Delta \mathcal{L} =& \frac{g_\times}{\Lambda^2} \bigl( \bar{b} \gamma^\mu P_L t \bigr)
\Bigl[ \bigl( \bar{u}_k \gamma_\mu P_L d_k \bigr) + \bigl( \bar{\nu}_k \gamma_\mu P_L e_k \bigr) \Bigr] + \text{h.c.} \\
&\qquad\text{with } g_\times = g\,\text{Re}\,C_{qW} \nonumber
\end{align}
(cf.~\cite{Grzadkowski:1997cj,Cao:2007ea,AguilarSaavedra:2010zi}),
giving a relation of coefficients
\begin{equation}
 V_L^\text{off} = \frac{\upsilon^2}{2\Lambda^2}\,g_\times~.
\end{equation}
Of course, one might ask at this point if such a coupling structure should
be counted among the anomalous $tbW$ sector, but then again it must
be noted that
as a consequence of the common operator basis, the trilinear couplings
are related to $V_L^\text{off}$ through the underlying operator coefficients.
Specifically, Eq.~\eqref{cpl} illustrates that a limit on $\delta V_L$ cannot be
unambiguously mapped onto a limit on the operator coefficient
$C_{\phi q}^{(3)}$ without also bounding $\delta V_L^\text{off}$
(or the anomalous NC sector, see below).
Moreover,
the operator basis~\eqref{op_basis} and the corresponding set
of~couplings~\eqref{L_tbW} parametrise \emph{all}
anomalous diagram insertions which can interfere with the SM~diagram
in a minimal way,
making this approach consistent at the amplitude level.
Finally,
as pointed out in Sec.~\ref{VO}, the inclusion of
the additional coupling also affects the interpretation of current and
upcoming experimental results at the LHC.

Before moving on to the phenomenological implications, let us discuss briefly
the issue of anomalous bottom couplings within the effective theory approach:
Since the original effective operators by construction respect the
full electroweak gauge symmetry $\mathbf{SU}(2)_L\times\mathbf{U}(1)_Y$,
it is no surprise to find certain relations within the set of anomalous
electroweak couplings of the heavy doublet $(t,b)$ after spontaneous
symmetry breaking. For example, an anomalous CC contribution $\delta V_L$
is directly related to the anomalous left-handed NC vector couplings
$ttZ$ and $bbZ$, the latter one stringently constrained by LEP data,
so turning on $\delta V_L$ while respecting all existing bounds
necessarily implies a non-vanishing anomalous contribution to the left-handed $ttZ$
vector coupling~\cite{AguilarSaavedra:2008zc,Berger:2009hi},
or a fine-tuned relation with $\delta V_L^\text{off}$, cf.~Eq.~\eqref{cpl}.
Similarly, $\delta g_R$ is directly related to the anomalous $ttZ$/$tt\gamma$ tensor
couplings, just like $\delta g_L$ is to the $bbZ$/$bb\gamma$ ones
(cf.~e.g.,~\cite{AguilarSaavedra:2008zc,Aguilar-Saavedra2012} for details).
In short, it is impossible to vary the anomalous CC couplings in a
consistent way within the effective operator approach without either getting
anomalous NC couplings or including additional operators to fine-tune
these effects away.
Although these relations basically have no effect on a purely CC single top study,
one should bear them in mind when addressing anomalous CC couplings
(the \textsc{Whizard} implementation contains the option to automatically enforce
these relations).

\section{LHC phenomenology}\label{pheno}

Apart from indirect searches using low energy observables, e.g.,~in
flavor physics~\cite{Grzadkowski:2008mf,Drobnak:2011aa,Drobnak:2011wj}
or SM precision observables~\cite{Zhang2012},
there are basically two different classes of direct observables
for top quark properties at the current collider experiments, namely those
related to top production or top decays. While it is clear that only a
combination of all available observables will deliver the best bounds on anomalous
contributions, it is crucial to understand each analysis separately before
the combination step. Therefore, we will focus here on the discussion of
single top production cross sections, citing and using results from top
decay studies to derive estimates for the most stringent bounds on the
full anomalous parameter space at the end of the article.

\begin{figure*}
 \includegraphics[trim = 99mm 216mm 98mm 40mm, clip, scale=1]{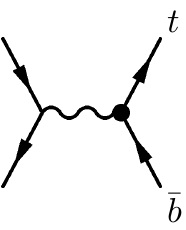}
 \hspace{1.4cm}
 \includegraphics[trim = 81mm 216mm 80mm 40mm, clip, scale=1]{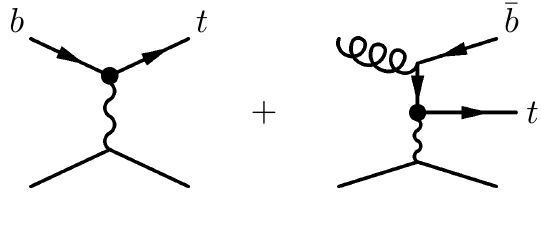}
 \hspace{1.4cm}
 \includegraphics[trim = 81mm 216mm 78mm 40mm, clip, scale=1]{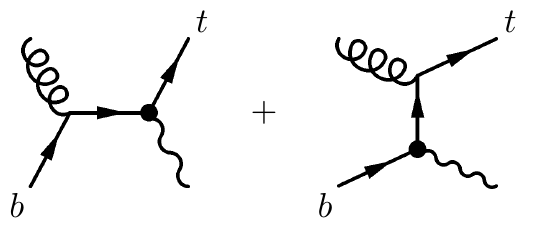}
 \caption{
Diagrams contributing to LO on-shell single top production
(anomalous $tbW$ vertex marked by a dot):
$s$~channel $tb$~production (left diagram), $t$~channel $tj+tbj$ production
(center diagrams) and associated $tW$~production (right diagrams). 
\label{diags}}
\end{figure*}

Single tops are produced at the LHC (and Tevatron) in three different channels,
namely $s$~channel $tb$ production, $t$~channel $tj$
production (where $j$ denotes a light hadronic jet),
and associated $tW$ production, cf.~Fig.~\ref{diags}.
While experimentalists
are struggling to identify and discriminate these channels at the detector
level with suitable selection criteria, the theoretical question is how the
corresponding measured cross sections $\sigma^\text{det}_i$ (for final states
\mbox{$i=tb,tj,tW$}) are represented as functions on the
anomalous parameter space, \mbox{i.~e.} how the measurement can be converted
into bounds on the parameters. In this respect, a first step may be to separate
the detector response from the hard production cross section:
\begin{equation}\label{xsec}
 \sigma^\text{det}_i(\vec{g}) = \sum_j \varepsilon_{ij} \cdot \sigma^\text{part}_j(\vec{g})~,
\end{equation}
summing over partonic production processes $j$ as functions of the parameter
point $\vec{g}$. $\varepsilon_{ij}$~denotes
the detector efficiency matrix mapping the process~$j$ onto the final state
selection~$i$, which can be retrieved with a detector simulation.
Once the functions $\sigma^\text{part}_i(\vec{g})$ are known,
experimentally measured confidence intervals for $\sigma^\text{det}_i$ can
be mapped onto confidence intervals for $\vec{g}$ by formal inversion of
Eq.~\eqref{xsec}.

However, the remaining question to be addressed in this approach is: Where did
we put the detector acceptance~$\Phi$, into $\varepsilon$ or
into $\sigma^\text{part}_i$? The significance of this question is obvious,
since anomalous couplings might very well affect the differential distributions,
thus making~$\Phi$ a function of~$\vec{g}$. Therefore, the
answer to that question influences the strategy as well as the efforts
necessary to compute the function $\sigma^\text{part}_i(\vec{g})$, and potentially
also the bounds derived from it,
as we will show in the following section.

\subsection{Technical setup}\label{on-off}

\subsubsection{Kinematics in the on-shell limit}\label{on}

The simplest approach is to neglect the $\vec{g}$-dependence of the acceptance
entirely and pull it into $\varepsilon$, implying that the
$\sigma^\text{part}_i(\vec{g})$ in Eq.~\eqref{xsec} represent the set of
total partonic cross sections integrated over the full phase space.
Further neglecting finite width and
interference effects with irreducible backgrounds enables one to decompose
$\sigma^\text{part}_i(\vec{g})$ as
\begin{equation}\label{onshell}
 \sigma^\text{part}_i(\vec{g}) = \sigma^\text{prod}_i(\vec{g}) \times \prod \mathcal{BR}~,
\end{equation}
where $\sigma^\text{prod}_i(\vec{g})$ denotes the full on-shell single top
production cross sections, and the product of branching ratios accounts for
the decays of the heavy particles, namely $t$ and one or two $W$s, depending
on the production channel. Since all the on-shell production diagrams
can contain only one anomalous $tbW$ vertex insertion, it is argued
in~\cite{AguilarSaavedra:2008gt} that $\sigma^\text{prod}_i(\vec{g})$
may be written as a polynomial up to second order in $\vec{g}$:
\begin{equation}\label{poly}
 \sigma^\text{prod}_i(\vec{g}) = \sigma^\text{SM}_i \sum_{k,l} \kappa^i_{\,kl}\, g_k\, g_l~,
\end{equation}
where the $\sigma^\text{SM}_i$ are the total SM cross sections, and the $\kappa^i_{\,kl}$
denote the integrated products of diagrams with one insertion of $g_k$ and
$g_l$ each, normalized to the SM point in each production channel~$i$.
Plugging~\eqref{onshell} and~\eqref{poly} into~\eqref{xsec},
one arrives at the ansatz employed in~\cite{AguilarSaavedra:2008gt}:
\begin{align}\label{xsecAS}
 \sigma^\text{det}_i(\vec{g}) &= \sum_{j,k,l} \left[\varepsilon\times\prod\mathcal{BR}\right]_{ij} \cdot \sigma^\text{SM}_j\cdot\kappa^j_{\,kl}\, g_k\, g_l\nonumber \\
     &\equiv \sum_{j,k,l} \varepsilon_{ij} \cdot \sigma^\text{SM}_j\cdot \kappa^j_{\,kl}\, g_k\, g_l~.
\end{align}
For brevity, this will be referred to as \textit{on-shell approach}
from here on. The advantage of the
formula is obvious: once $\varepsilon$ and the constant $\kappa$s are known, the
conversion of measured results into bounds on $\vec{g}$ becomes very simple
and efficient. However, the validity of the assumptions leading to this
result shall be addressed now.

\subsubsection{Full matrix elements and acceptances including
  anomalous couplings}\label{off}

Eq.~\eqref{xsecAS} tells us that one should be able to vary the coupling
point~$\vec{g}$ within the ranges relevant for the study,
with only minor effects on the detector response $\varepsilon$
in the phase space window which corresponds to a given final state
selection.
However, Eq.~\eqref{xsecAS} implies even more, namely that retrieving the
matrix element response as a function of $\vec{g}$ and applying
acceptance cuts on the phase space should approximately commute,
or equivalently,
Eq.~\eqref{xsecAS} should give the
same results as~e.g.,
\begin{equation}\label{xsecW2}
 \sigma^\text{det}_i(\vec{g}) = \sum_j \varepsilon_{ij}^\prime \cdot \left[ \Phi^\text{part} \times \sigma^\text{part}\right]_j(\vec{g})~,
\end{equation}
where the basic detector acceptance cuts such as $p_T$ and $\eta$ cuts on the
partons and leptons, represented by $\Phi^\text{part}$, are applied directly
to the phase space integration and hence formally included in the
$\vec{g}$-dependent part of the formula, while the matrix $\varepsilon^\prime$,
assumed to be constant in $\vec{g}$,
denotes the efficiency of mapping the partonic final states
\emph{at the acceptance level}~$\Phi^\text{part}$ onto the final state selections at detector level.

To be more explicit, the idea is to accommodate as much of the acceptance
cuts as possible within the $\vec{g}$-dependent part without becoming
exclusive to any of the different final state selections, which are
still contained in the $\vec{g}$-independent $\varepsilon^\prime$.
This obviously implies that the phase space window covered
by~$\varepsilon^\prime$ must be fully contained within the acceptance
window~$\Phi^\text{part}$,
leading to the notion that partonic acceptance and final state
selection cuts should be adapted to each other as closely as possible.

Assuming leptonic $t$ decay,
we therefore apply the following acceptance cuts on the partonic
phase space integration:
\begin{subequations}\label{acc}
\begin{align}
 \Phi^\text{part}:\qquad\qquad
   p_T\left(\ell,\nu\right) > \unit[25]{GeV}\quad & \text{and}\quad \left| \eta\left(\ell\right) \right| < 3\,, \label{phi_l} \\
   p_T\left(j,b\right) > \unit[30]{GeV}\quad & \text{and}\quad \left| \eta\left(j,b\right) \right| < 5\,, \label{phi_j} \\
   \unit[150]{GeV} < m_{b\ell\nu} & < \unit[225]{GeV}\,, \label{phi_mt}
\end{align}
\end{subequations}
where Eq.~\eqref{phi_j} is required for only one of the two $b$s in the
$tbj$~process to be inclusive\footnotemark,
and all the cuts are in correspondence to the
detector-level selection criteria stated below.
\footnotetext{
Clearly, for full inclusiveness one would have to entirely drop the distinction
of light and $b$ flavors at partonic level, but the actual chance of mistagging the
light forward jet is negligible once the full event topology (cf. $tj$ selection
below) is taken into account.
}
Associated $tW$ production is entirely omitted for the time being,
because modelling this process within its detector acceptance window
while at the same time remaining inclusive with respect to the other processes
is highly nontrivial, and only marginally affects our following
statements (the main effect being the neglected contamination of the other
final states at the detector level, which amounts to $\lesssim$\unit[10]{\%}
in the $tb$~channel and practically vanishes in the dominant
$tj$~channel)\footnotemark.
Clearly, this is still not the fully correct answer at detector level,
but it should be closer to the truth than entirely neglecting the
$\vec{g}$-dependence of~$\varepsilon$, and the consistency of the two approaches
can be checked.
\footnotetext{
Moreover, the suppression of the huge irreducible $t\bar{t}$ background
in the radiative correction diagram~$tWb$ still is a~topic of vivid
discussion (cf. e.g.,~\cite{Schilling:2012dx}),
a~problem which again does not affect the main statements
of this paper. Still, it is clear that in the end also this channel should
be addressed and included in a complete study.
}

Although it is clear that the object
$\left[ \Phi^\text{part} \times \sigma^\text{part}\right]_i(\vec{g})$ to
be computed is much harder to handle than the constant $\kappa$s (even more so
if the full matrix element response including all off-shell and interference
effects is to be taken into account), it is basically just a technical
issue which can be tackled with appropriate Monte Carlo machinery and
respective CPU time.
For brevity, we will refer to this approach as
\textit{full matrix element (ME) approach} from now on.
In the following section, we compare the results of
Eq.~\eqref{xsecAS} and Eq.~\eqref{xsecW2}, and see if the effort is
justified.

\subsection{Comparison of the results in the on-shell limit with the
  full results}\label{comp}

For the measurement of the total cross section of a given final state at the
detector level, the experimental sensitivity is given in terms of
a~measure for $\Delta\sigma/\sigma$,
where estimations for total uncertainties are adopted
from~\cite{AguilarSaavedra:2008gt} for consistency, amounting to
$\unit[20.8]{\%}$ ($tb$~sel.) and $\unit[13.5]{\%}$ ($tj$~sel.)
for $\unit[10]{fb^{-1}}$ of LHC data at $\sqrt{s}=\unit[14]{TeV}$.
Therefore, we separate the overall normalization, which is basically given by
higher-order SM results for the total production cross sections, from
the modelling of the normalized LO matrix element response as
a~function of the anomalous coupling set~$\vec{g}$, i.\,e.~the
$\kappa_{\text{on}}$ coefficients in
the on-shell approach or, more generally, a function
$\Delta\sigma/\sigma(\vec{g})\equiv\kappa(\vec{g})$ for each partonic
input process~$i$, where
\begin{subequations}\label{k_g}
 \begin{align}
  \text{on-shell:}&&\kappa^i_\text{on}\left(\vec{g}\right) & = \sum_{k,l}  \kappa^i_{\,kl}\, g_k\, g_l \label{k_on} \,, \\
  \text{full ME:}&&\kappa^i_\text{full}\left(\vec{g}\right) & = \frac{\left[ \Phi^\text{part} \times \sigma^\text{part}\right]_i(\vec{g})}{\left[ \Phi^\text{part} \times \sigma^\text{part}\right]_i|_\text{SM}} \,, \label{k_off}
 \end{align}
\end{subequations}
cf. Eqs.~\eqref{xsecAS} and~\eqref{xsecW2}. Since the aim is to
accommodate \emph{all} coupling dependence therein, it is fruitful to
first set $\varepsilon\equiv 1$ and compare this function for the different approaches
at the partonic level.

\subsubsection{Partonic level}

To be self-consistent, we essentially redo the analysis
procedure presented in~\cite{AguilarSaavedra:2008gt} for the on-shell approach,
employing \textsc{Whizard} to compute the coefficients of $\kappa_\text{on}$ as well as produce parton-level
samples, which are then processed with \textsc{Pythia} and \textsc{Delphes} to
retrieve~$\varepsilon$.
The quadratic coefficients (i.~e. $\sim g_i^2$) are obtained in each production
channel, cf.~Fig.~\ref{diags}, by separately setting each $g_i=1$, integrating the total cross
section for on-shell single top production and finally normalizing to the
SM point ($V_L=1$, $V_R=g_{L,R}=0$). The interference terms are computed similarly, setting
always two couplings to~1 and subtracting the quadratic parts from the result
before normalizing to the SM. The implementation of the vertices and phase
space integration has been checked by switching off the pdfs and comparing
\textsc{Whizard} to analytical results.
Using the pdfs and parameter setup quoted in~\cite{AguilarSaavedra:2008gt},
\textsc{Whizard} also reproduces the on-shell $\kappa$ coefficients stated there
within numerical uncertainties. For all further \textsc{Whizard} results,
we set $m_t=\unit[173.1]{GeV}$, $m_b=\unit[4.2]{GeV}$, $m_W=\unit[80.42]{GeV}$
and choose CTEQ6L1~\cite{Pumplin:2002vw} for pdfs.

\begin{figure*}
 \includegraphics[scale=0.6]{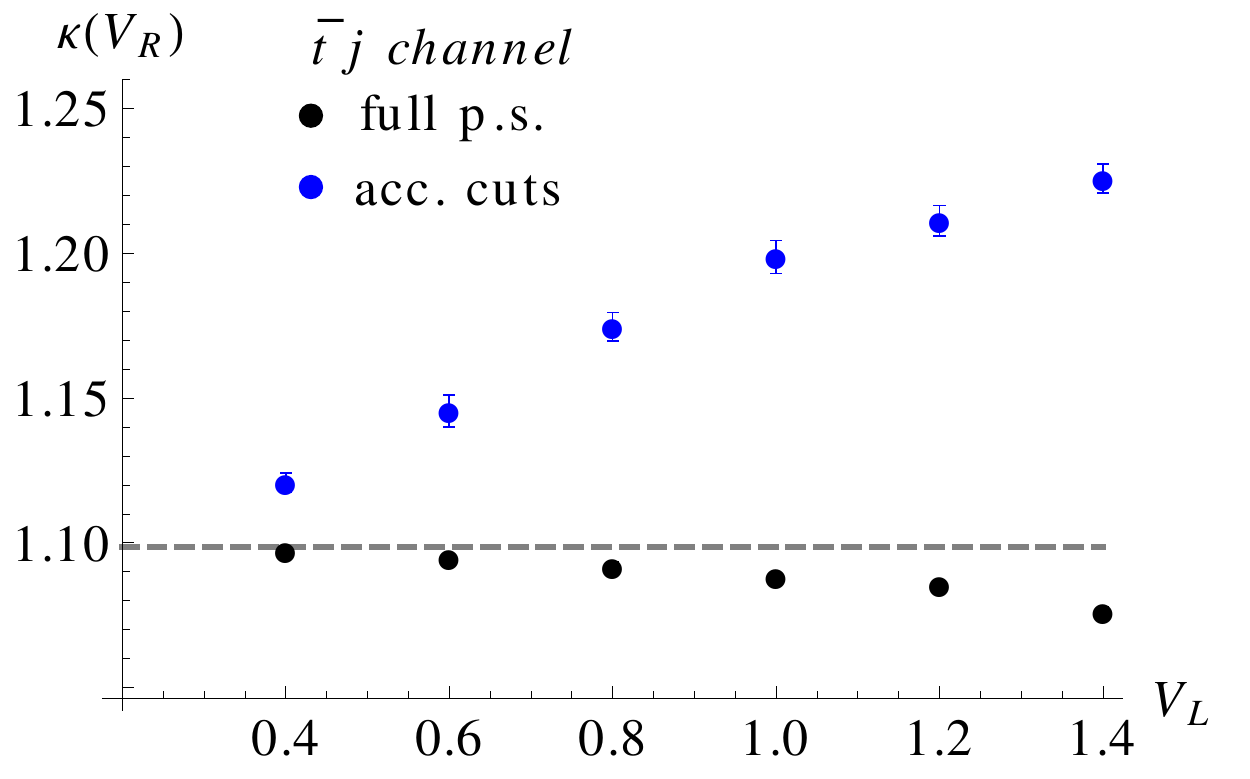}
 \hspace{0.5cm}
 \vspace{0.5cm}
 \includegraphics[scale=0.6]{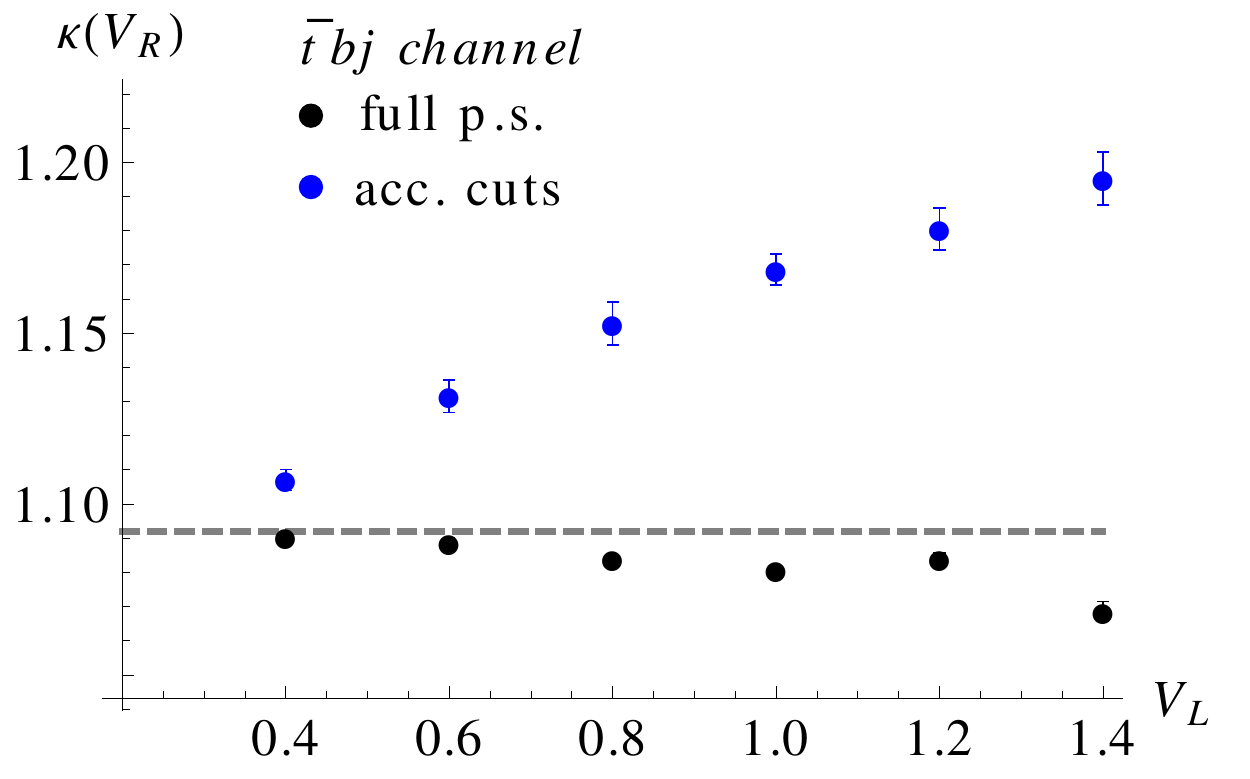}
 \vspace{0.5cm}
 \includegraphics[scale=0.6]{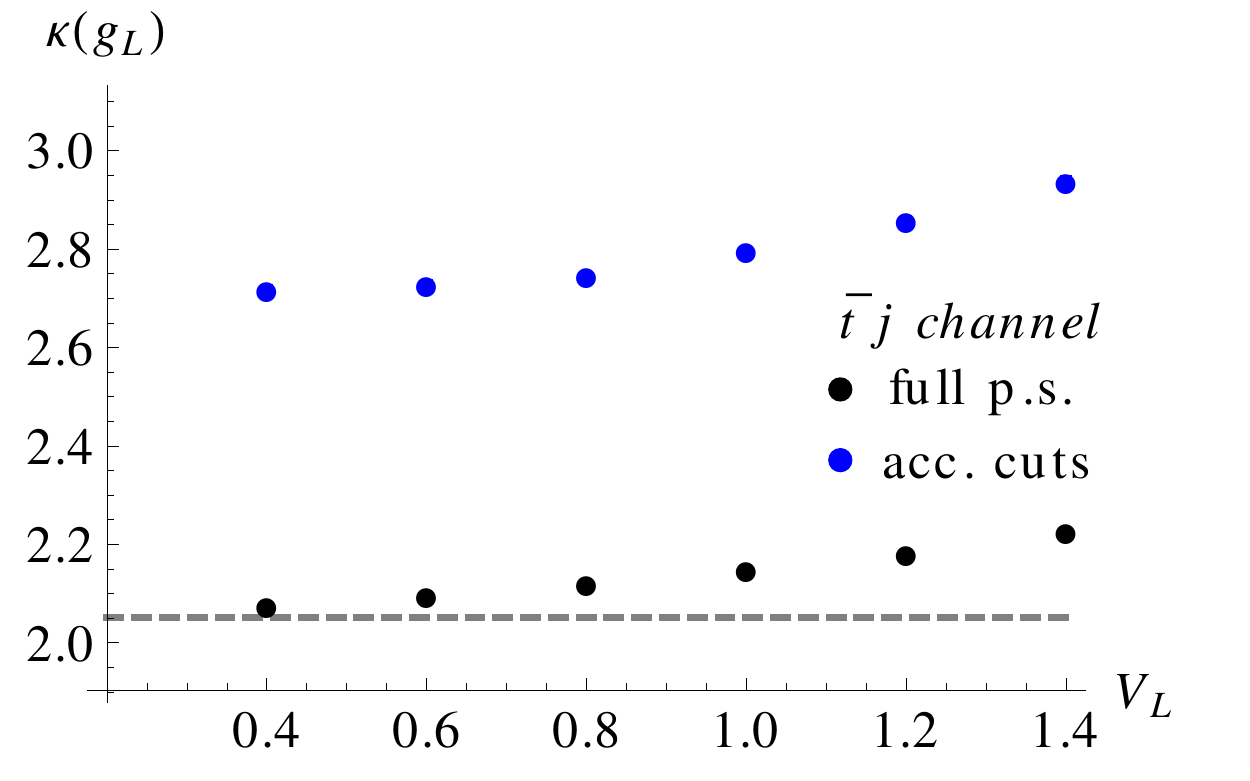}
 \hspace{0.5cm}
 \includegraphics[scale=0.6]{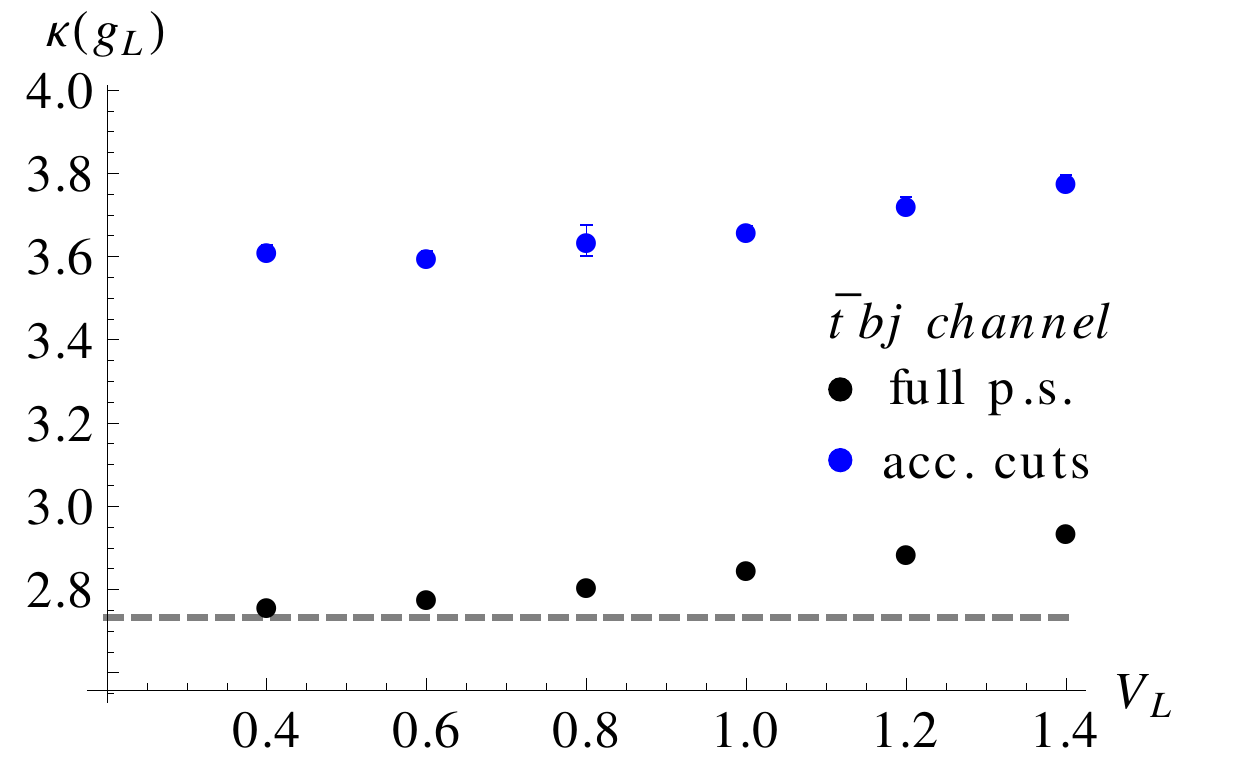}
 \vspace{0.5cm}
 \includegraphics[scale=0.6]{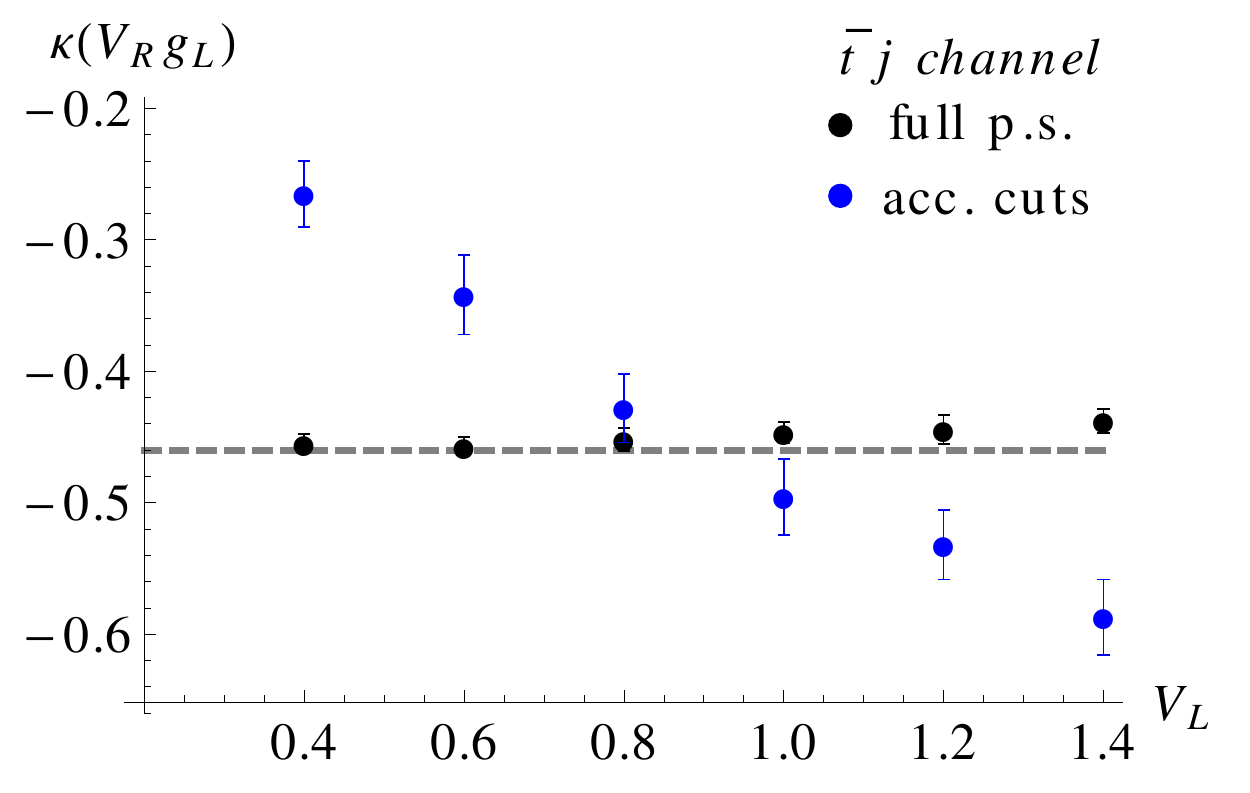}
 \hspace{0.5cm}
 \includegraphics[scale=0.6]{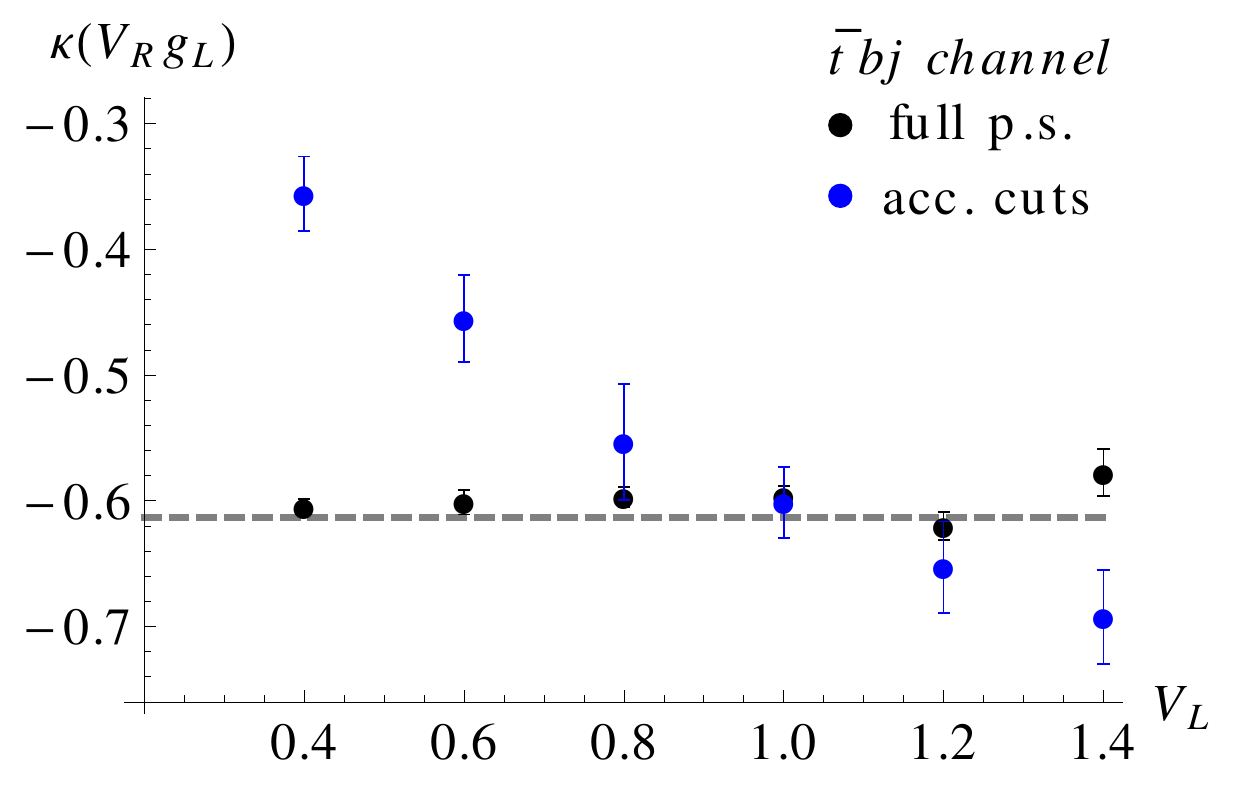}
 \caption{Numerical results for various $\kappa$s in the processes $\bar{t}j$ (left)
and $\bar{t}bj$ (right), as extracted from quadratic fits to the normalized
cross sections. The dashed line indicates the value of the on-shell $\kappa$
in each case.
\label{k_tjc}}
\end{figure*}

In the full ME approach, the matrix element response is modelled
according to Eq.~\eqref{k_off} by applying the acceptance cuts $\Phi^\text{part}$,
Eq.~\eqref{acc}, to the full partonic off-shell matrix elements. In this
approach, taking all finite width and interference effects into account, it
is \emph{a priori} not clear that the function~$\kappa_\text{full}\left(\vec{g}\right)$
obeys a simple polynomial expansion in~$\vec{g}$, so rather than
assuming a specific functional form,
we use the \textsc{Whizard} machinery to scan the entire \mbox{4-dimensional}
parameter space $\vec{g}=\left(V_L,V_R,g_L,g_R\right)$
(effects of $V_L^\text{off}$ will be addressed separately in Sec.~\ref{VO})
within the relevant numerical ranges $0<V_L<1.2$, $-1.2<V_R<1.2$ and
$-0.6<g_{L,R}<0.6$ (cf.~\cite{AguilarSaavedra:2008gt}),
also including the full dependence of the top width on the couplings
$\Gamma_t(\vec{g})$.
Since $\Gamma_t$ has already been measured, the most recent bound
from D$\slashed 0$ being $\Gamma_t=\unit[2.00^{+0.47}_{-0.43}]{GeV}$~\cite{Abazov:2012vd},
it is included in our analysis as an additional observable\footnote{
Clearly the experimental analysis performed in~\cite{Abazov:2012vd}
to extract $\Gamma_t$ from data will itself also be affected by $\vec g$~dependent
acceptances as discussed in the course of this paper.
However, since we do not aim at a reassessment of their analysis in this respect,
yet also want to exclude regions in parameter space which are
completely unphysical with respect to $\Gamma_t$, we still include the observable,
inferring the $\Gamma_t(\vec g)$~dependence over the full phase space.
This approach is conservative, because our results indicate that including the
full acceptance dependence generally tends to improve the sensitivities,
cf.~e.g.,~Fig.~\ref{k_tjc}.
\emph{A posteriori}, we find that, due to the still large error bars, the
current measurement of $\Gamma_t$ constrains $\vec{g}$ 
much less than the cross sections.  The limits
on~$\vec{g}$ would therefore not be affected substantially by such a reassessment.
}.
The numerical results can then be used to test the validity of the polynomial
parametrisation, Eq.~\eqref{k_on}, in the following way:
The normalized matrix element response $\kappa$ may always
be expanded as
\begin{equation}
 \kappa\left(\vec{g}\right) = \sum_i \kappa_1\left(g_i\right) + \sum_{i,j} \kappa_2\left(g_i,g_j\right) + \sum_{i,j,k}\kappa_3\left(g_i,g_j,g_k\right) + \text{...}\,,
\end{equation}
where the $\kappa_i$ are polynomials in their respective arguments.
Offsets $\kappa_0$ from squared irreducible background diagrams could be
considered, but are obviously independent of $\vec{g}$ and merely add to the
background normalization, so we just subtract them
from the scans, while keeping all interference effects (appearing as terms linear
in the $g_i$ in $\kappa_1$) for completeness.
Usually this series terminates after $\kappa_2$, which becomes obvious when
applying the narrow width approximation, where additional coupling effects
cancel in the interplay of the decay vertex insertion, width dependence and phase space
integration. This basically leads to the quadratic form in the on-shell approach.
However, in the special case of
single tops, production and decay are interrelated via the same set of
CC~couplings, thus affecting production as well as decay distributions,
which in combination with the detector acceptance cuts might lead to
deviations from the on-shell result in some regions of the parameter
space\footnotemark.
\footnotetext{
Note that this is a~qualitative difference to QCD (LHC) or NC (ILC)
$t\bar{t}$~production, where production and decay are affected by different
sets of anomalous couplings.
}

\begin{figure*}
 \includegraphics[scale=0.62]{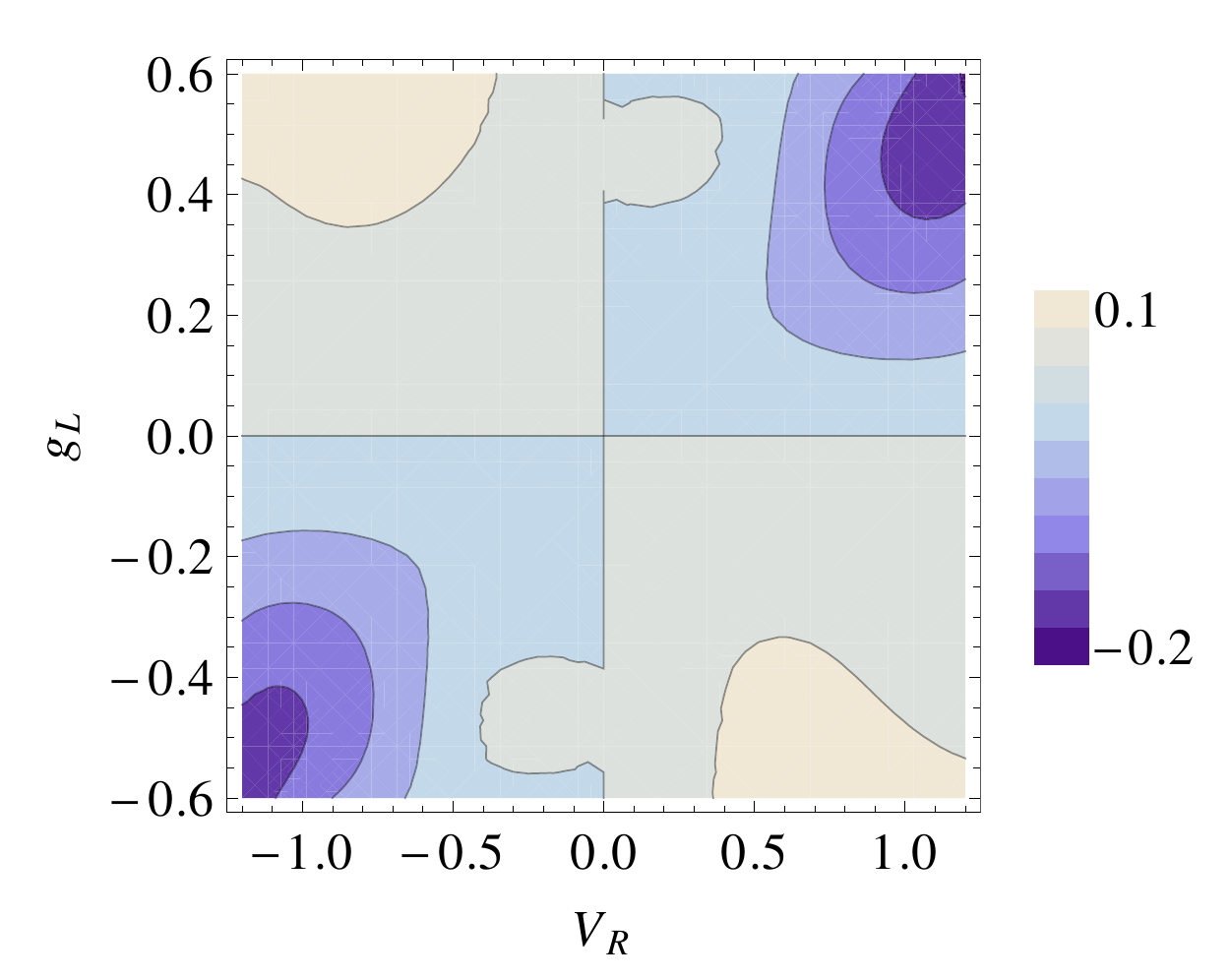}
 \includegraphics[scale=0.62]{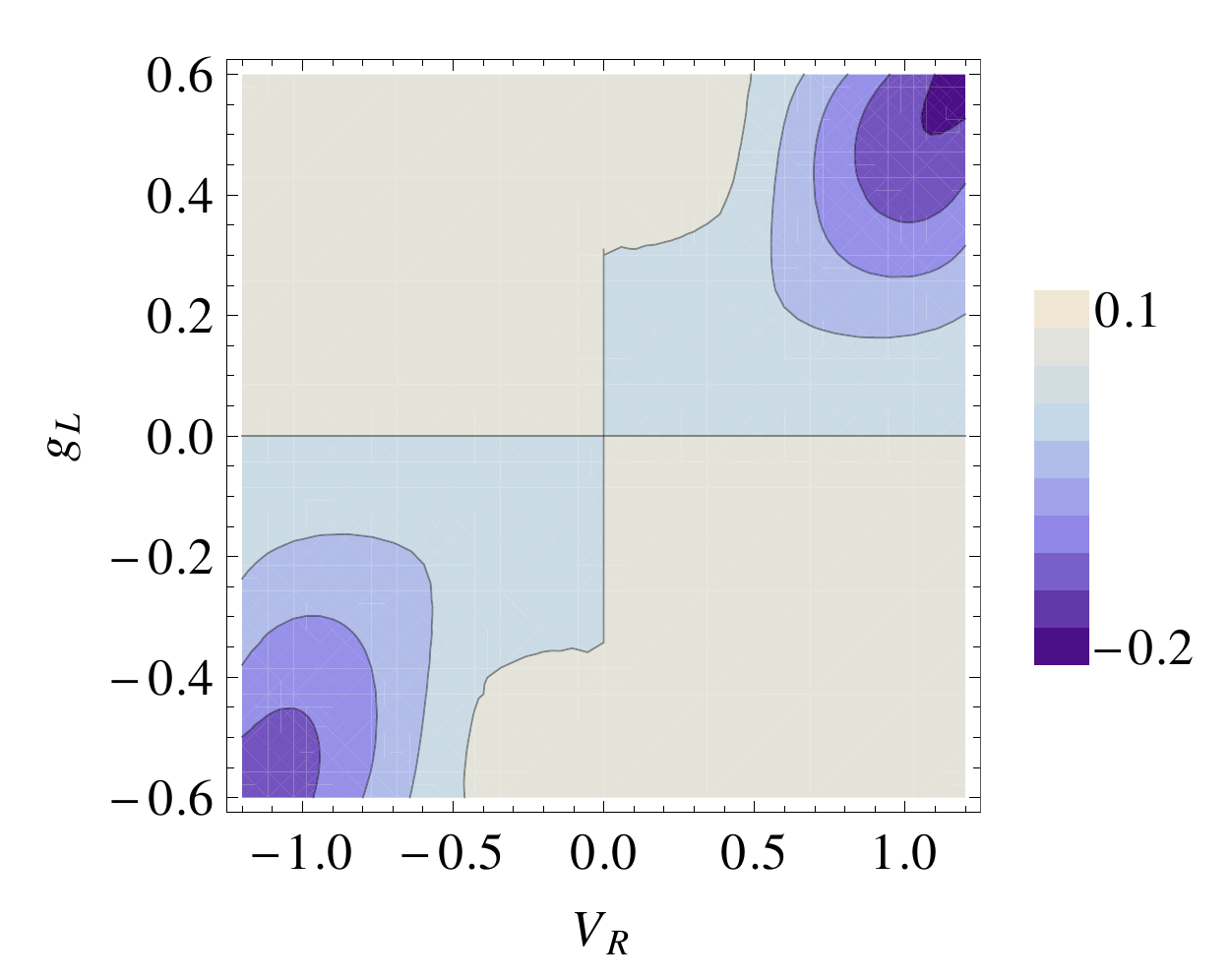}
 \includegraphics[scale=0.62]{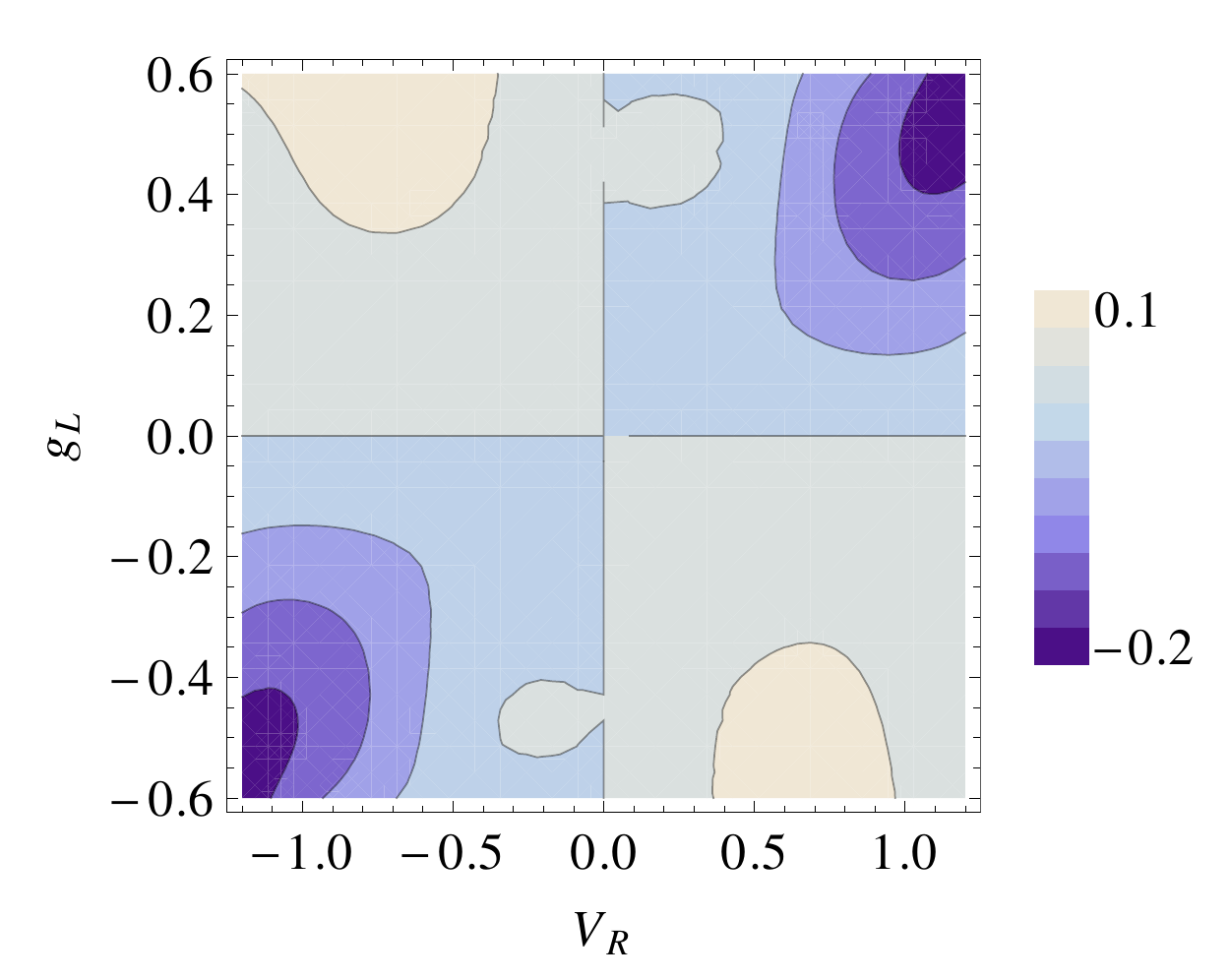}
 \includegraphics[scale=0.62]{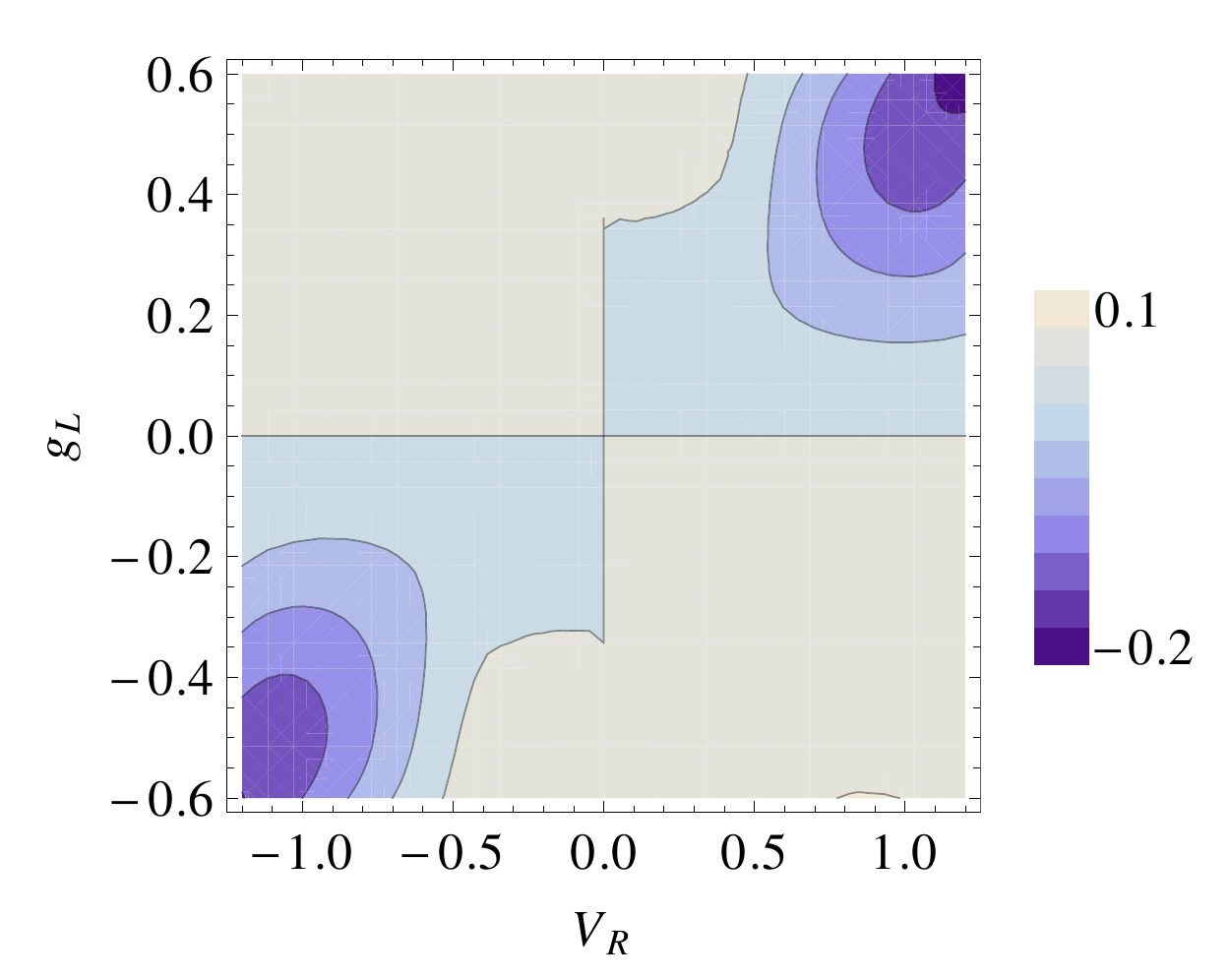}
 \caption{$\Delta\kappa|_{V_L=1.2}-\Delta\kappa|_{V_L=0.2}$
in the $\bar{t}j$ process (left) and $\bar{t}bj$ process (right),
based on full matrix elements (top) and resonant diagrams (bottom).
\label{dk_VRgL}}
\end{figure*}

In order to estimate the size of the moments $\kappa_3$,
in our cross section scans
we consider \mbox{2-dimensional} subplanes $\left(g_i,g_j\right)$
among the anomalous couplings $(V_R,g_L,g_R)$
for different fixed values of $V_L$. After subtracting all the lowest moments,
\begin{equation}
 \Delta\kappa(\vec{g}) = \kappa(\vec{g}) - \sum_{k=i,j,V_L} \kappa_1(g_k) - \sum_{k=i,j} \kappa_2(g_k,V_L) \,,
\end{equation}
within the on-shell picture
the remaining contribution $\Delta\kappa(\vec{g})\sim\kappa_2(g_i,g_j)$
should then be independent of the value of $V_L$,
otherwise it would by definition contain some $\kappa_3(g_i,g_j,V_L)$.
For illustration, we choose the coupling
subspace~$(g_i,g_j)=(V_R,g_L)$, because it is one of
the dominant interference contributions to all production processes,
and evaluate $\Delta\kappa|_{V_L=1.2}-\Delta\kappa|_{V_L=0.2}$.
The resulting picture is
mixed: while in the $s$~channel the result is practically 0 all over
the $V_R$--$g_L$ plane, in the $t$ channel
process~$\bar{t}j$, which plays a~central role because of its
comparably large cross section, this difference amounts to $\sim -0.2$
at $V_R\simeq\pm1$ and $g_L\simeq\pm0.5$ along the
interference direction $V_R\sim 2g_L$ (cf.~Fig.~\ref{dk_VRgL}), which
is of the same size as the respective on-shell contribution $\sim-0.5\times V_R\times g_L$,
(the same is true for the $V_Lg_R$ interference in the $tj$ channel).
For comparison,
we repeat the whole procedure selecting only the resonant
single top diagrams for the scan (including
the full top width dependence on~$\vec{g}$), finding that background
interference only plays a minor role.

Furthermore, note that there
are other interference directions also showing substantial effects,
e.g., in the $g_L$--$g_R$ plane of the $s$ channel, but the $V_R g_L$ interference
is the most interesting one because it is large in all channels,
and respective bounds are expected to remain rather weak also from
other experiments along the considered direction
$V_R\sim 2g_L$~\cite{AguilarSaavedra:2008gt}.
All in all, these numbers indicate strongly that the interference and
off-shell effects contained in the full matrix elements might become
important at the level of the coupling limits to be expected experimentally,
at least in a stand-alone single top cross section study at the LHC,
and should be checked in any case.

\begin{figure*}
 \includegraphics[scale=0.6]{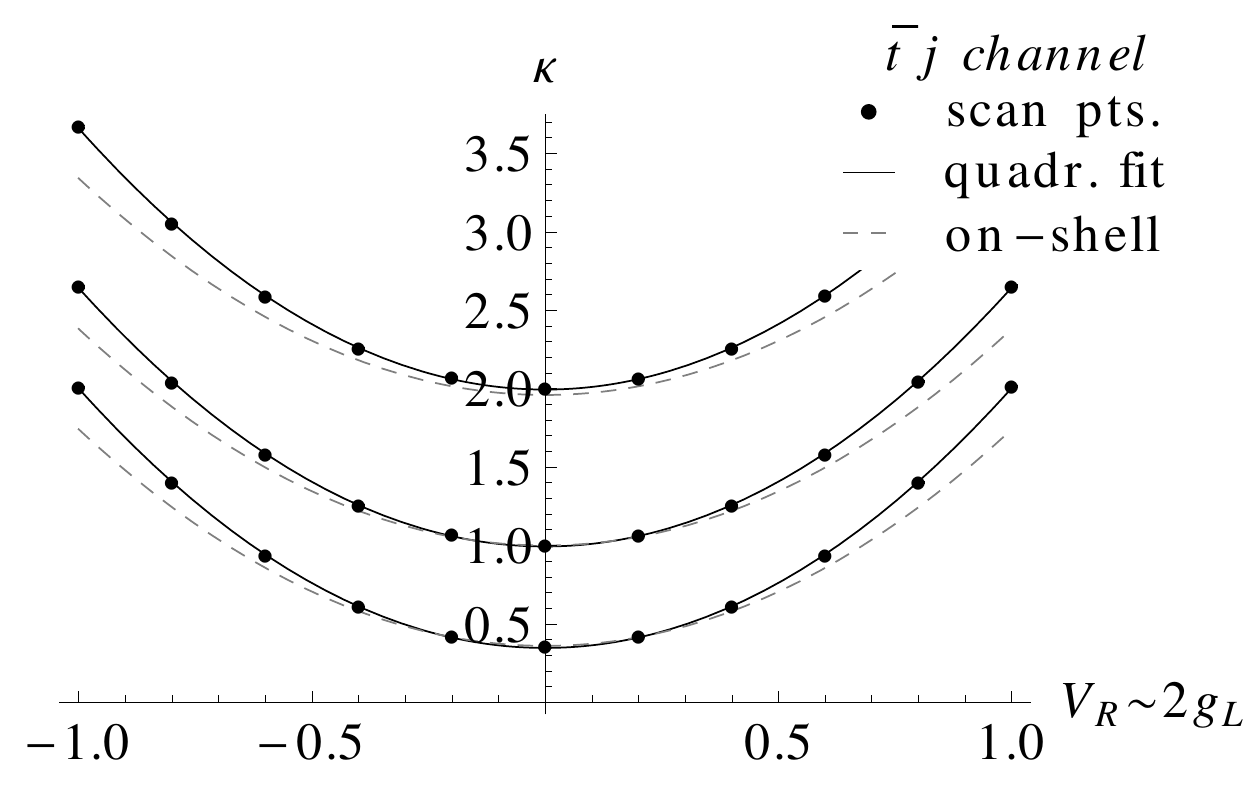}
 \hspace{0.5cm}
 \includegraphics[scale=0.6]{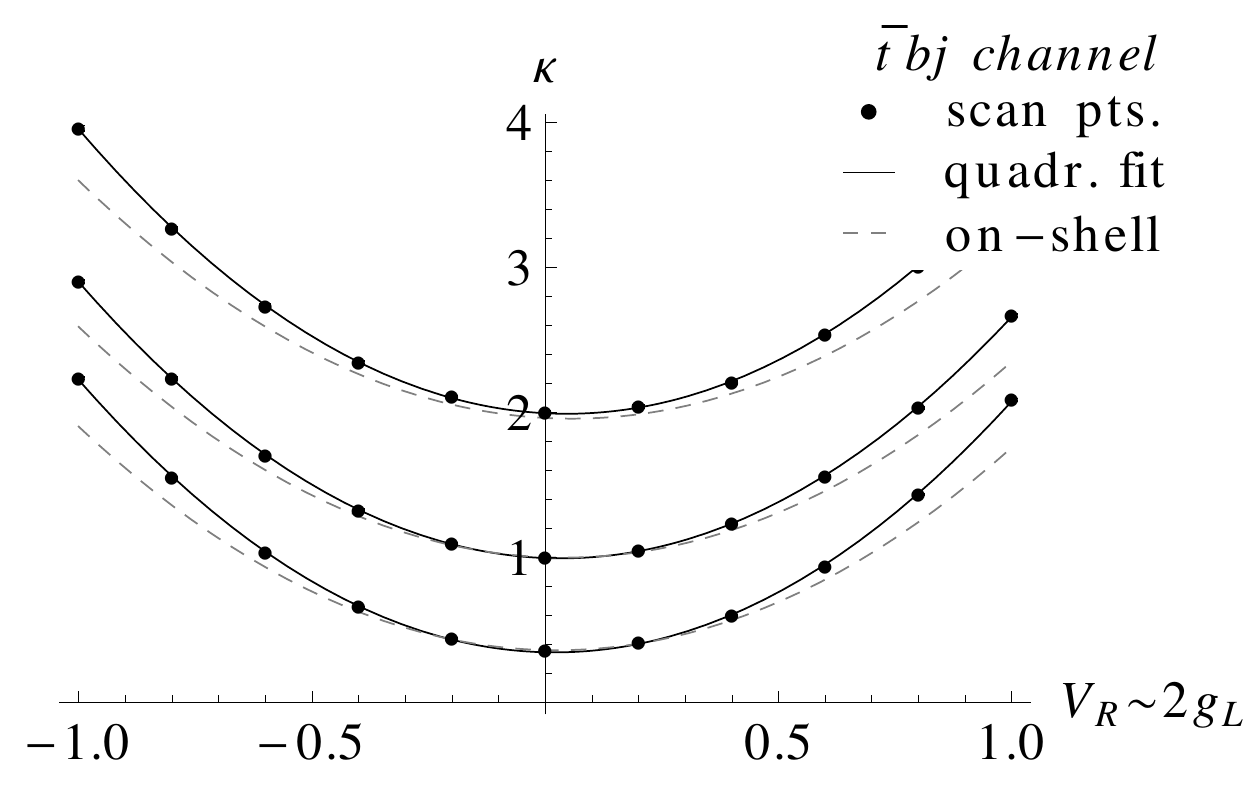}
 \caption{
$\kappa$ scan results and quadratic fits of resonant matrix elements
inside the acceptance region along the coupling direction $V_R\sim2g_L$,
for $V_L=0.6$, $1.0$ and $1.4$ (bottom to top) in the
$\bar{t}j$~channel (left) and the $\bar{t}bj$~channel (right).
Grey dashed lines indicate the on-shell $\kappa$ function.
\label{k_VR_2gL}}
\end{figure*}

In order to further quantify this effect and exclude potential artefacts from
unphysical regions in the parameter space, we now systematically scan the $V_R g_L$
interference along the direction $V_R\sim 2g_L$ as a function of $V_L$,
including resonant off-shell diagrams for anomalous single top production and decay.
We observe that the scans do show a quadratic dependence on $V_R$ and $g_L$
to a very good approximation at any value of $V_L$ (cf.~Fig.~\ref{k_VR_2gL}),
so the respective coefficients $\kappa_{V_R}\sim V_R^2$, $\kappa_{g_L}\sim g_L^2$ and
$\kappa_{V_R g_L}\sim V_R\,g_L$ can be extracted from quadratic fits along the
axes $V_R$, $g_L$ and $V_R=2g_L$, for each fixed value of $V_L$
and $g_R=0$.
This is done for the full phase space integration as well as for the
acceptance region defined in Eq.~\eqref{acc}.
As illustrated in Fig.~\ref{k_tjc}
for the $t$ channel processes, when integrating over the full phase space the
deviations from the on-shell result are very small as expected from the narrow width
approximation, whereas application of the acceptance cuts leads to substantially
different fit coefficients which also depend on the value of~$V_L$.

\begin{figure*}
 \includegraphics[scale=0.6]{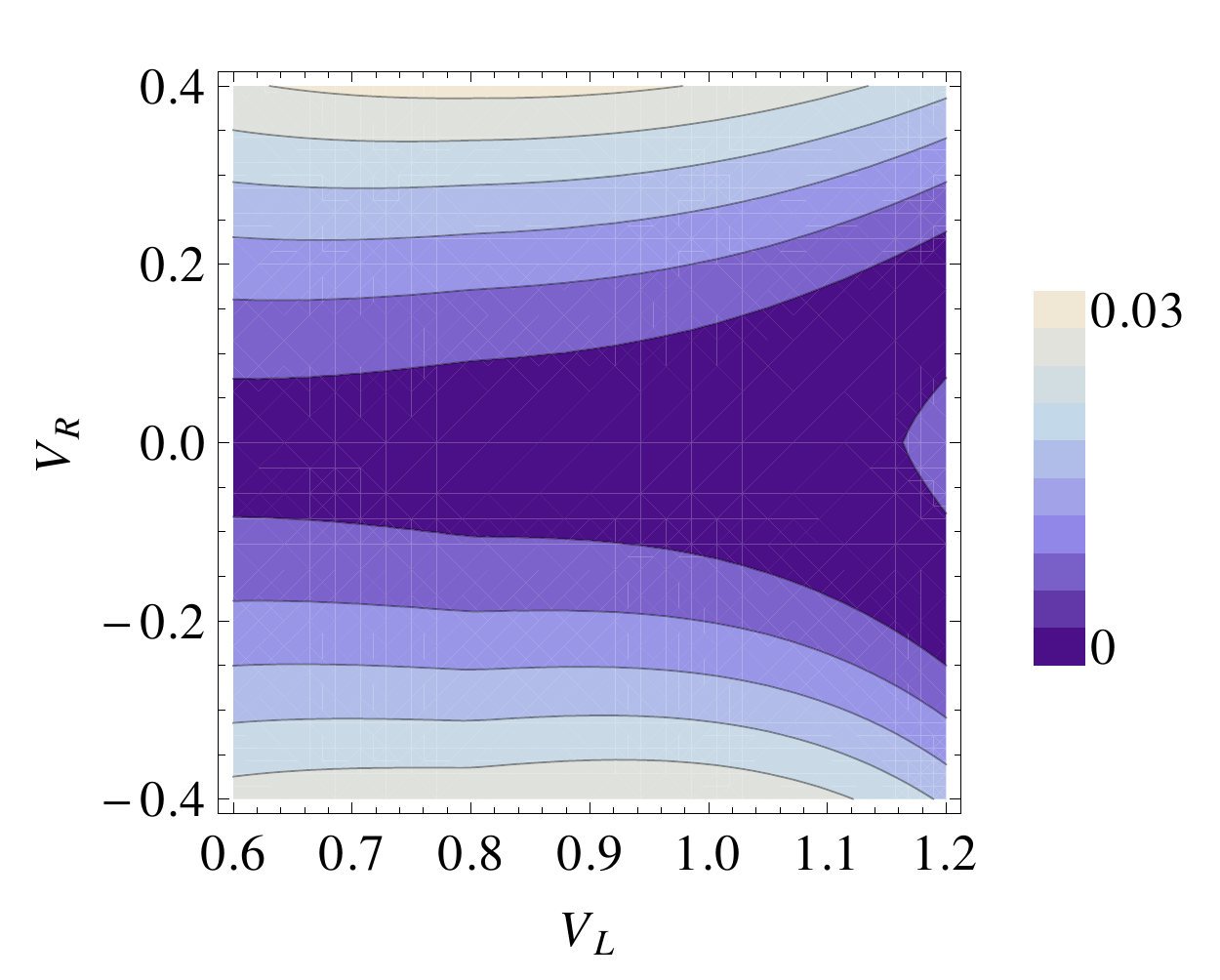}
 \hspace{0.5cm}
 \vspace{0.5cm}
 \includegraphics[scale=0.6]{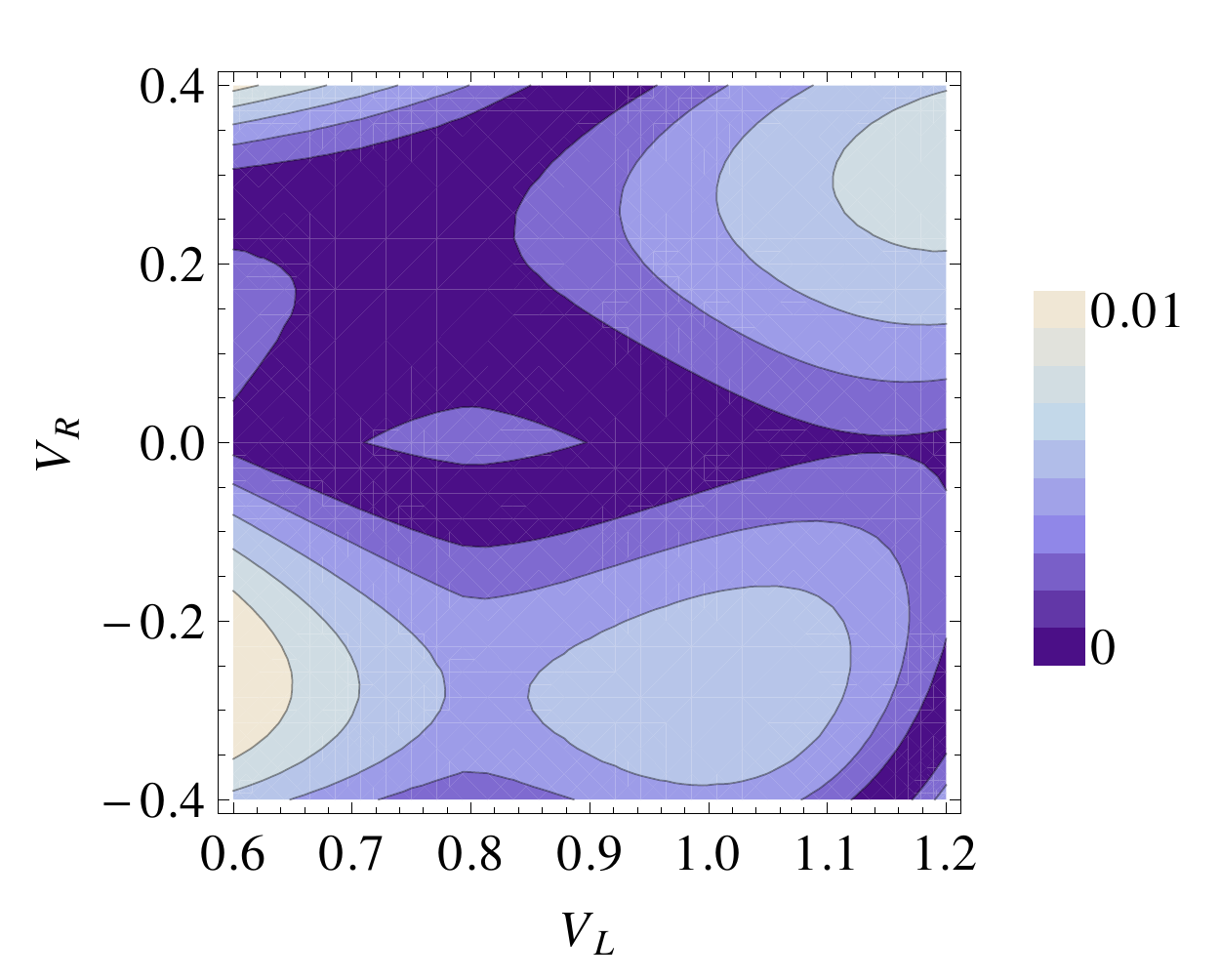}
 \vspace{0.5cm}
 \includegraphics[scale=0.6]{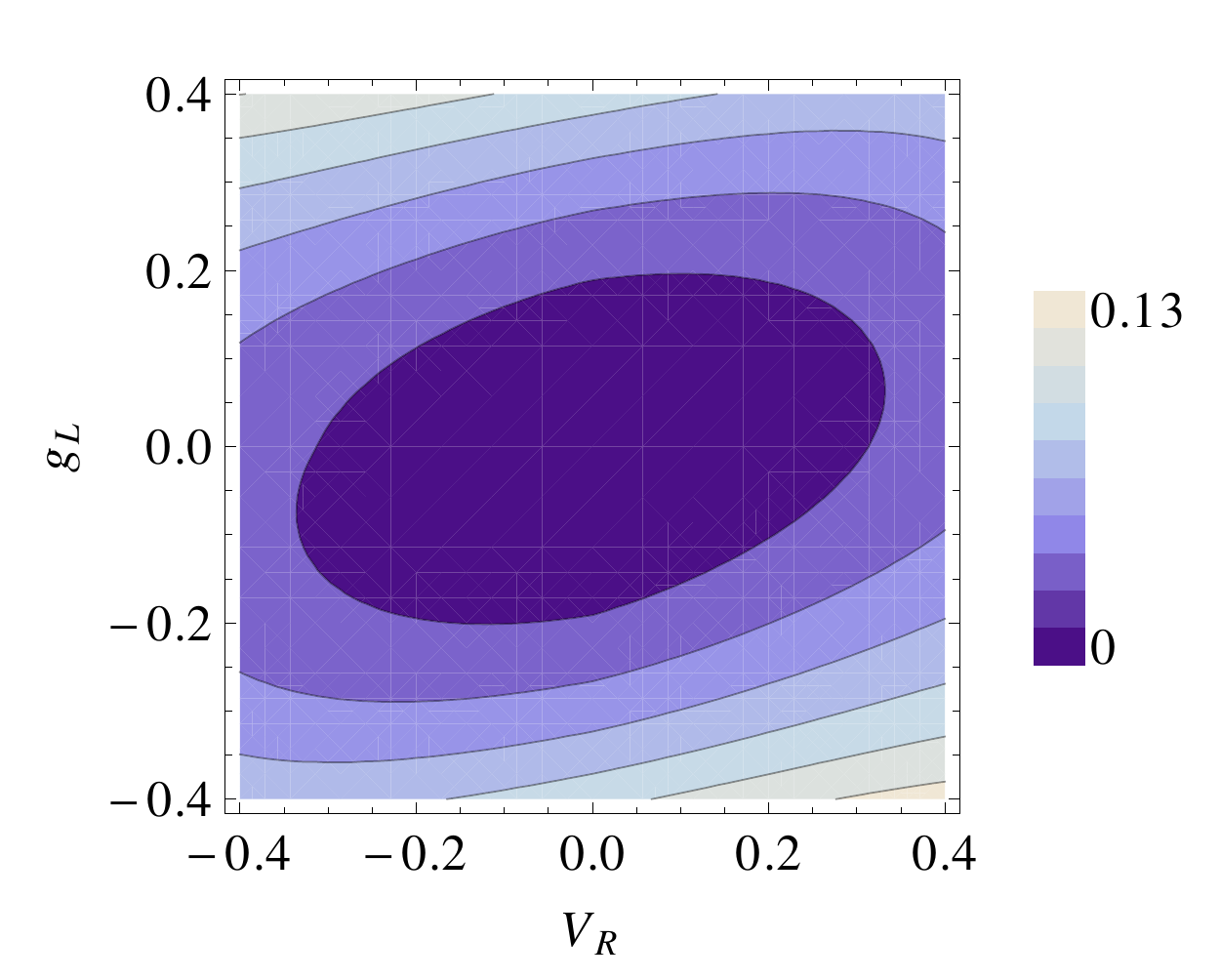}
 \hspace{0.5cm}
 \includegraphics[scale=0.6]{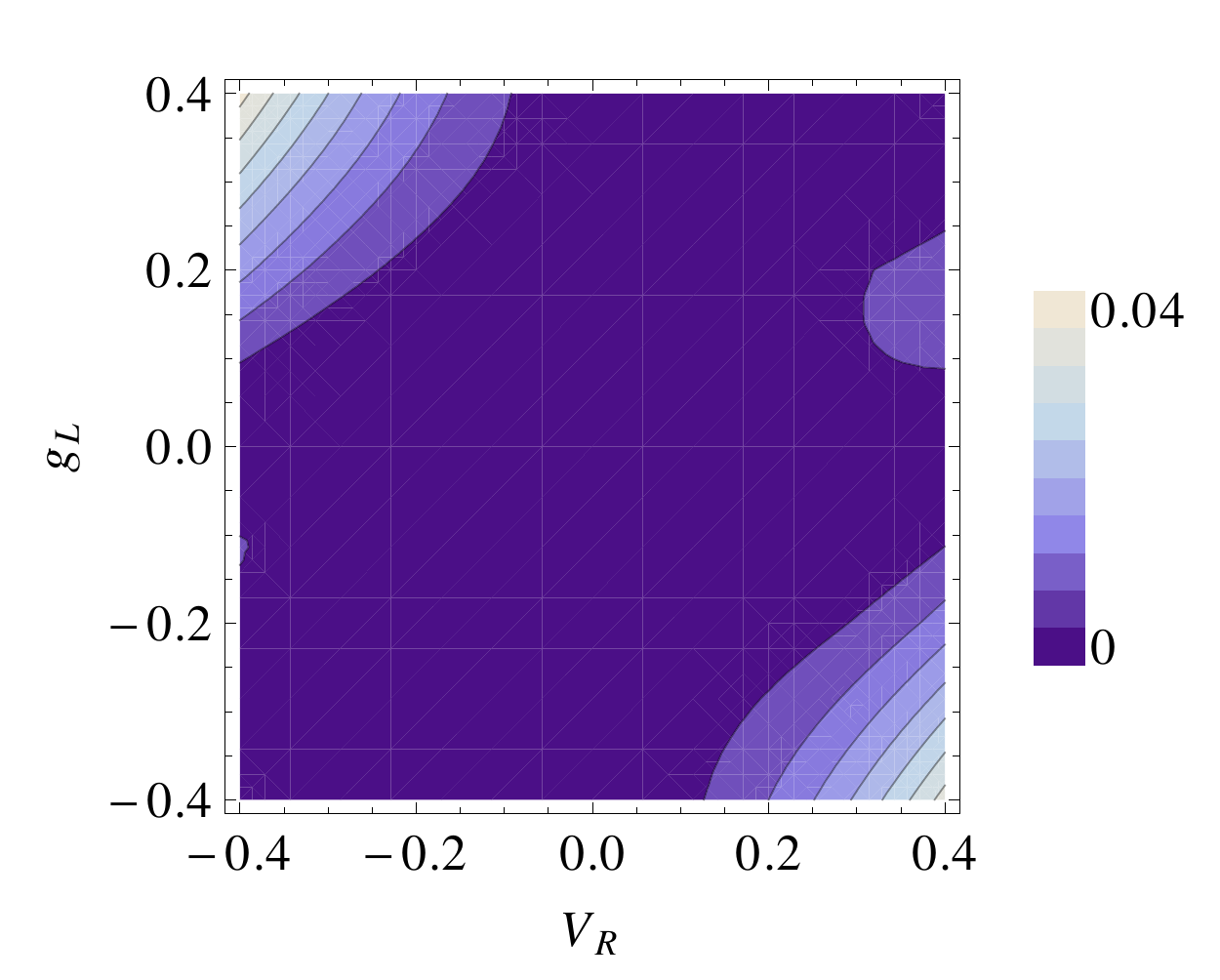}
 \includegraphics[scale=0.6]{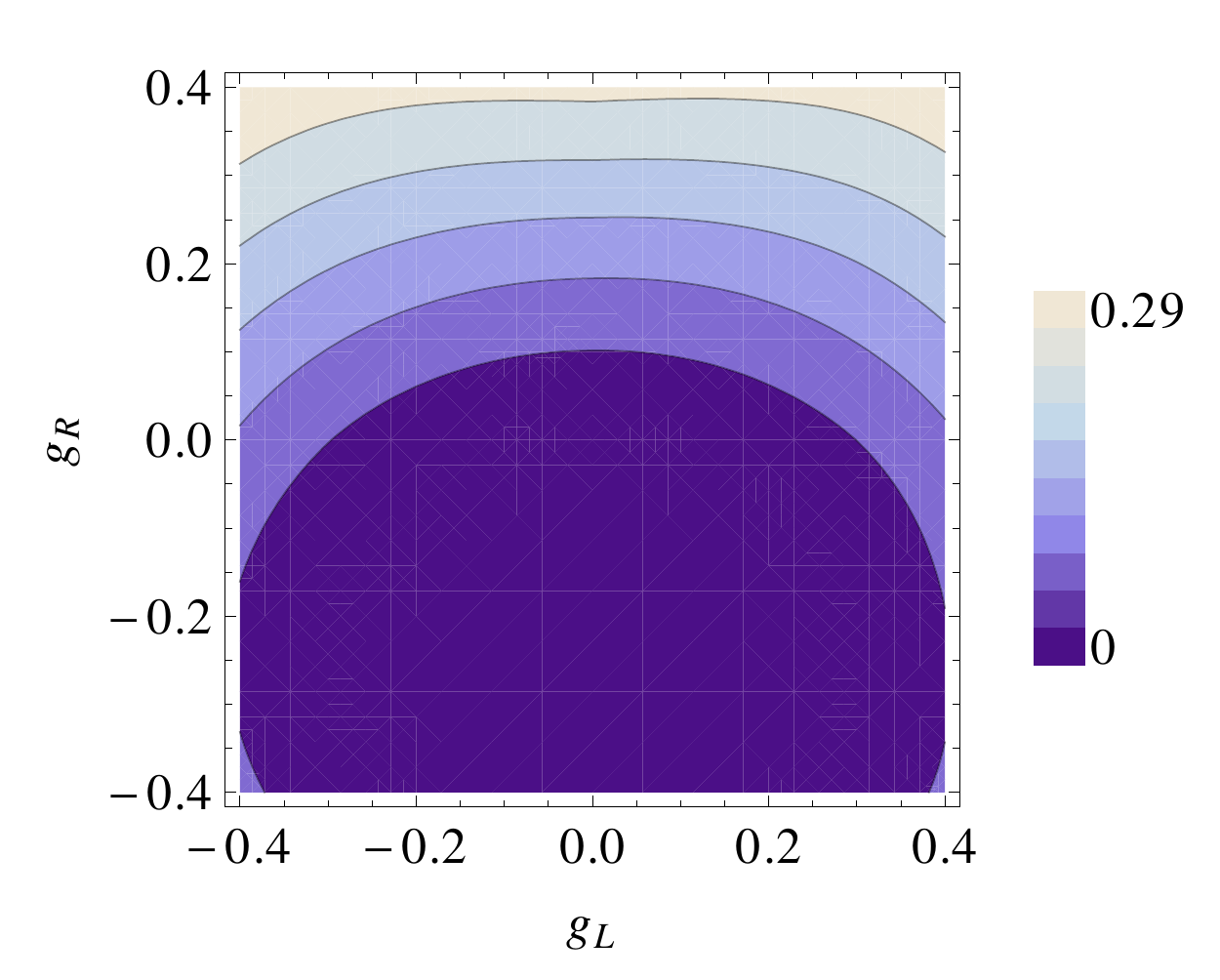}
 \hspace{0.5cm}
 \includegraphics[scale=0.6]{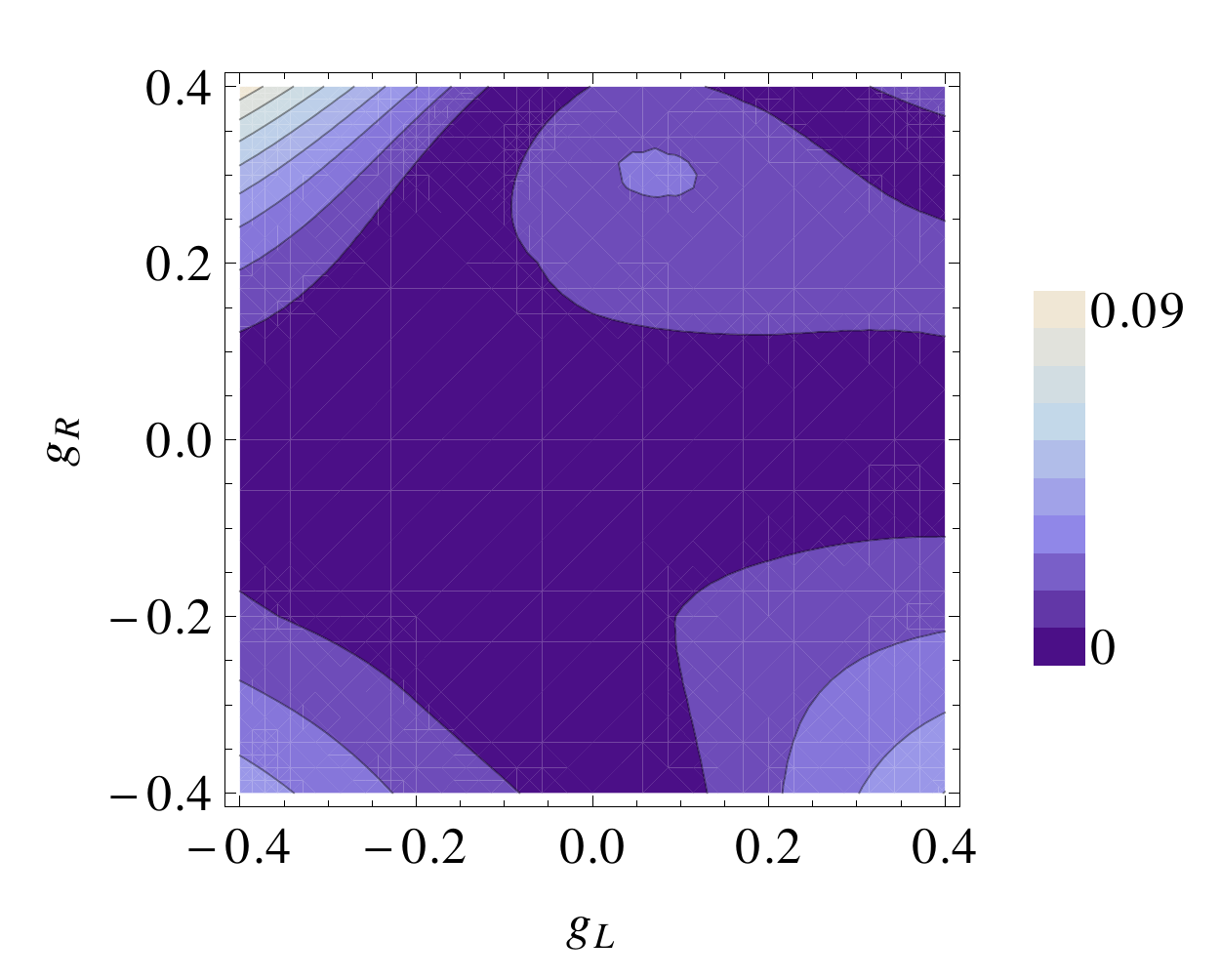}
 \caption{
The differences $\left|\kappa_\text{full}-\kappa_\text{on}\right|$
(left) and $\left|\kappa_\text{full}-\kappa_\text{fit}\right|$ (right)
in various coupling planes of the $\bar{t}j$ process.
Note that
the heat scales on the left are of the size of the expected experimental
sensitivity to the $t$~channel ($\sim\unit[13]{\%}$),
and significantly decrease on the right.
\label{tjc_on_off}}
\end{figure*}

\begin{figure*}
 \includegraphics[scale=0.6]{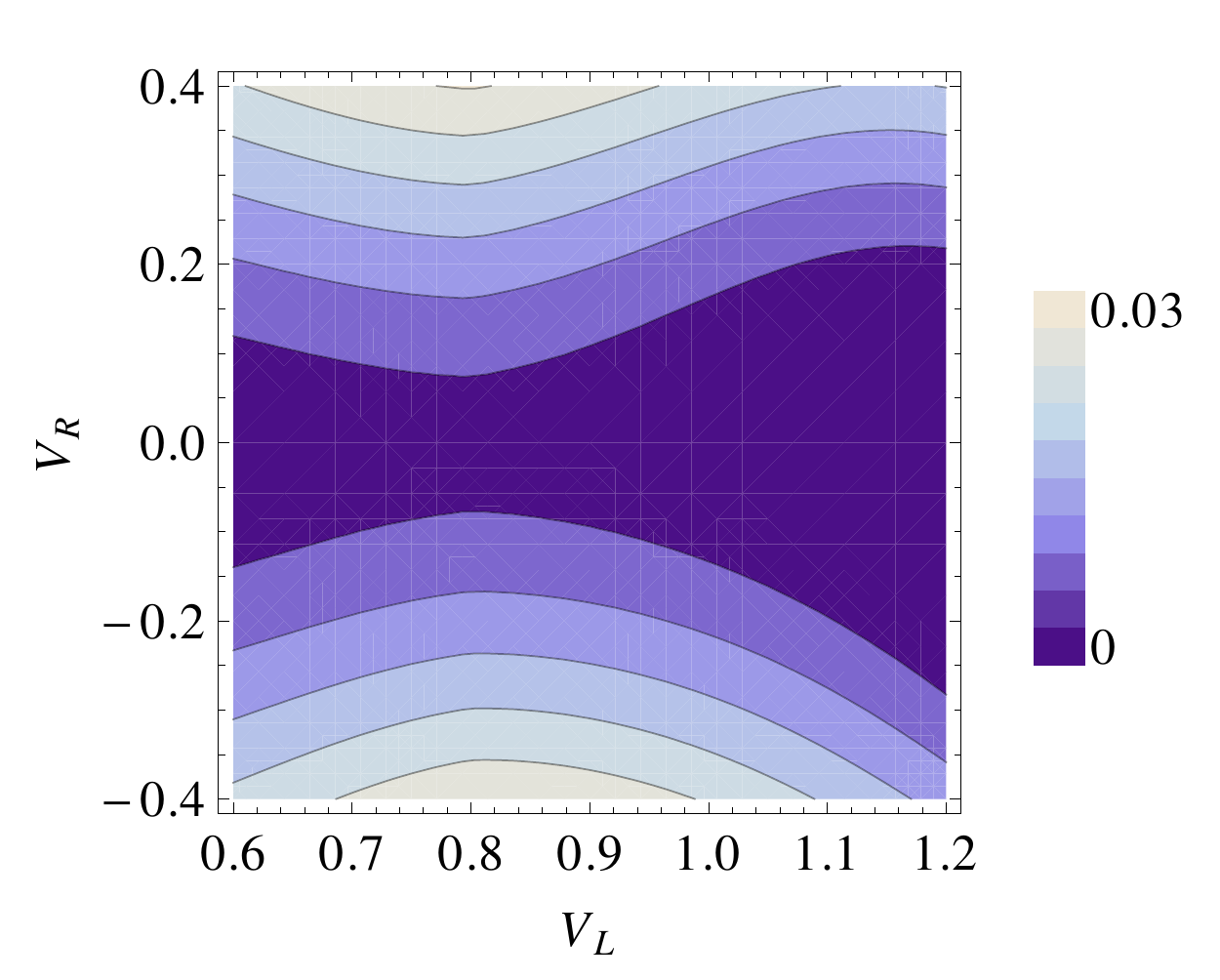}
 \hspace{0.5cm}
 \vspace{0.5cm}
 \includegraphics[scale=0.6]{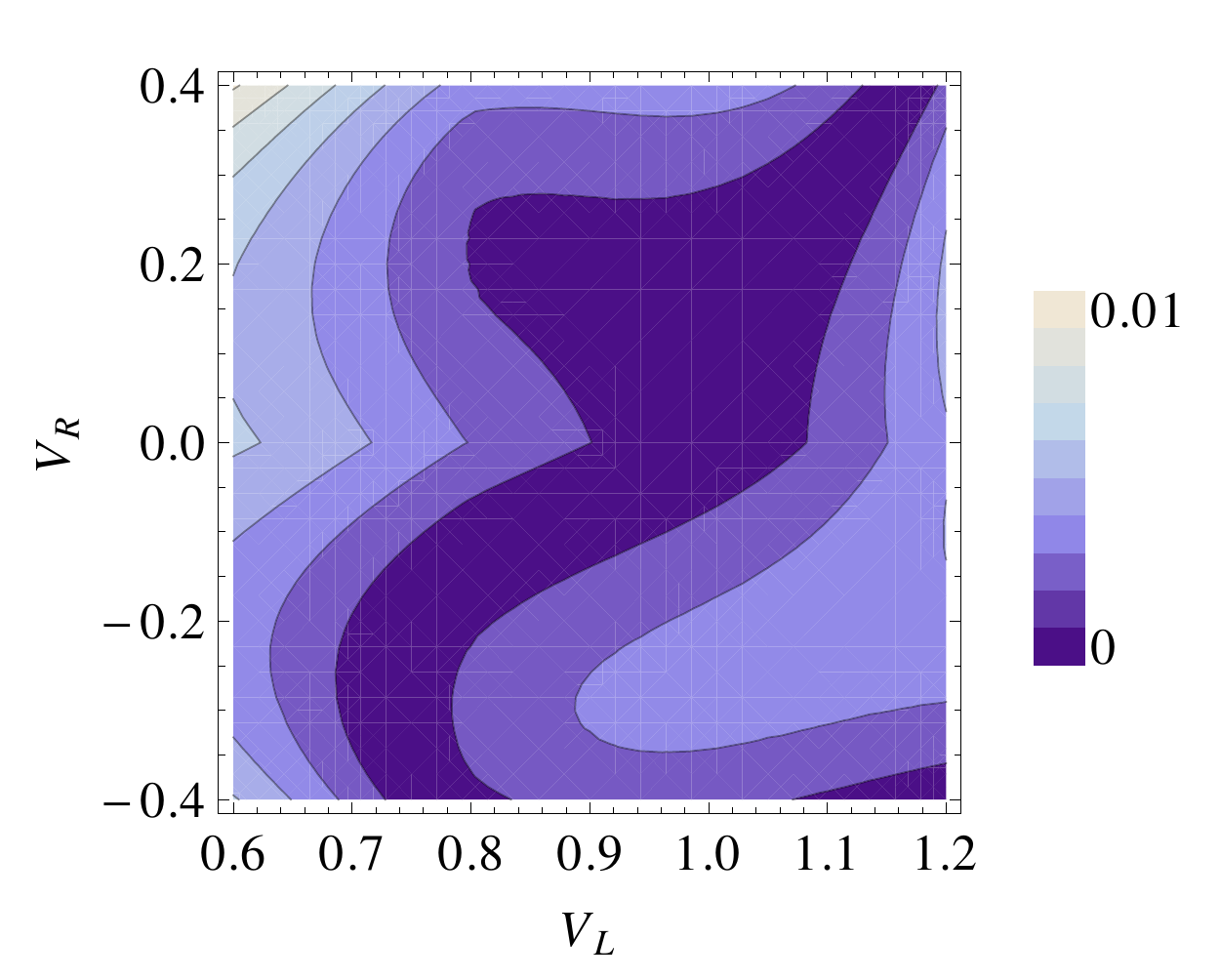}
 \vspace{0.5cm}
 \includegraphics[scale=0.6]{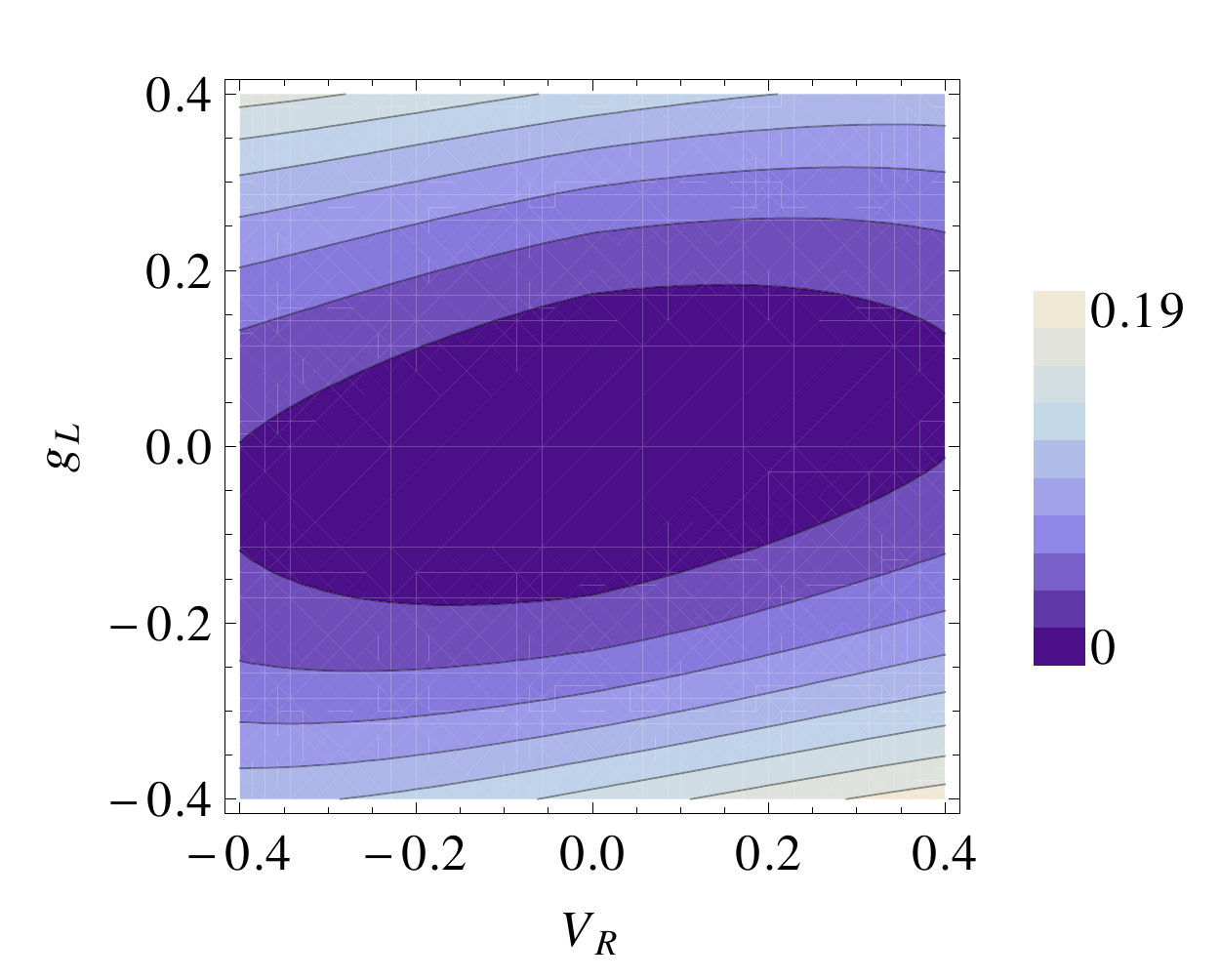}
 \hspace{0.5cm}
 \includegraphics[scale=0.6]{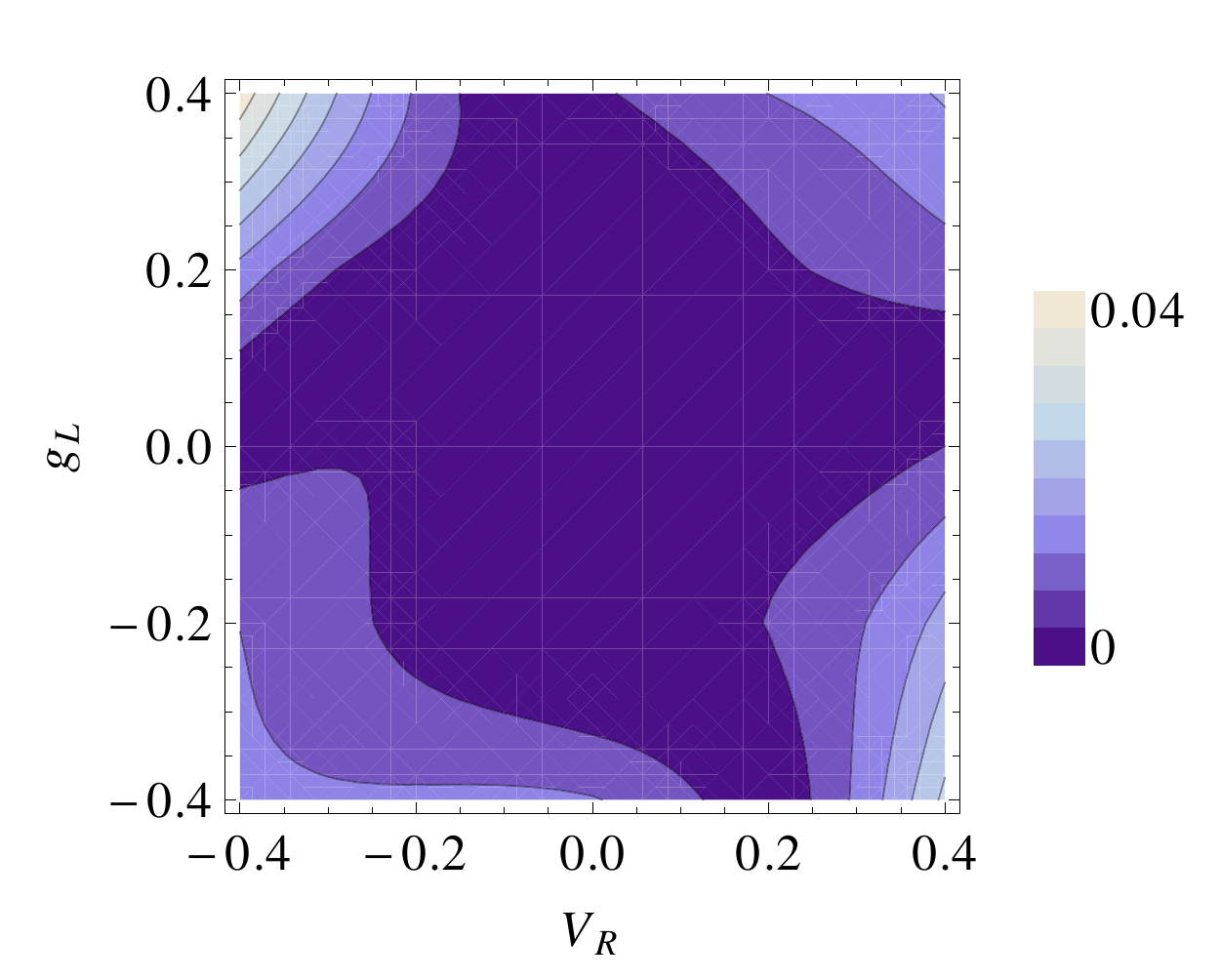}
 \includegraphics[scale=0.6]{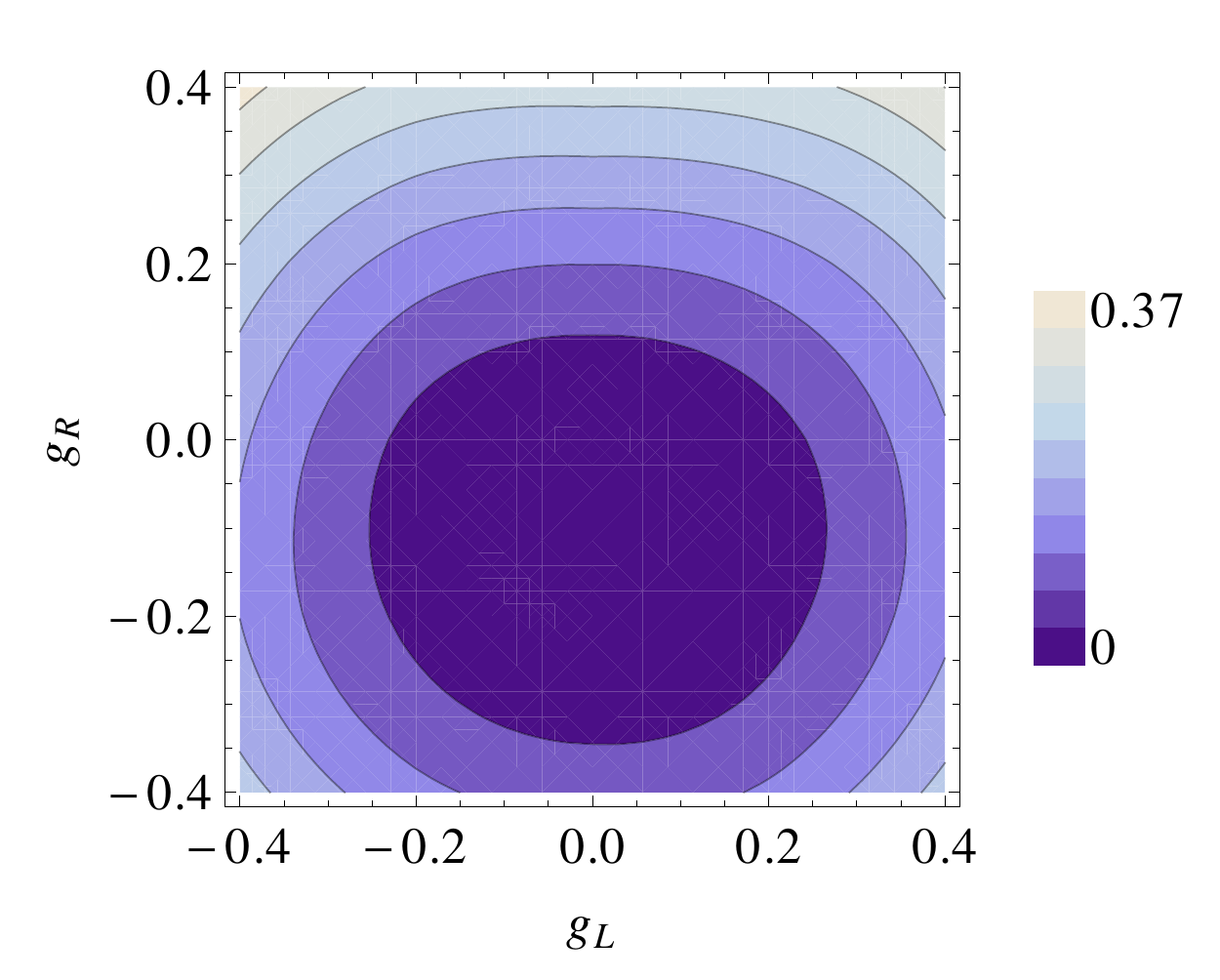}
 \hspace{0.5cm}
 \includegraphics[scale=0.6]{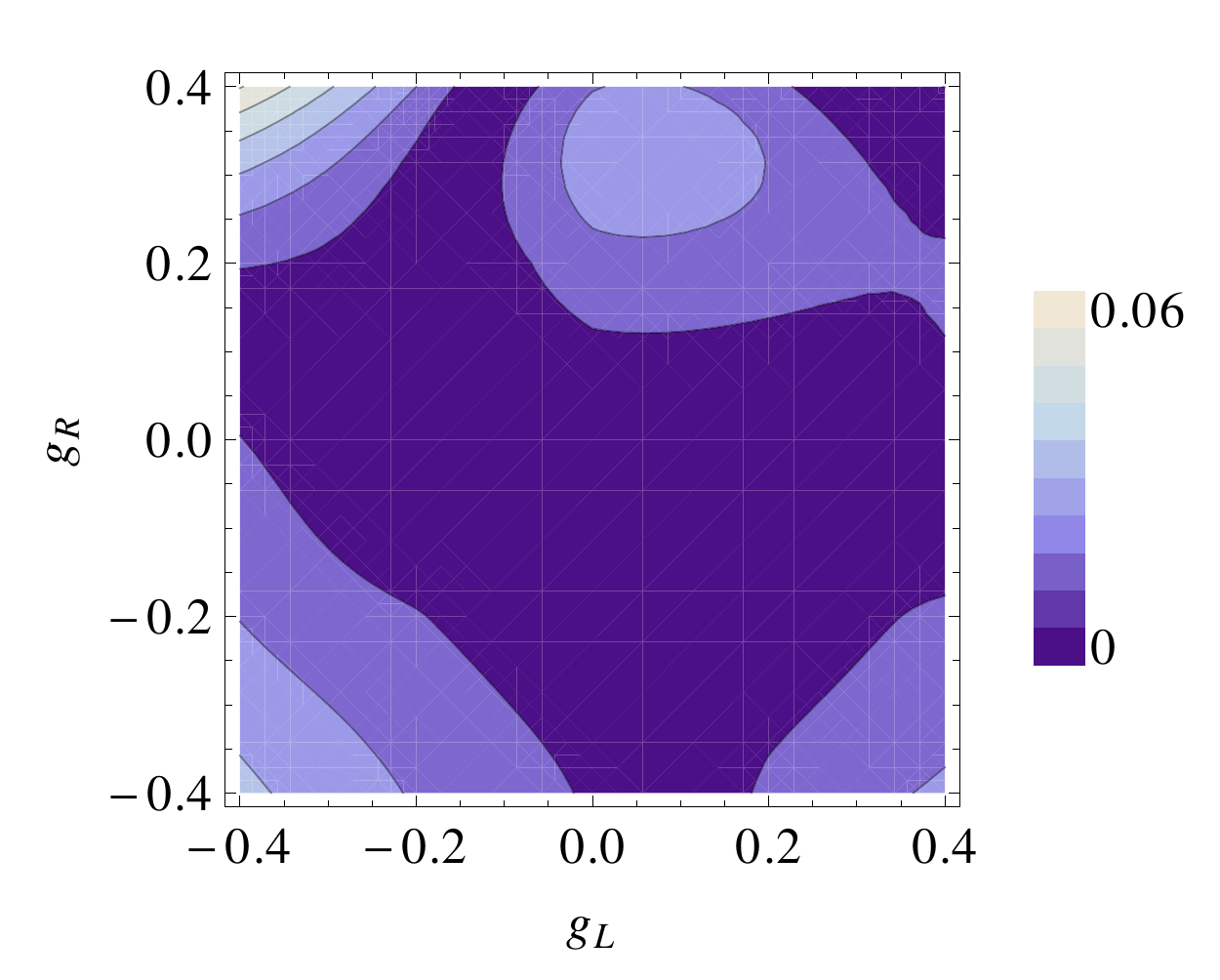}
 \caption{
The differences $\left|\kappa_\text{full}-\kappa_\text{on}\right|$
(left) and $\left|\kappa_\text{full}-\kappa_\text{fit}\right|$ (right)
in various coupling planes of the $\bar{t}bj$ process
(note the heat scales, cf.~discussion in~Fig.~\ref{tjc_on_off}).
\label{tbjc_on_off}}
\end{figure*}

\begin{figure*}
 \includegraphics[scale=0.55]{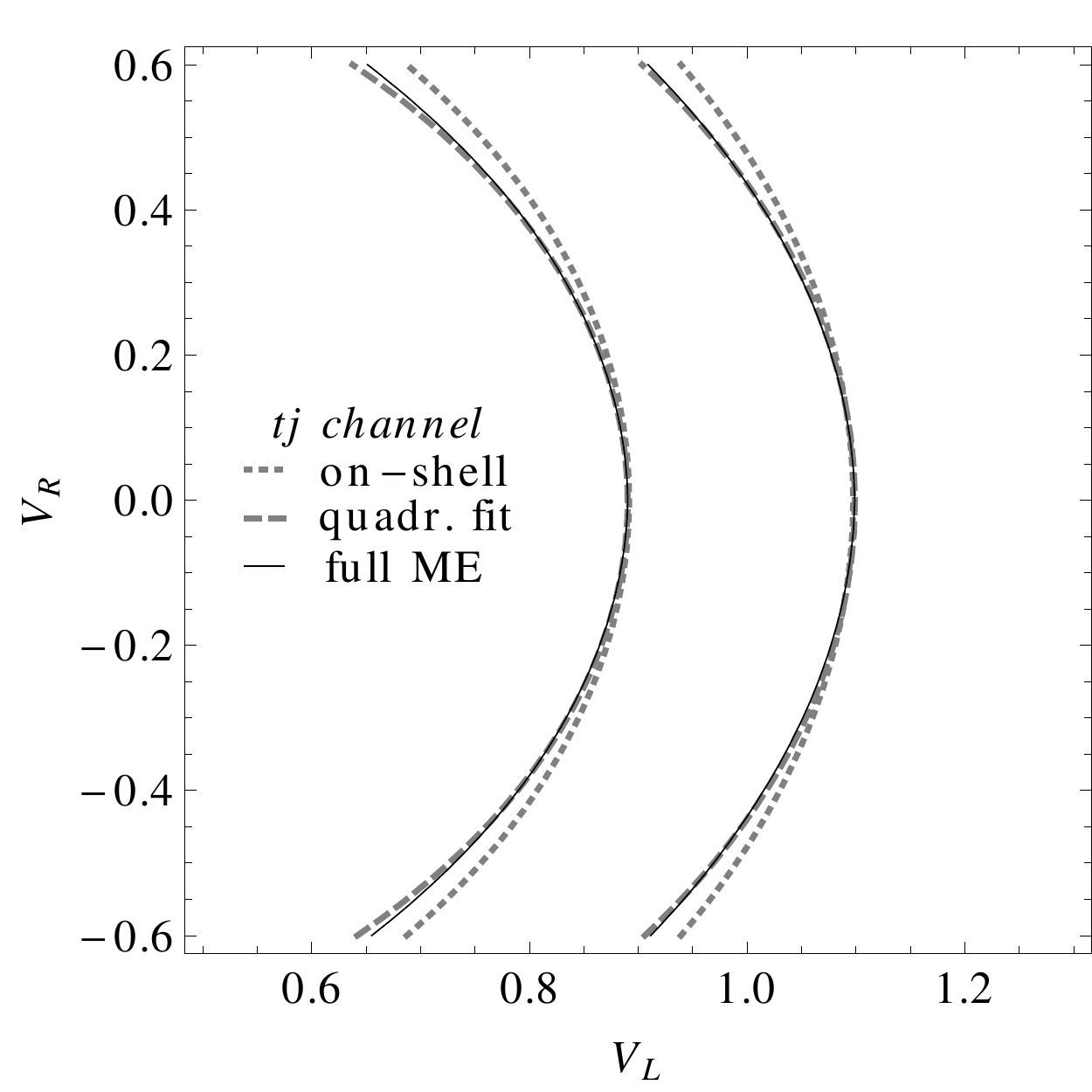}
 \hspace{0.3cm}
 \includegraphics[scale=0.55]{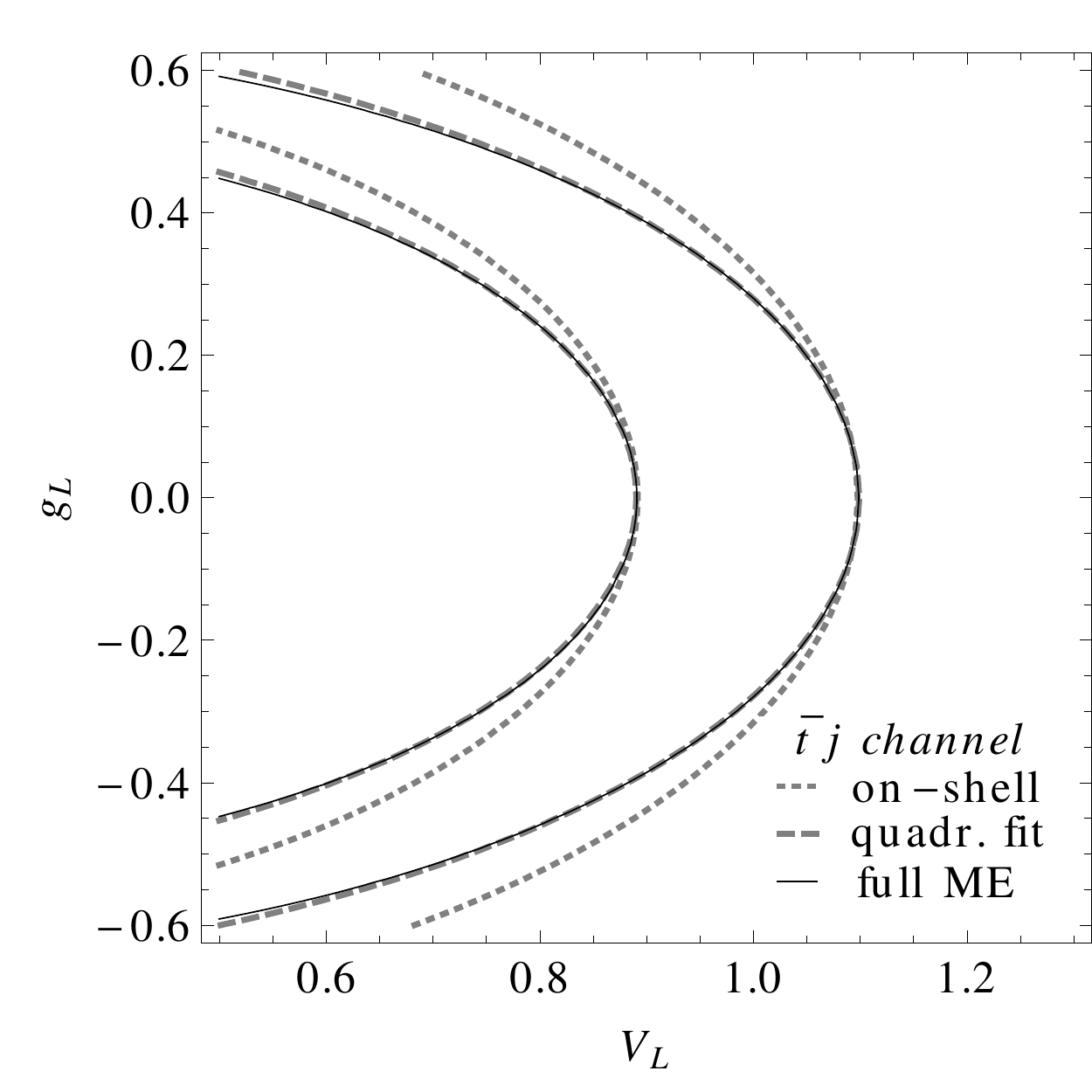}
 \includegraphics[scale=0.55]{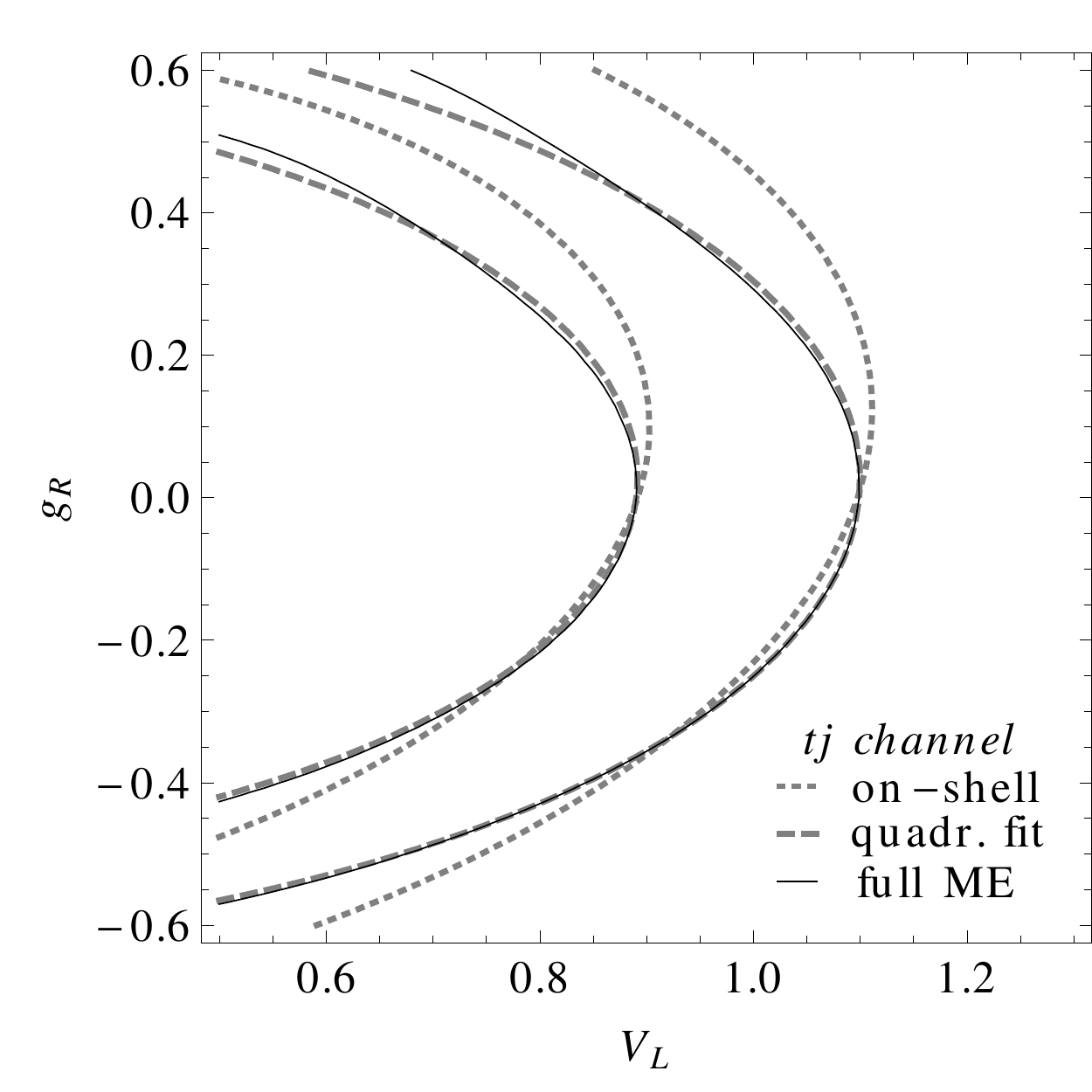}
 \hspace{0.3cm}
 \includegraphics[scale=0.55]{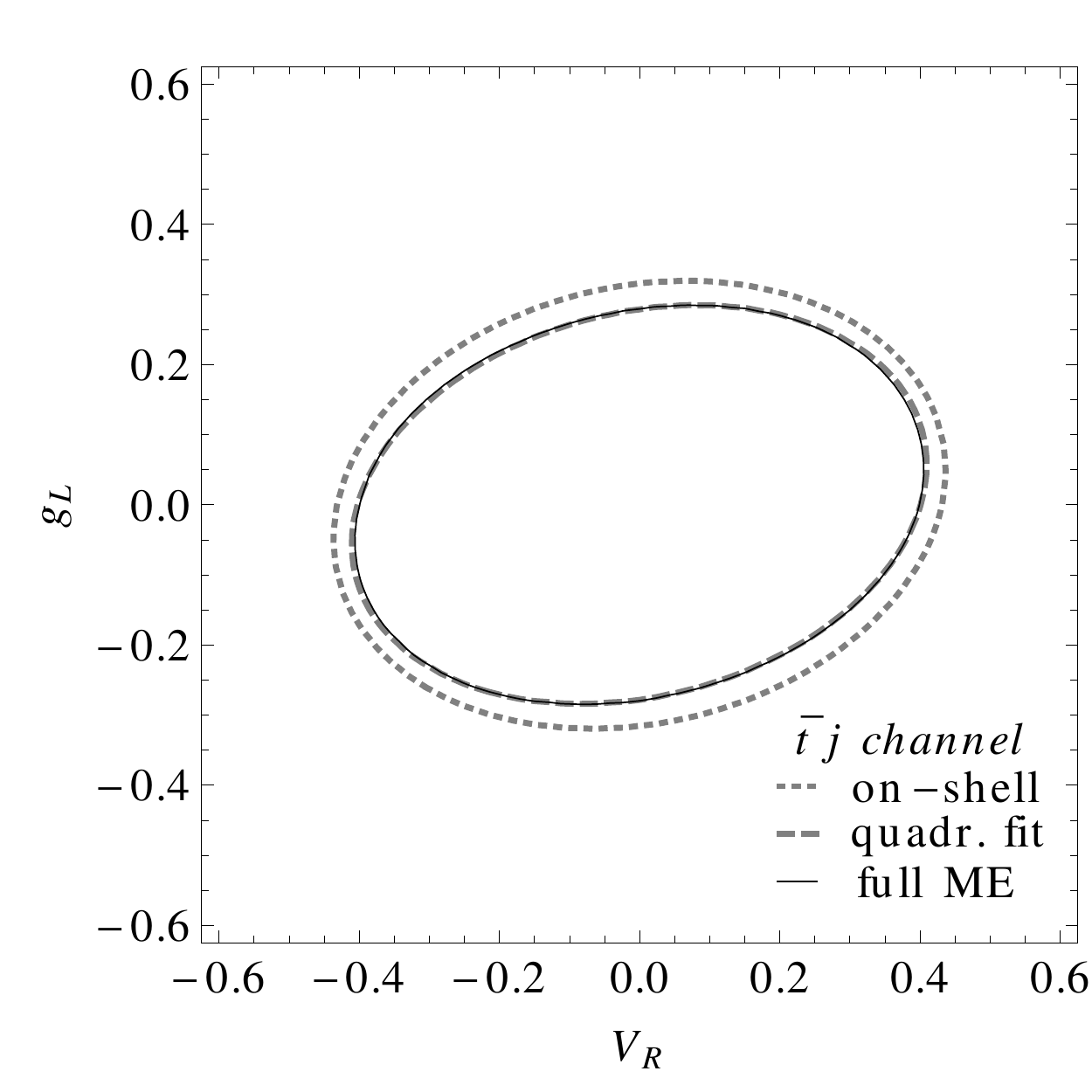}
 \includegraphics[scale=0.55]{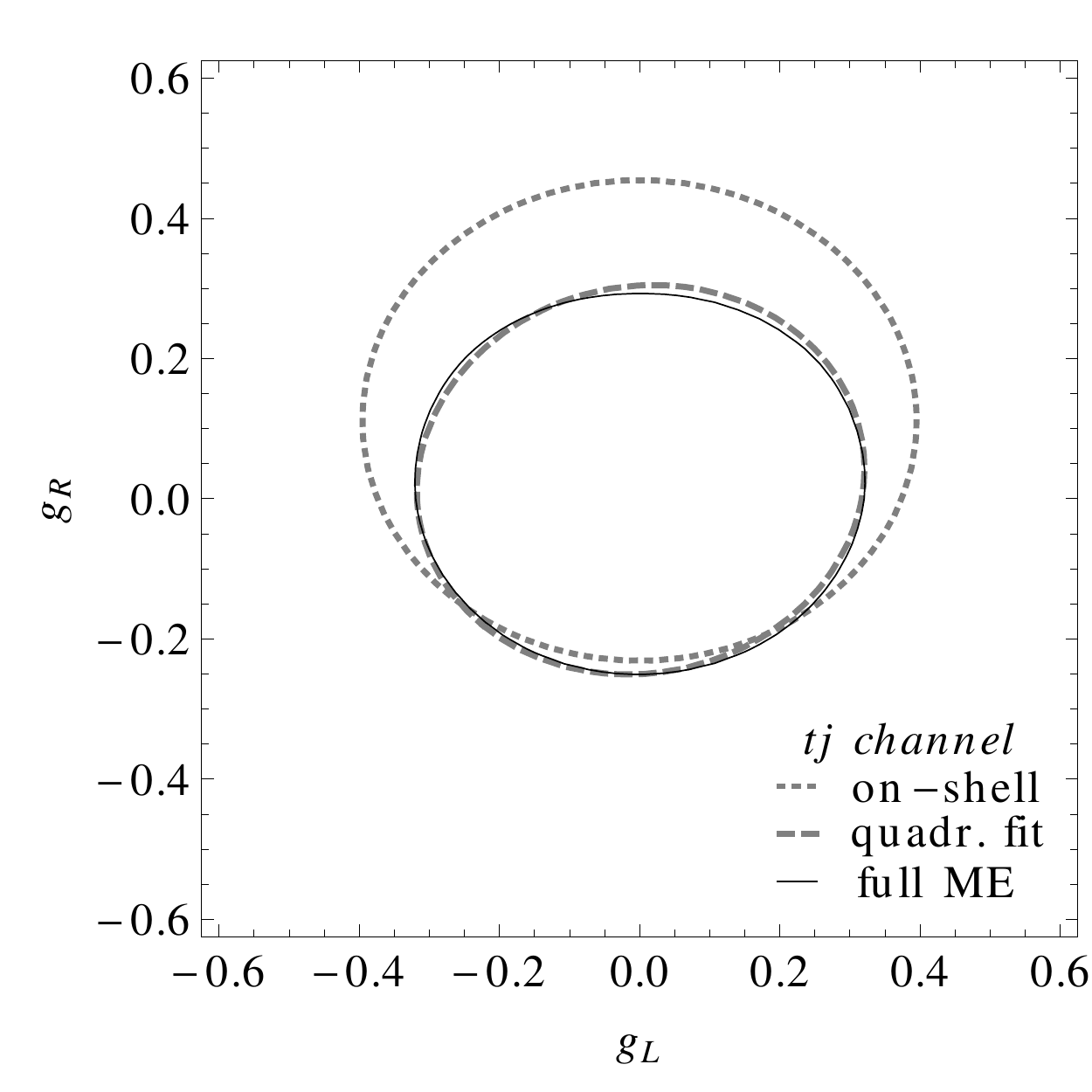}
 \hspace{0.3cm}
 \includegraphics[scale=0.55]{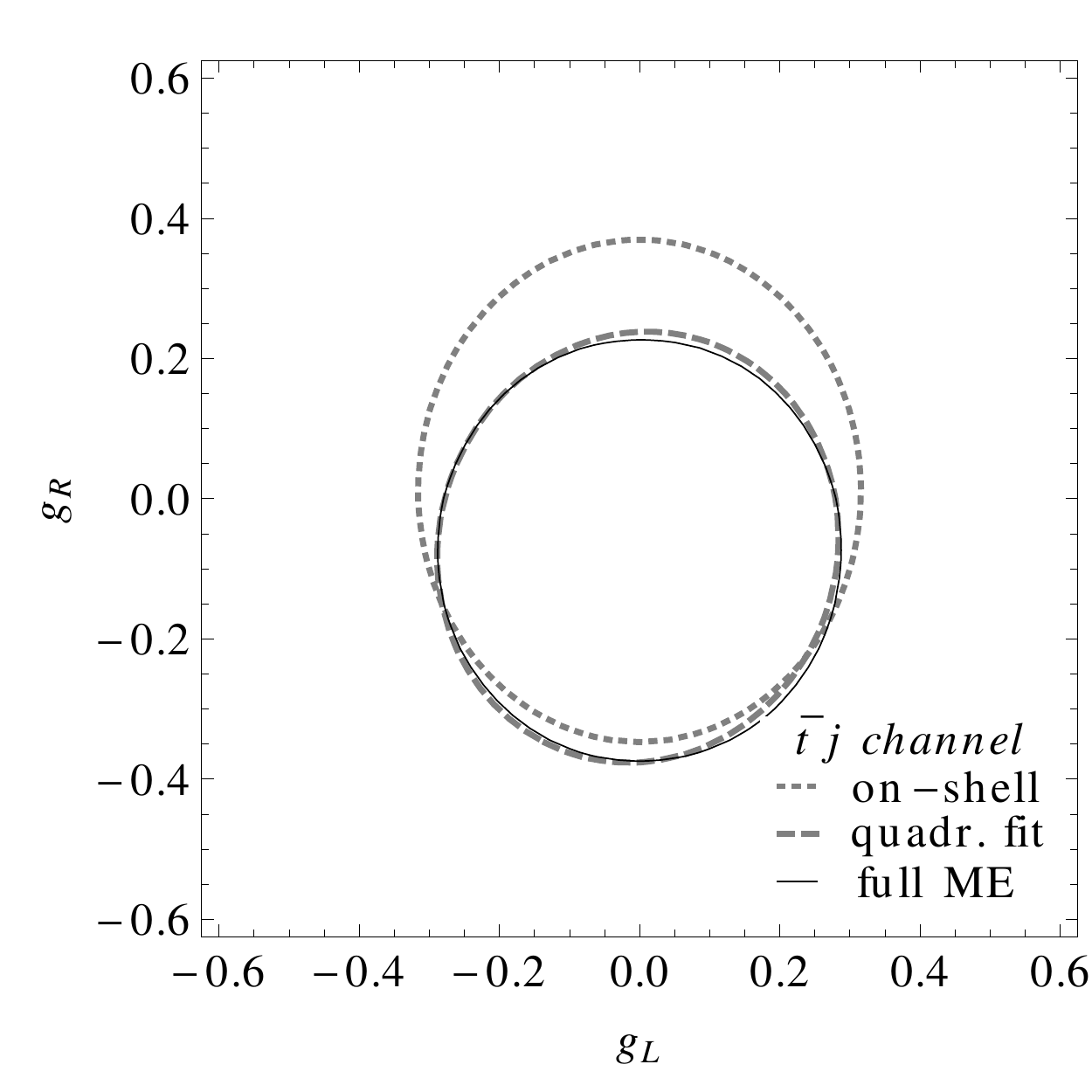}
 \caption{Comparison of the $1\sigma$ contours for the various matrix element
response functions $\kappa_\text{on}$, $\kappa_\text{fit}$ and
$\kappa_\text{full}$ at parton level, for $tj$ and $\bar{t}j$ production processes
in different coupling planes (setting the others to their SM values).
\label{tj_1sigma_part}}
\end{figure*}

\begin{figure*}
 \includegraphics[scale=0.55]{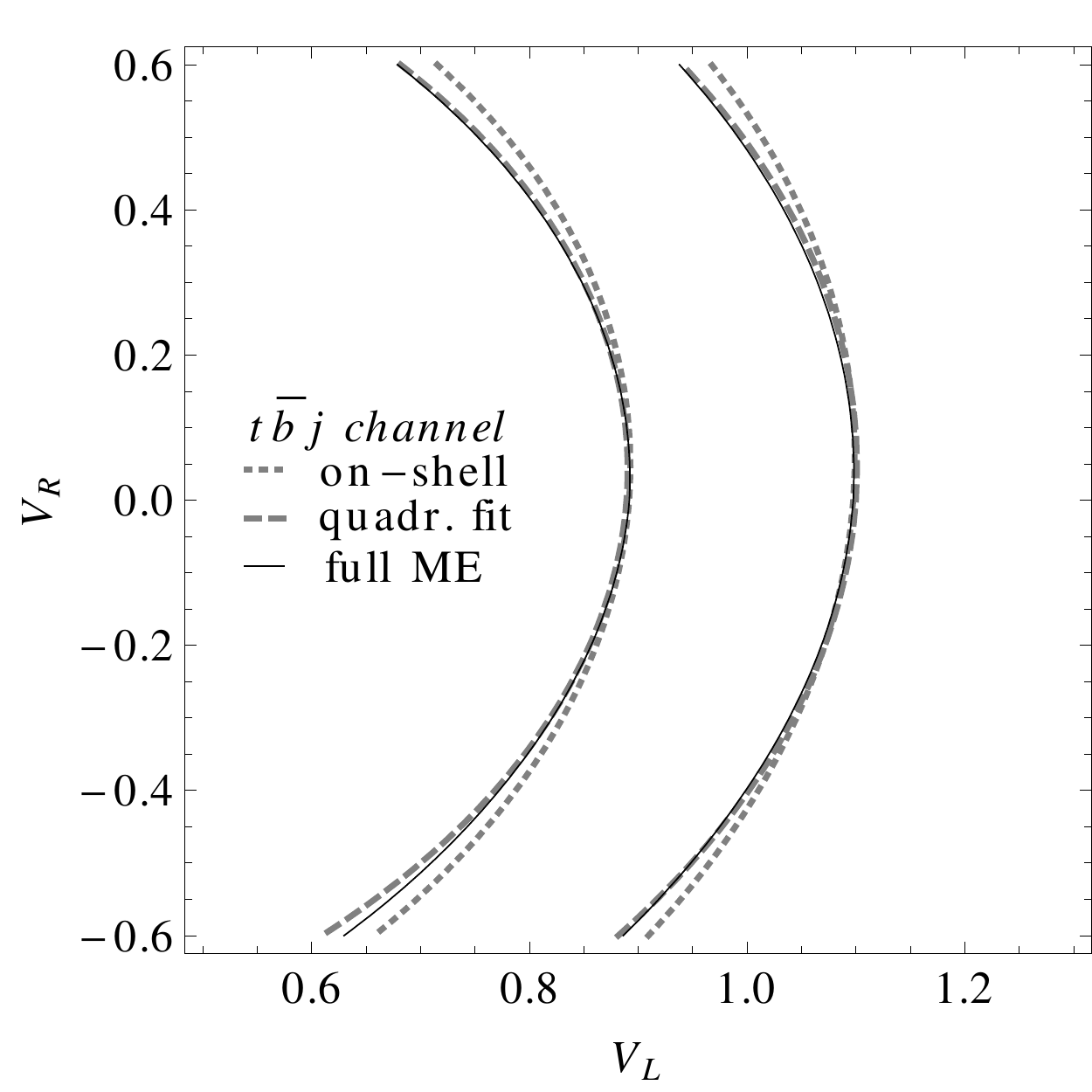}
 \hspace{0.3cm}
 \includegraphics[scale=0.55]{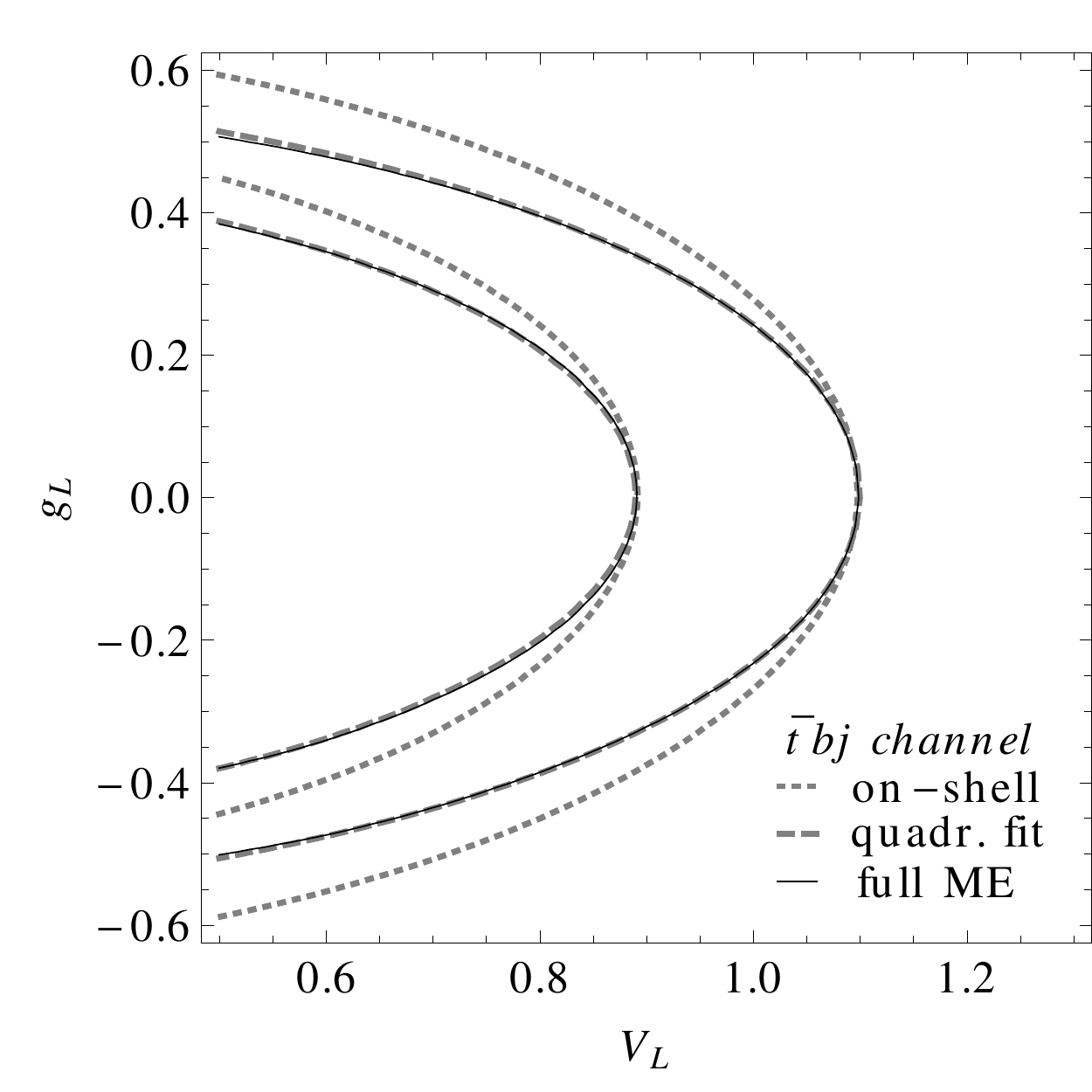}
 \includegraphics[scale=0.55]{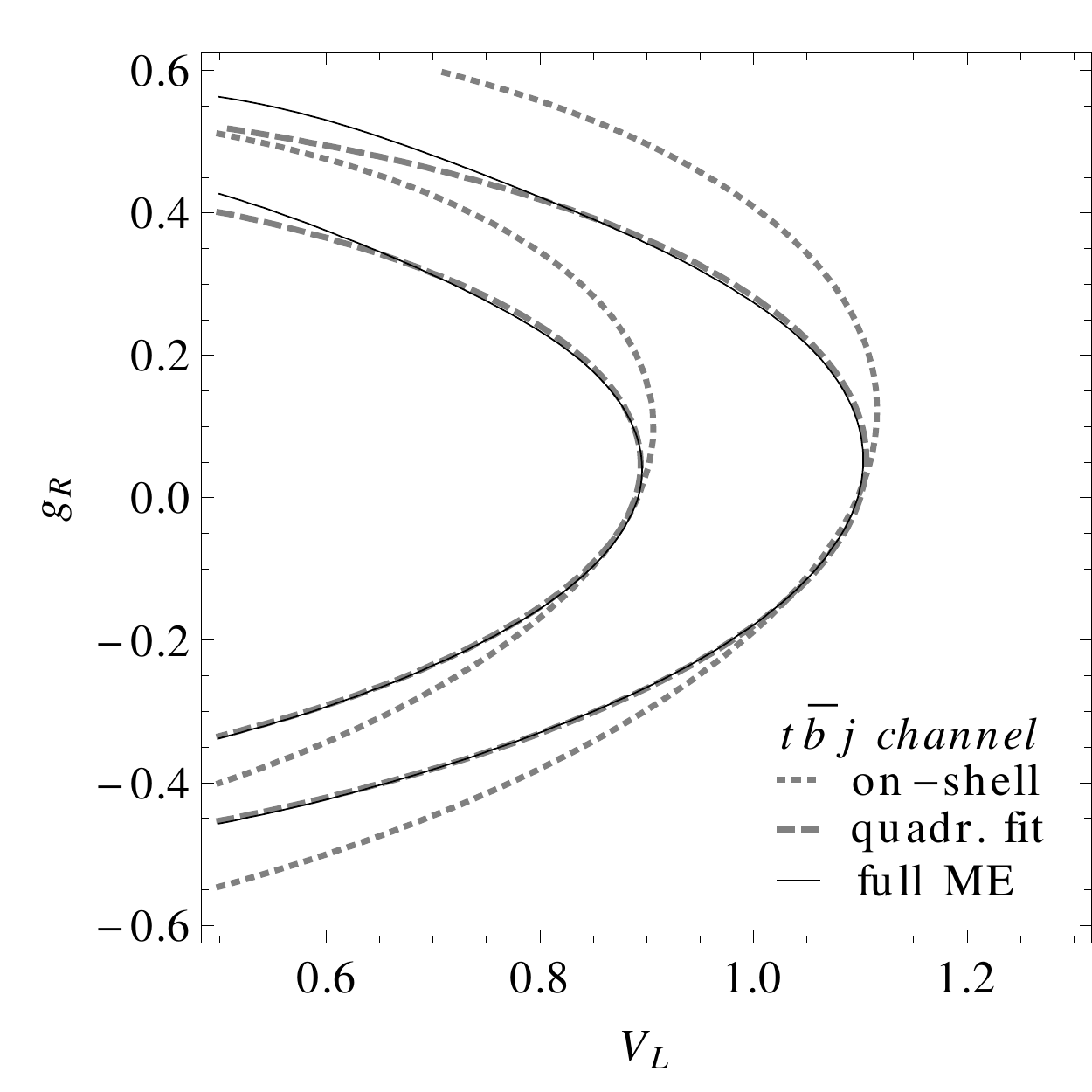}
 \hspace{0.3cm}
 \includegraphics[scale=0.55]{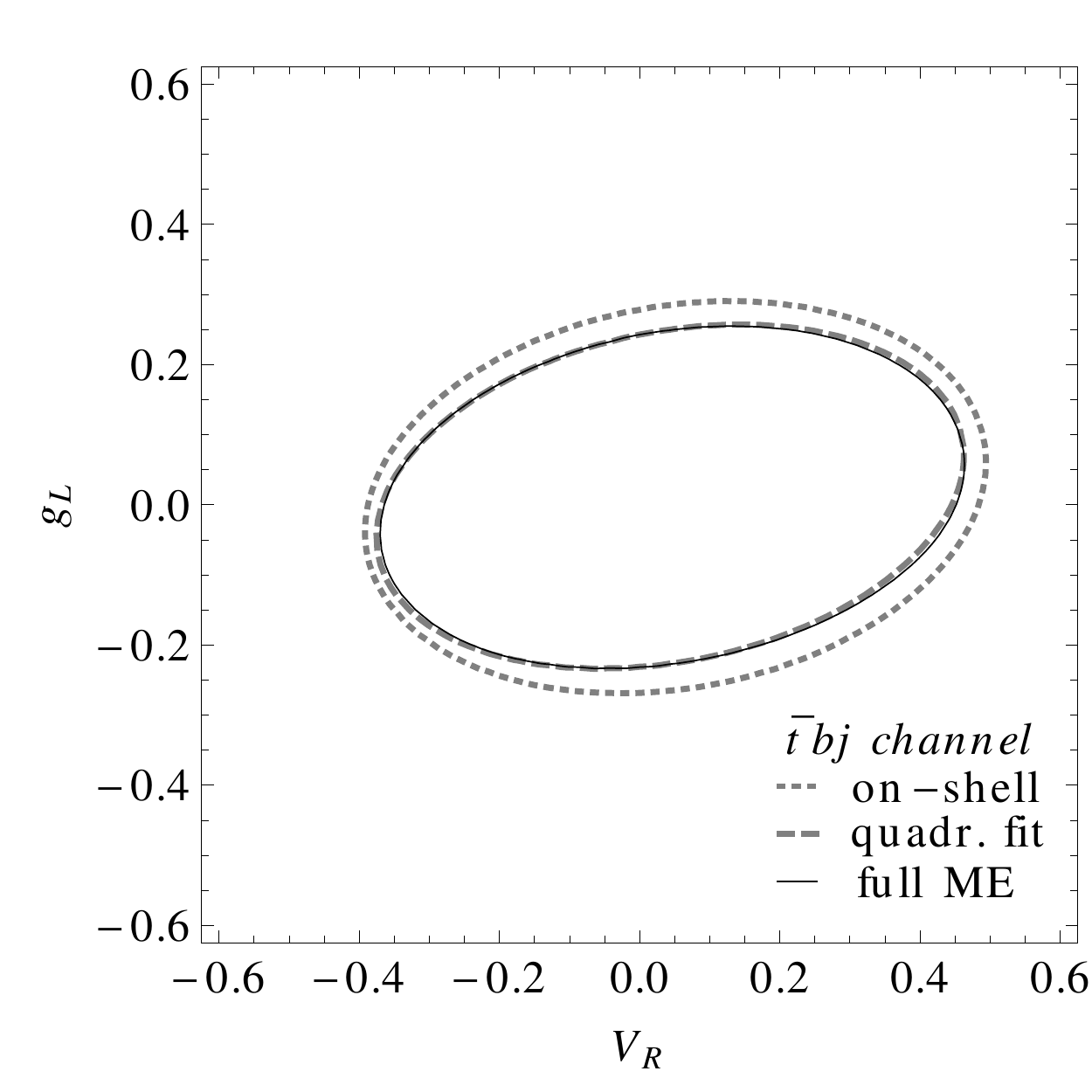}
 \includegraphics[scale=0.55]{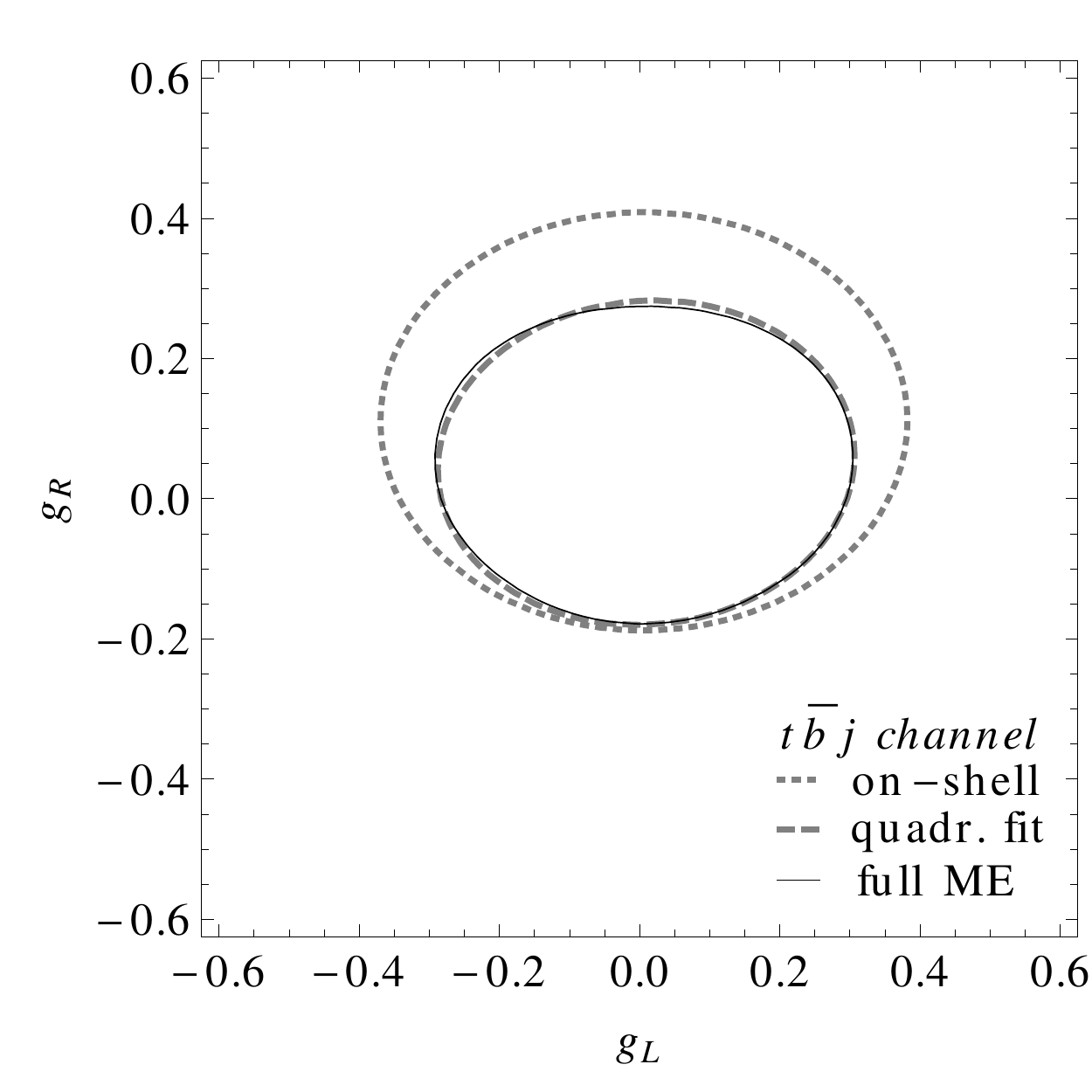}
 \hspace{0.3cm}
 \includegraphics[scale=0.55]{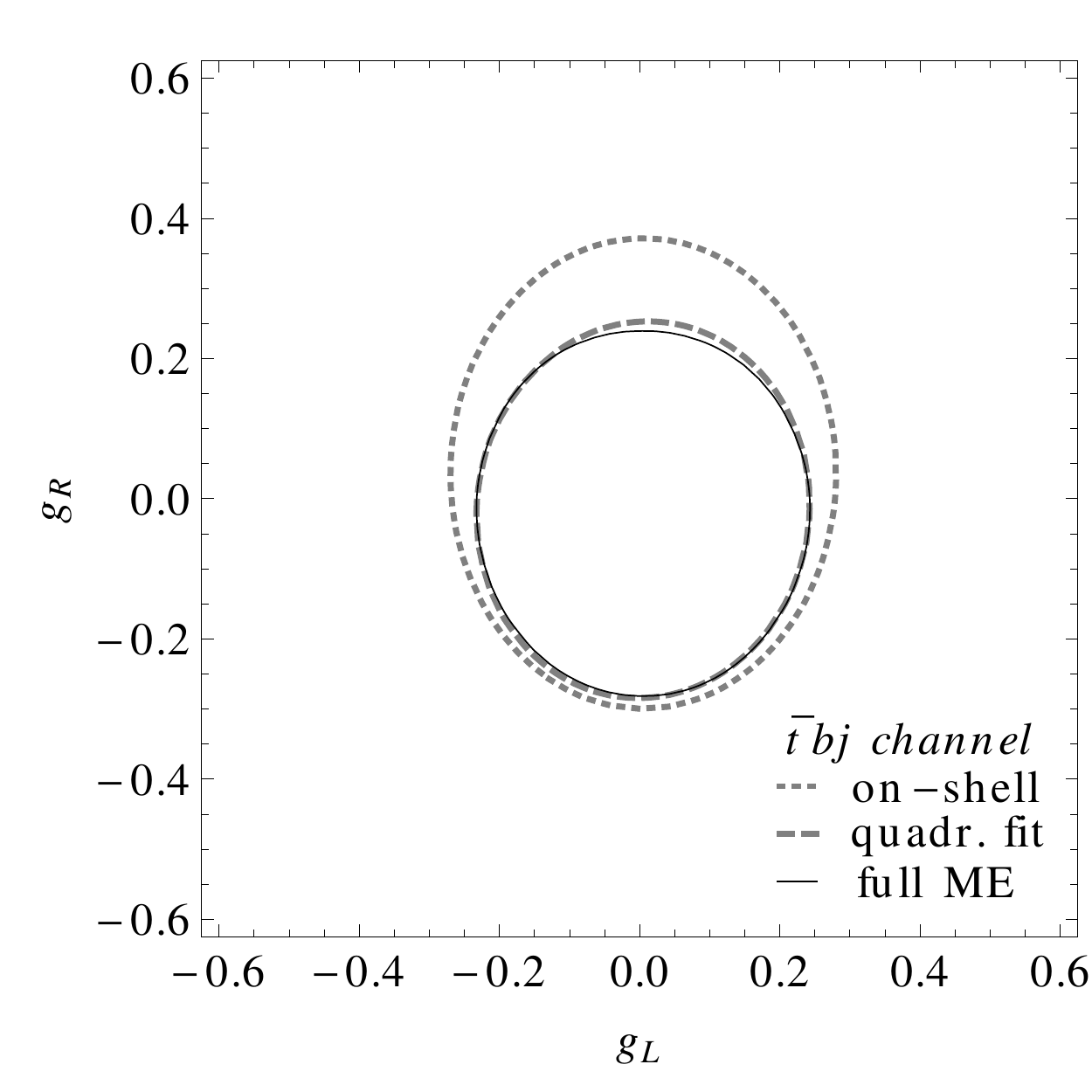}
 \caption{Comparison of the $1\sigma$ contours for the various matrix element
response functions $\kappa_\text{on}$, $\kappa_\text{fit}$ and
$\kappa_\text{full}$ at parton level, for $t\bar{b}j$ and $\bar{t}bj$ processes
in different coupling planes (setting the others to their SM values).
\label{tbj_1sigma_part}}
\end{figure*}

Particularly this dependence of the magnitude of the $V_R g_L$ interference on $V_L$
is an example for the breakdown of the quadratic on-shell approach. However, this
dependence is rather small in $\kappa_{V_R}$ and $\kappa_{g_L}$ which come with squares of
the respective couplings and therefore dominate the sensitivity of a given process
to these couplings, so it might still be a good approach to expand the full
$\kappa_\text{full}$ as a quadratic form in \emph{small} anomalous deviations from the SM point
$V_L=1$, namely by choosing $V_L=1$ as an origin.
As explained before, the numerical values are extracted from quadratic fits
to \mbox{1-dimensional} coupling scans (including acceptance cuts) in all directions of $\vec{g}$,
and all interference directions $g_i g_j$.
A major difference to the original on-shell quadratic form is the appearance
of large linear terms in the couplings, which now encode the interference with the SM.
In Figs.~\ref{tjc_on_off} and~\ref{tbjc_on_off},
we compare the different quadratic parametrisations, namely $\kappa_\text{on}$
from the on-shell approach and $\kappa_\text{fit}$ inferred from the fits,
to the full ME response $\kappa_\text{full}$,
illustrating two statements:
Firstly, $\kappa_\text{on}$ significantly deviates from $\kappa_\text{full}$
in various parts of the parameter space relevant to the analysis, particularly
in the $g_L$--$g_R$ and $V_R$--$g_L$ planes.
Secondly, although still neglecting the higher coupling dependences,
$\kappa_\text{fit}$ inferred from quadratic fits to the
full scan does show a significantly improved agreement with the full scan
while still being fast and efficient.
This is further illustrated in Fig.~\ref{tj_1sigma_part}
and~\ref{tbj_1sigma_part}
showing $\pm1\sigma$ sensitivity contours
around the SM strength $\kappa=1$ for various anomalous coupling combinations
and production processes: Especially when the momentum-dependent couplings $g_{L,R}$
are involved, there are remarkable shifts of the contours when going from the
on-shell to the full ME approach, but generally these effects are modelled
very well by the adapted quadratic parametrisation $\kappa_\text{fit}(\vec{g})$,
while the machine cost reduces from a 4D scan over $\vec{g}$ to
a set of 1D scans along all axes $g_i$ and interference directions $g_i g_j$
for each input process.
We now go on to the detector level to quantify the impact
of these different approaches on exclusion bounds on anomalous
couplings from combined cross section measurements.

\subsubsection{Detector level}

In order to obtain a detector efficiency matrix in the various approaches,
samples of $\unit[500]{k}$ events are produced in each partonic production process,
once integrating the full off-shell matrix elements over the acceptance
region, Eq.~\eqref{acc}, and
once integrating the resonant matrix elements over
the full phase space,
letting the tops decay off-shell into a $b$~quark, a~charged lepton and
a~neutrino, analogously to~\cite{AguilarSaavedra:2008gt}.
All the parton-level samples
are processed with \textsc{Pythia} and \textsc{Delphes} to obtain events
at detector level. This is done for the SM point as well as the coupling
configurations
\begin{equation}\label{cpl_points}
 \begin{array}{rlll}
  \text{A:}\hspace{30pt} & V_L=1\,,\; & V_R=0.3\,,\; & g_L=0.15\,, \\
  \text{B:}\hspace{30pt} & V_L=1\,,\; & g_R=0.024\,, & 
 \end{array}
\end{equation}
taken from~\cite{AguilarSaavedra:2008gt} to facilitate comparison. Note
that in each case a consistent top width~$\Gamma_t\left(\vec{g}\right)$
is calculated beforehand and the result checked to comply with
experimental constraints~\cite{Abazov:2012vd}).
Again for comparison reasons, we also adopt the final state selection cuts
stated in~\cite{AguilarSaavedra:2008gt} which define the various components
of~$\varepsilon$: apart from requiring an isolated lepton
(that is, $e$~or~$\mu$)
with $p_T>\unit[25]{GeV}$ and missing transverse energy
$\slashed E_T>\unit[25]{GeV}$,
the selection criteria for the three final state signatures are, respectively
\begin{enumerate}

 \item for $tb$ selection: exactly two $b$ tagged jets (assuming a tagging efficiency
of~0.6)
with $p_T>\unit[30]{GeV}$, and neither central nor forward light
jets with $p_T>\unit[15]{GeV}$. In addition, the top momentum $p_t$ is reconstructed
from one of the $b$s together with the charged lepton and~$\slashed E_T$
(to be identified with the neutrino~$p_T$), by applying the on-shell
constraint $\left( p_\ell+p_\nu \right)^2=m_W^2$ and picking the smaller of
the two solutions for the longitudinal component of $p_\nu$. Finally,
the resulting top mass must lie between 150 and $\unit[225]{GeV}$.

 \item for $tj$ selection: at least one $b$ jet with $p_T>\unit[30]{GeV}$
(one of them reconstructing $p_t$ together with the leptons as explained above),
one light forward jet with $p_T>\unit[50]{GeV}$ and $2.5<\left|\eta\right|<5$
and no more than one additional light central jet, which may have
$p_T<\unit[30]{GeV}$ only.

\end{enumerate}
By applying every final state
selection to each of the $\unit[500]{k}$ event samples corresponding
to the partonic input processes and averaging over lepton flavors and charge states,
we find
for the samples from resonant diagrams integrated over the full phase space
an efficiency matrix~$\varepsilon$~(in~$\%$) at the SM point:
\begin{equation}\label{eps_full_ps}
 \begin{array}{r|ccc}
                  & tb       & tj       & tbj      \\
  \hline
  tb\,\text{sel.} & 0.658(6) & 0.040(1) & 0.051(2) \\
  tj\,\text{sel.} & 0.165(3) & 0.647(6) & 0.531(5) \,.
 \end{array}
\end{equation}
For the full ME approach, we run the selection criteria described above on the
detector level samples from full matrix elements integrated over the acceptance region,
inferring $\varepsilon^\prime$ (in $\%$)
\begin{equation}\label{eps_acc}
 \begin{array}{r|ccc}
                  & tb       & tj       & tbj      \\
  \hline
  tb\,\text{sel.} & 1.28(1)  & 0.039(1) & 0.031(1) \\
  tj\,\text{sel.} & 0.282(4) & 1.52(1)  & 1.023(7) \,
 \end{array}
\end{equation}
at the SM point.

Before moving on to detector-level coupling limits, the detector efficiency matrices
can be compared between the various coupling points, Eq.~\eqref{cpl_points},
to test the assumption of a constant detector response. While the small admixture
of $g_R$ in point~B only has a negligible effect on the efficiencies, we observe
that there are sizeable changes in the efficiencies when going from the SM point
to point~A, amounting to some $\unit[15]{\%}$ ($\unit[8]{\%}$) of the original
values for $tb$ ($tj$) selection in the samples
with full phase space integration at parton level. However, when going close
to the detector acceptance region already with the partonic input, this
dependence is reduced to $\sim\unit[6]{\%}$ ($\unit[4]{\%}$),
thus improving on another source of systematic uncertainty.

Taking approximate NNLO on-shell $s$ and $t$ channel production cross sections
from the literature~\cite{Kidonakis:2010tc,Kidonakis:2011wy}
(multiplied by a partonic acceptance efficiency corresponding to the cuts
in~Eq.~\eqref{acc} in the full ME approach)
to normalize the SM reference point for each input process,
we now have all ingredients at hand to derive limits on~$\vec{g}$ from a set of
cross section measurements, modelled by Eqs.~\eqref{xsecAS} and \eqref{xsecW2},
and compare the results.
In the $t$ channel, the matrix element response for the detector-level analysis
is modelled using only $t\bar{b}j$
and $\bar{t}bj$ processes for simplicity, and because it was argued that
the corresponding distributions already describe the proper NLO behaviour
rather well~\cite{Sullivan:2004ie,Boos:2006af}.
Moreover, it was shown~\cite{Falgari:2011qa,Campbell:2012uf} that NLO
corrections affect the differential distributions in $s$ and $t$
channel single top production only marginally, at the few~$\%$ level,
and can thus be readily accounted for by channel specific overall $K$
factors, as in our analysis.
A more comprehensive analysis including coupling dependent $K$ factors,
similar to the case of anomalous flavor changing gluon
couplings~\cite{Gao:2011fx}
and extending an existing study on anomalous top decays at NLO~\cite{Drobnak:2010ej},
will moderately influence the numerical
values of the exclusion bounds.  However, our results concerning the
relative importance of the quartic couplings and the need to include
the full matrix elements remain valid.

\begin{figure*}
 \includegraphics[scale=0.55]{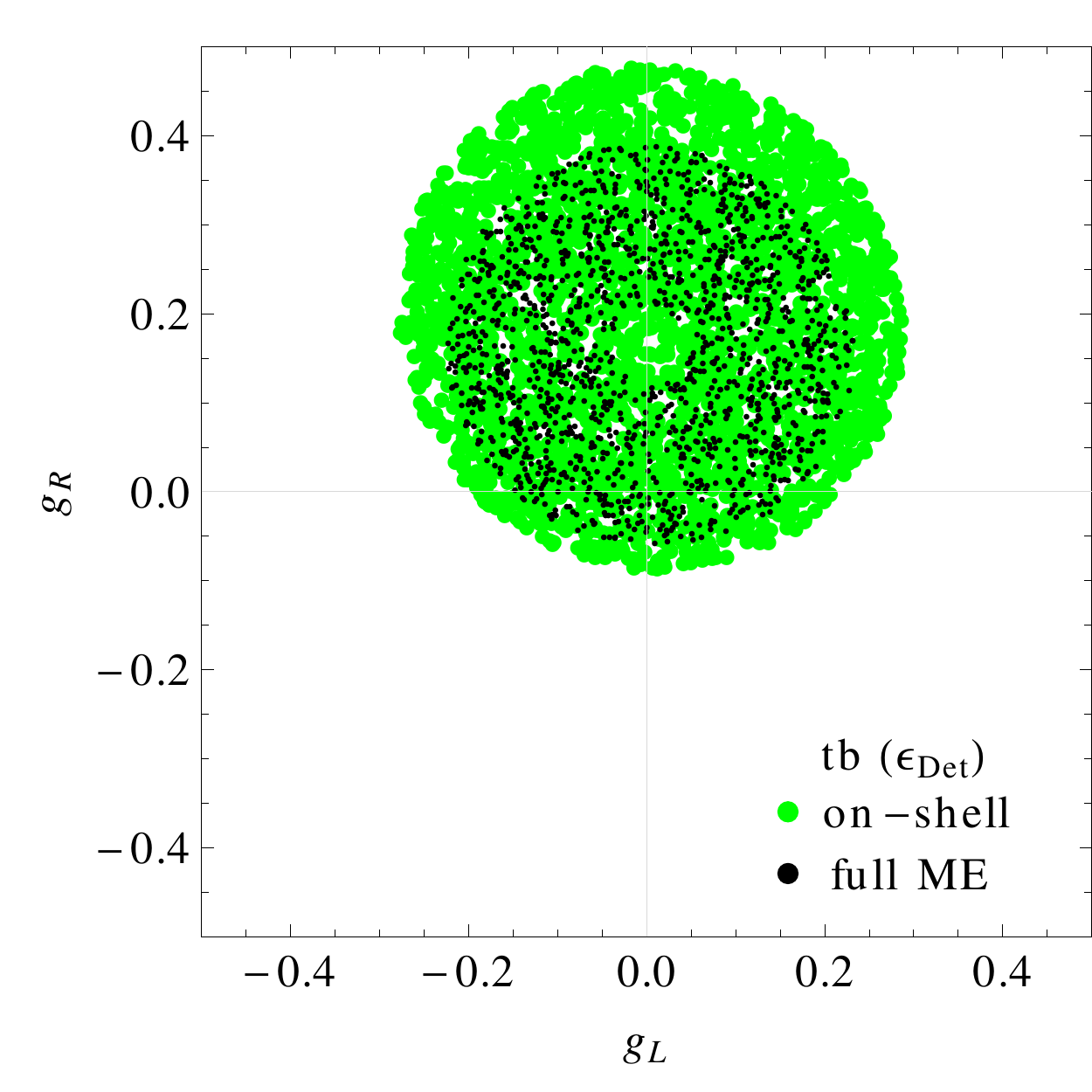}
 \hspace{0.3cm}
 \includegraphics[scale=0.55]{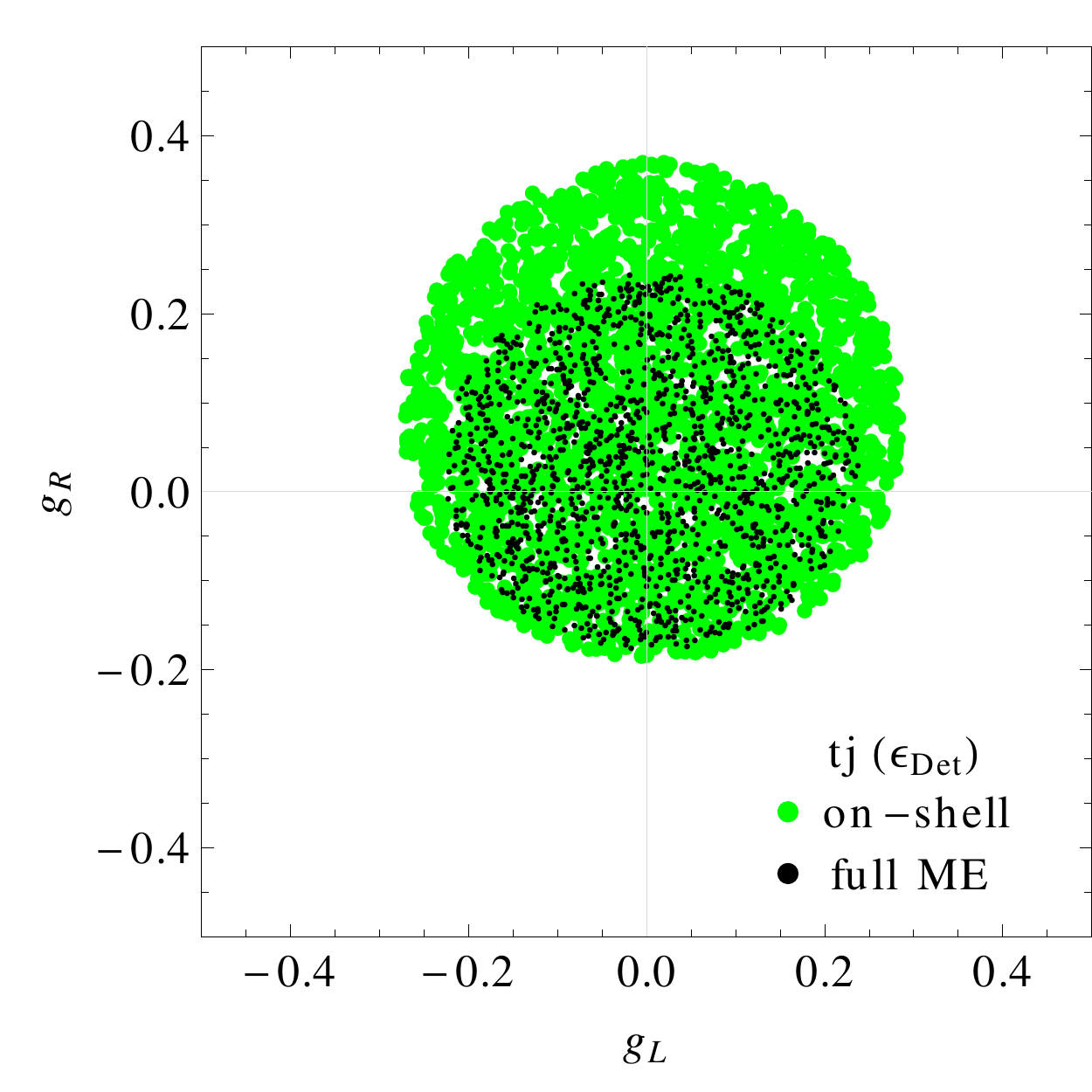}
 \includegraphics[scale=0.55]{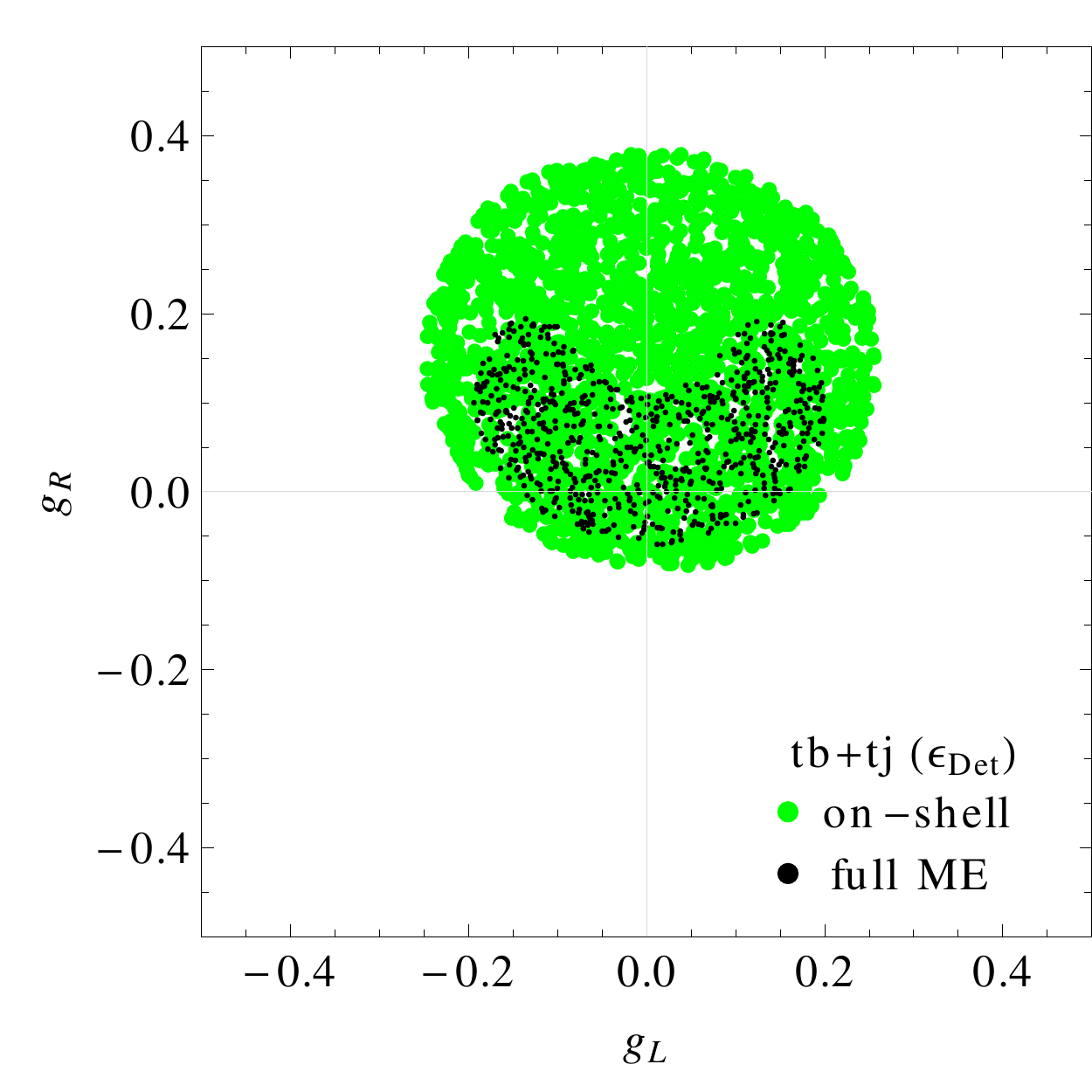}
 \hspace{0.3cm}
 \includegraphics[scale=0.55]{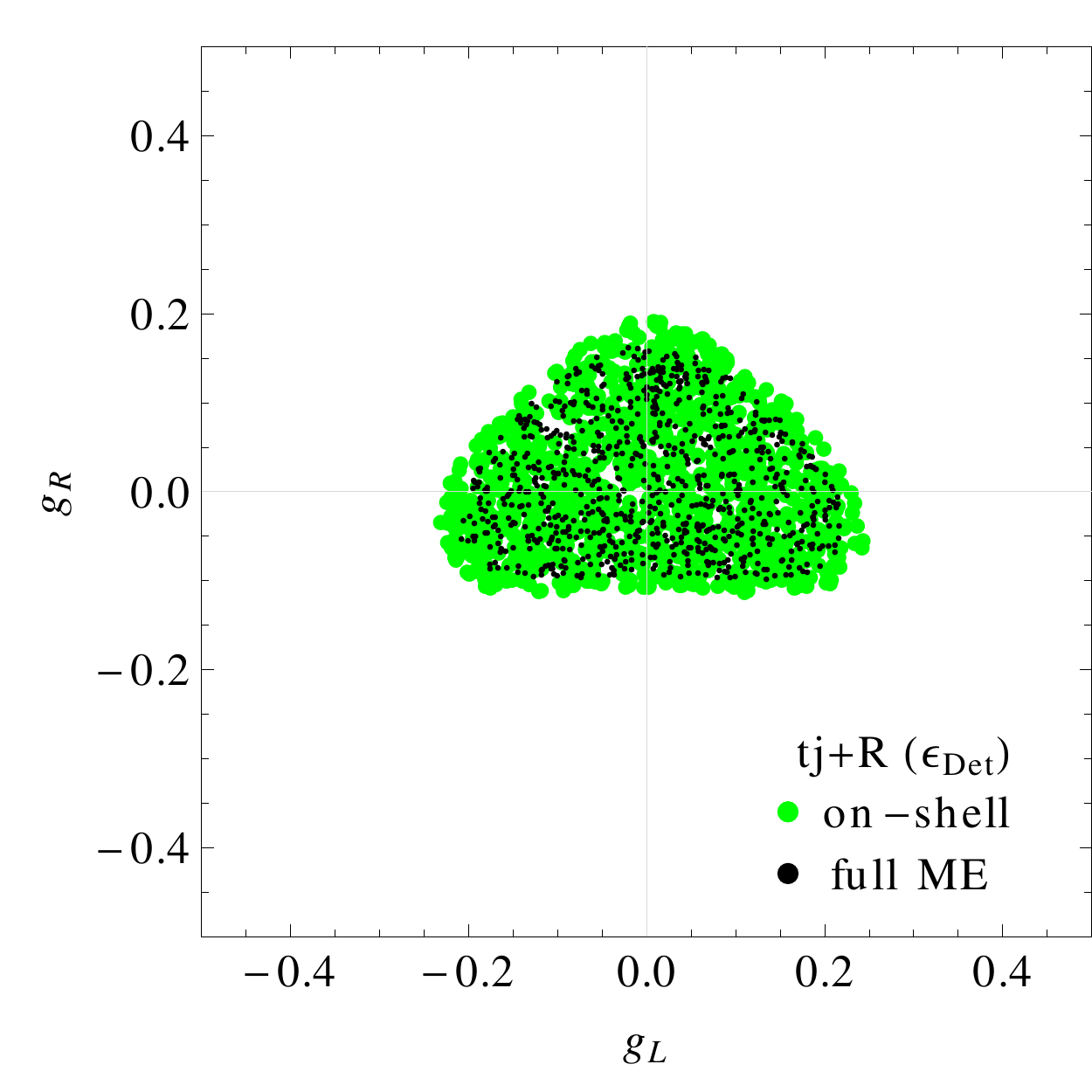}
 \includegraphics[scale=0.55]{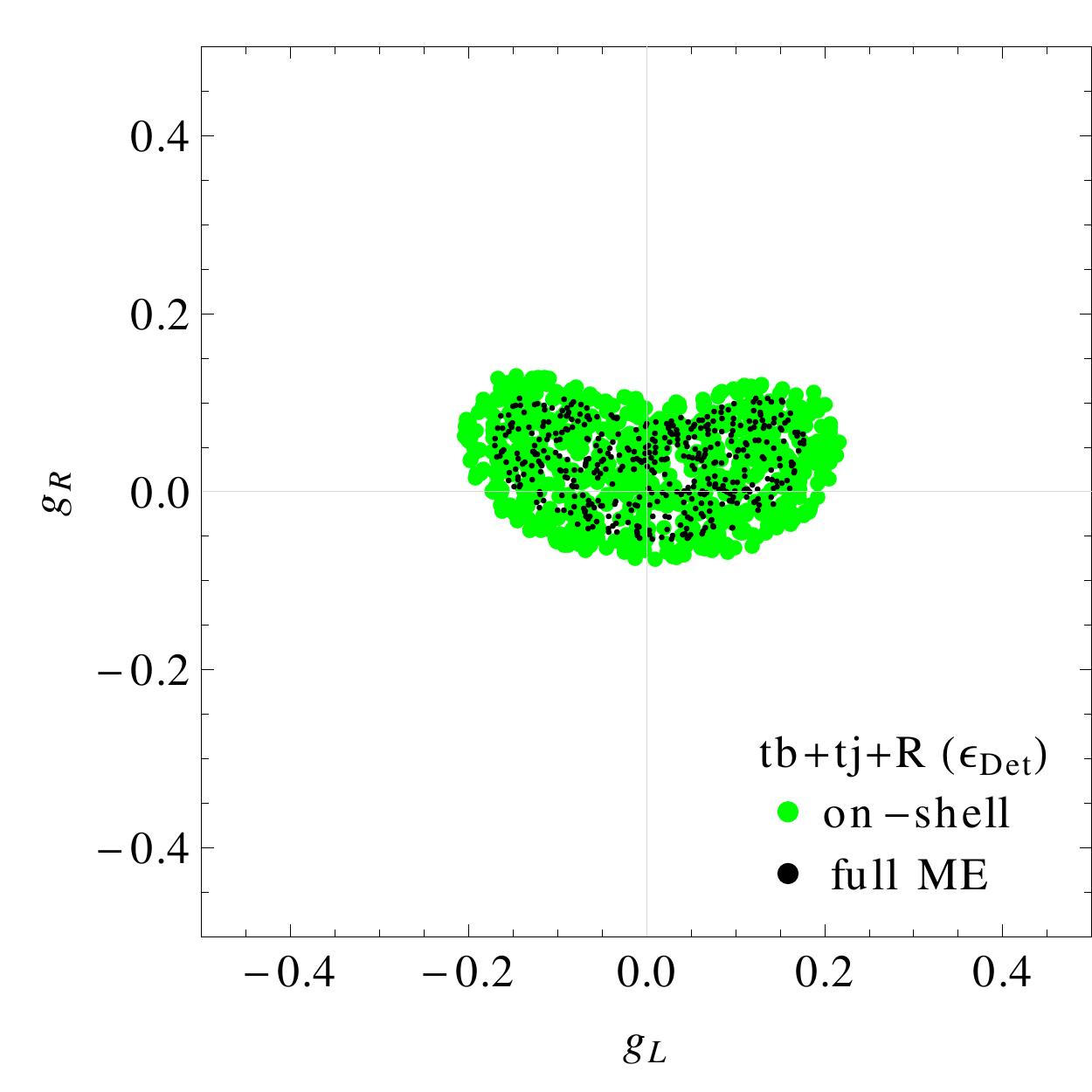}
 \hspace{0.3cm}
 \includegraphics[scale=0.55]{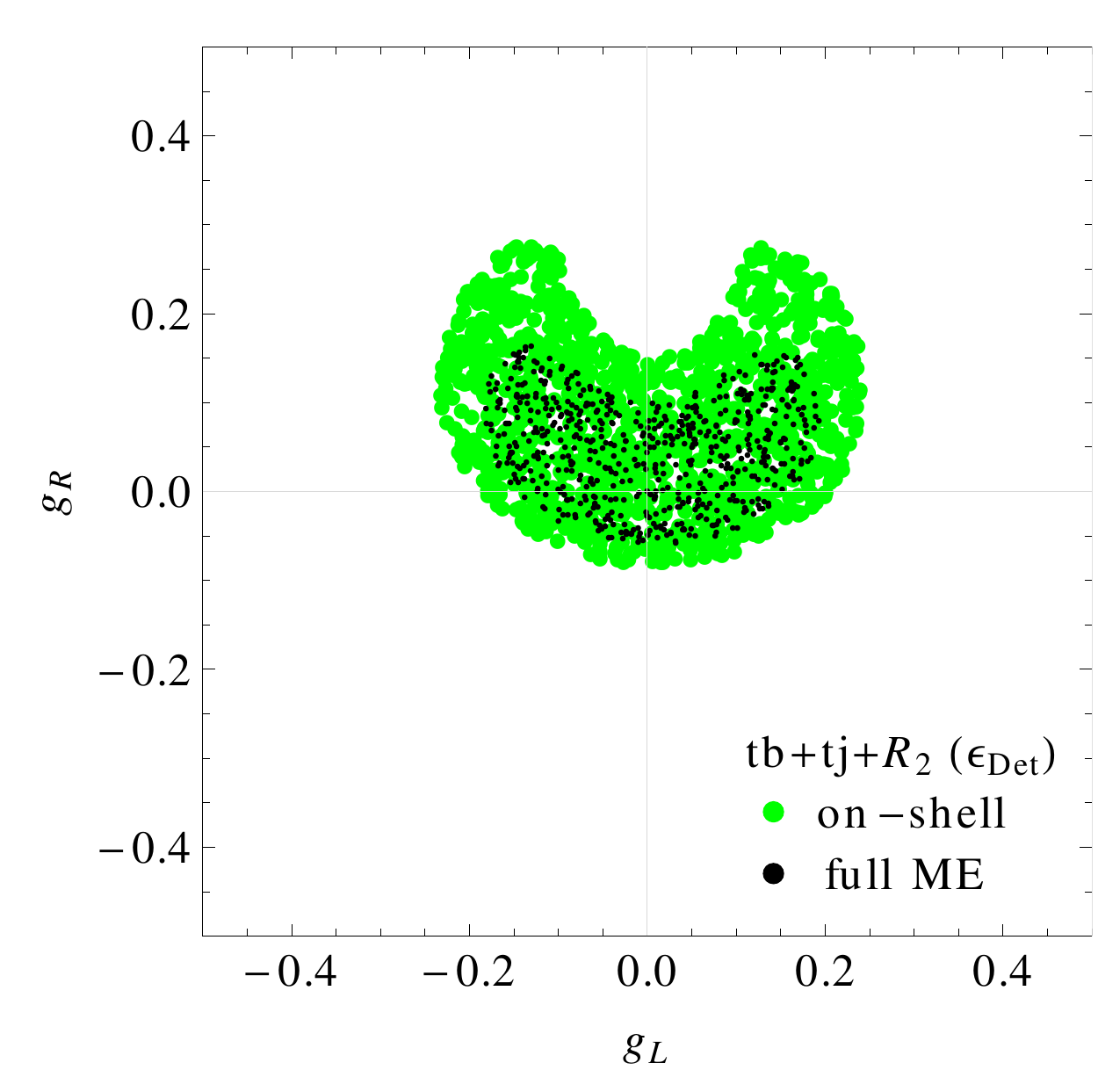}
 \caption{
Combined $1\sigma$ limits on $g_L$ and $g_R$ ($V_L=1$, $V_R=0$) from single
final states (top), combined final states (center left), and including
the observable $R(\bar{t}/t)$ (center right and bottom). ``$R_2$'' denotes
a factor 2 on the experimental resolution of $R$.
\label{1sigma_det}}
\end{figure*}

Now turning to the resulting bounds at the detector level,
as already anticipated from the $1\sigma$ contours
in~Fig.~\ref{tj_1sigma_part}
and~\ref{tbj_1sigma_part} the effects on $V_{L,R}$
remain small in general, while the largest differences are found when the momentum
dependent couplings $g_{L,R}$ are involved, particularly in the $g_L$--$g_R$ plane
illustrated in~Fig.~\ref{1sigma_det}.
In this case, when single channels and charge states are considered separately,
the different approaches tend to produce very different exclusion bounds.
Fig.~\ref{1sigma_det} might also suggest that after combining all channels and
considering the ratio $R(\bar{t}/t)$ of cross sections for $t$ and $\bar{t}$ production in the
$t$ channel as an additional observable
(tentatively assuming $\unit[2]{\%}$ statistical and $\unit[3]{\%}$ systematic
uncertainty as estimated for $\unit[10]{fb^{-1}}$ at $\sqrt s=\unit[14]{TeV}$
in~\cite{AguilarSaavedra:2008gt}, in the absence of a more detailed experimental assessment),
one might end up with the same exclusion
limits again, but indeed this depends heavily on the total uncertainty of $R$
in the actual experiment (cf. ``$R_2$'' in Fig.~\ref{1sigma_det}).
In any case, when the aim is to properly understand
and quantify the sensitivities to anomalous couplings of the various final states
separately, going from the on-shell approach to full matrix element responses inside
the selection acceptance region produces considerable effects that should not be
neglected. In that respect, the adapted quadratic parametrisation introduced above,
employing quadratic fits to off-shell scans inside the acceptance region,
represents a very good approximation to the full off-shell parameter scan
(cf.~Fig.~\ref{tjc_on_off}--\ref{tbj_1sigma_part}).

\subsection{Pinning down the off-shell coupling}\label{VO}

After discussing the technical issue of modelling the LO matrix element response
to anomalous top couplings at an experimentally relevant acceptance level,
and validating an adapted quadratic parametrisation which simultaneously meets the
demands of machine efficiency and good agreement with the full off-shell
coupling scan in the previous section, the closing section of the article is
devoted to the application of the new approach to a physical issue, namely
a possible admixture of the additional anomalous coupling $V_L^\text{off}$
introduced in Sec.~\ref{anom_top}, to the single top cross sections
(the total top width is also included as an observable, but its sensitivity
to $V_L^\text{off}$ is kinematically suppressed compared to the other
anomalous couplings, since the relevant scale $m_t$ is lower than
$\sqrt{\hat s}$).

Considering the experimental sensitivities to the anomalous couplings of the
total cross sections stated above for the LHC (which are already dominated
by systematics), it is clear that a stand-alone study of single top cross sections
alone will never provide the most stringent bounds on the complete parameter space
of anomalous CC couplings, including $V_L^\text{off}$ or not.
Therefore, rather than just adding another direction to~$\vec{g}$,
the focus shall be directed here to those regions of the parameter space where
single top cross sections actually become the crucial inputs to the combined limits.

\begin{figure*}
 \includegraphics[scale=0.55]{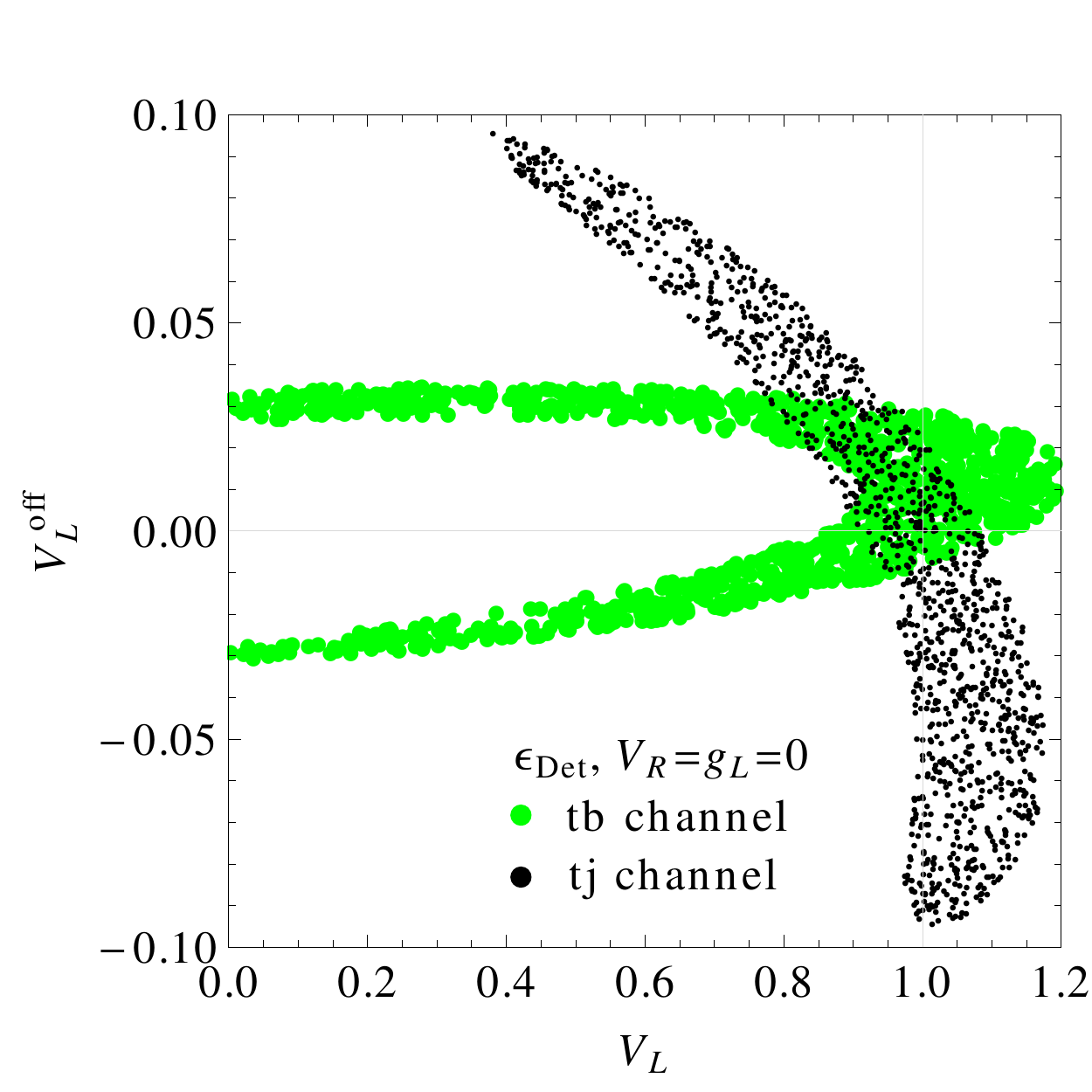}
 \hspace{0.3cm}
 \includegraphics[scale=0.55]{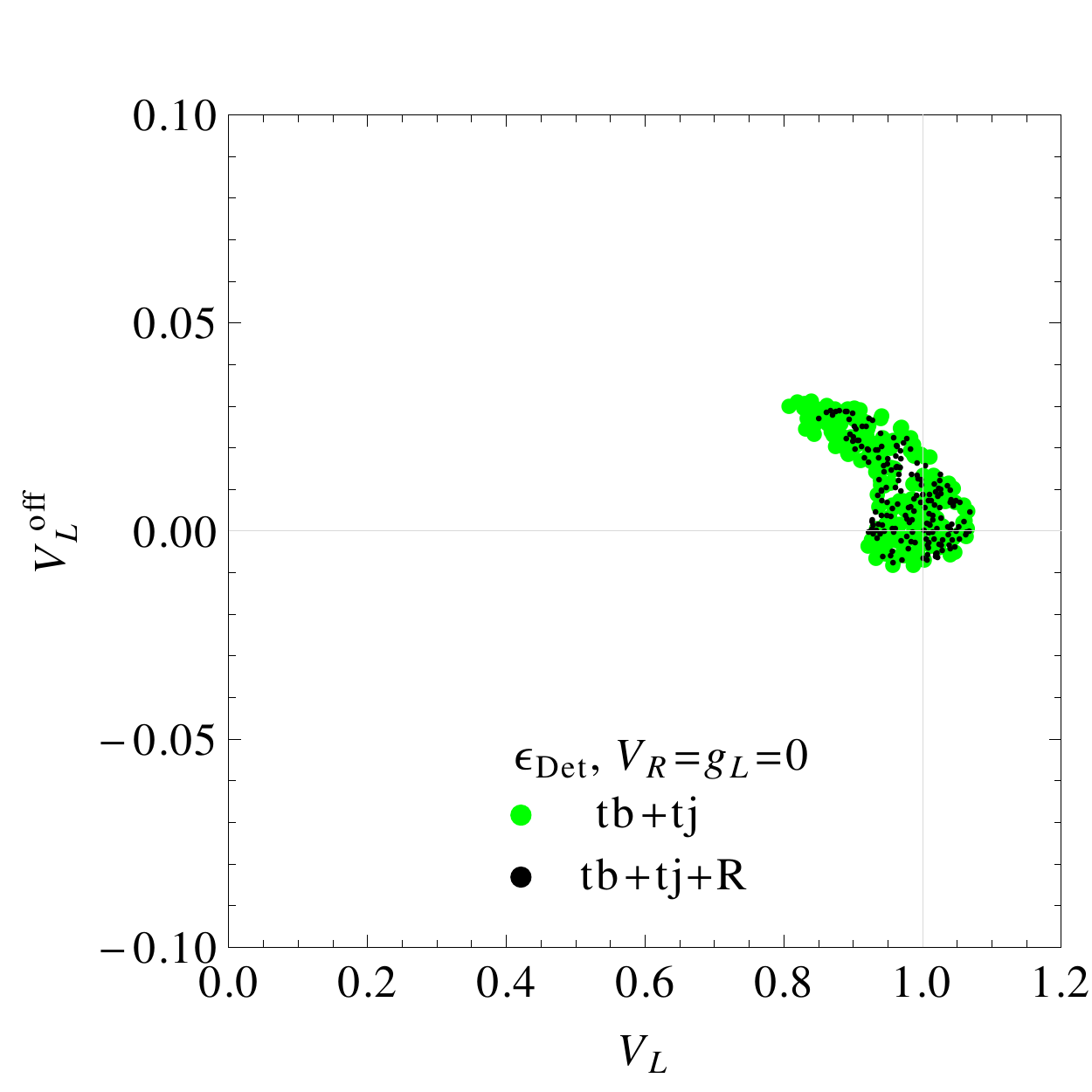}
 \includegraphics[scale=0.55]{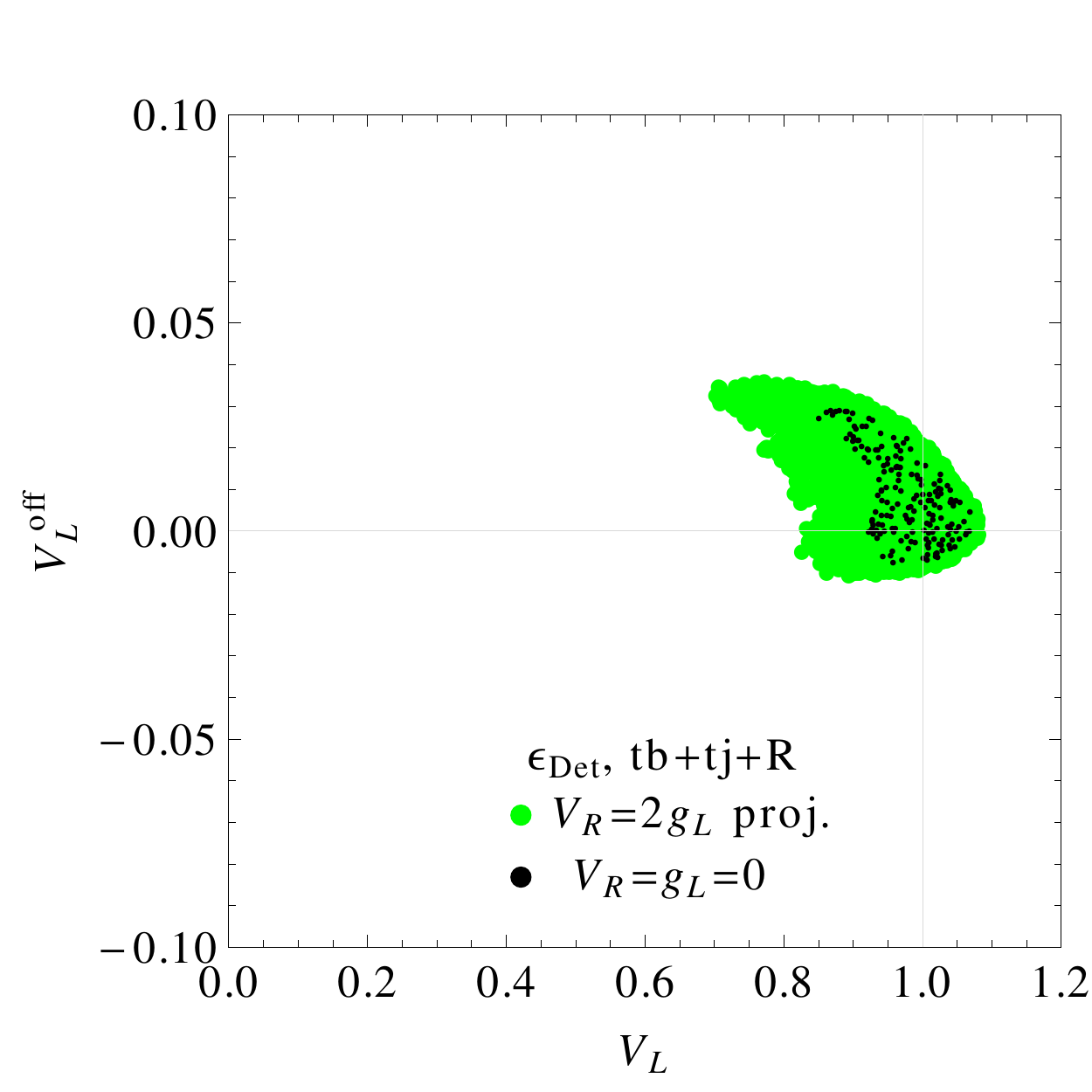}
 \caption{Combined $1\sigma$ contours in the $V_L$--$V_L^\text{off}$ plane,
setting $V_R=g_L=0$ (top) or projecting over the direction $V_R=2g_L$ (bottom).
\label{VL_VO}}
\end{figure*}

\begin{figure*}
 \includegraphics[scale=0.55]{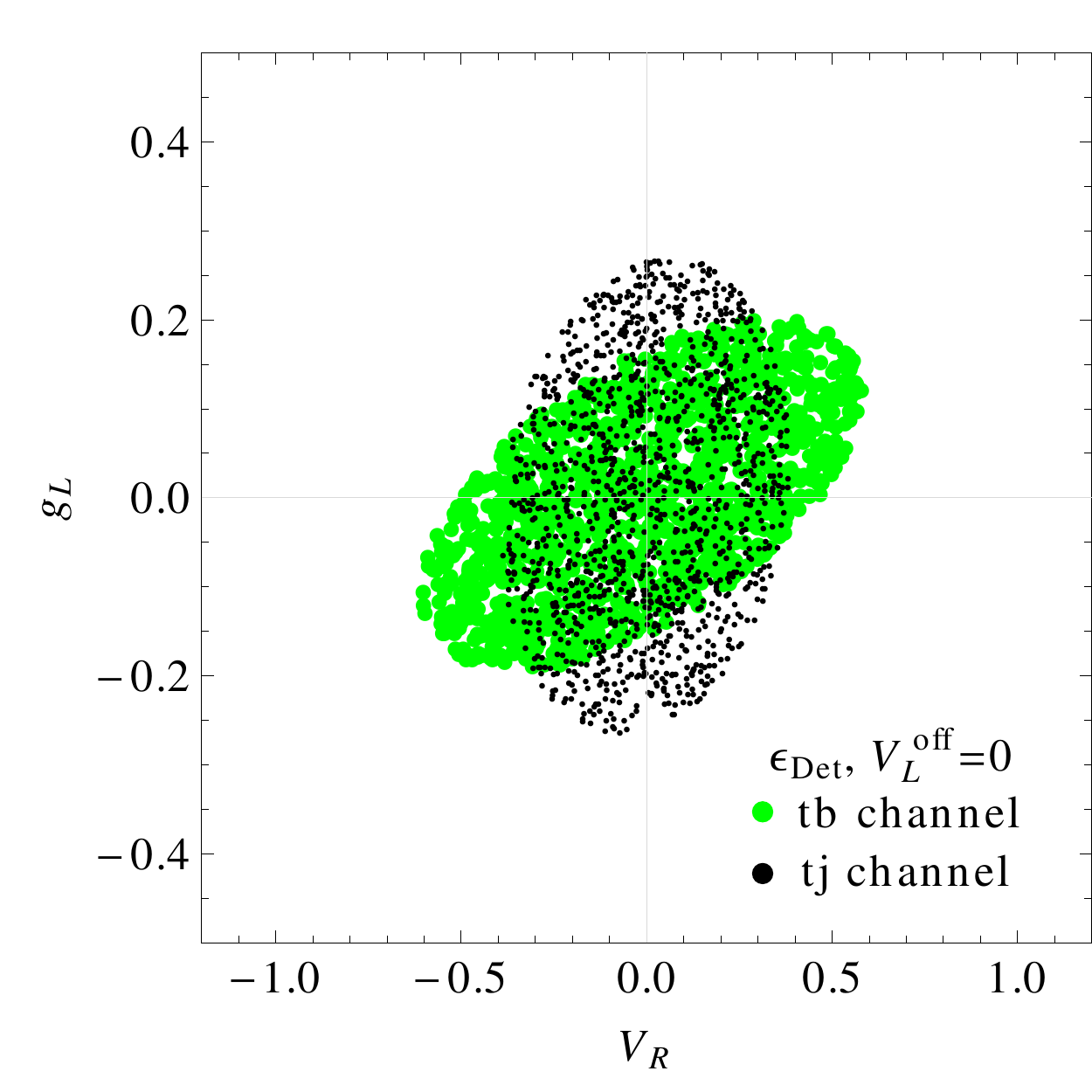}
 \hspace{0.3cm}
 \includegraphics[scale=0.55]{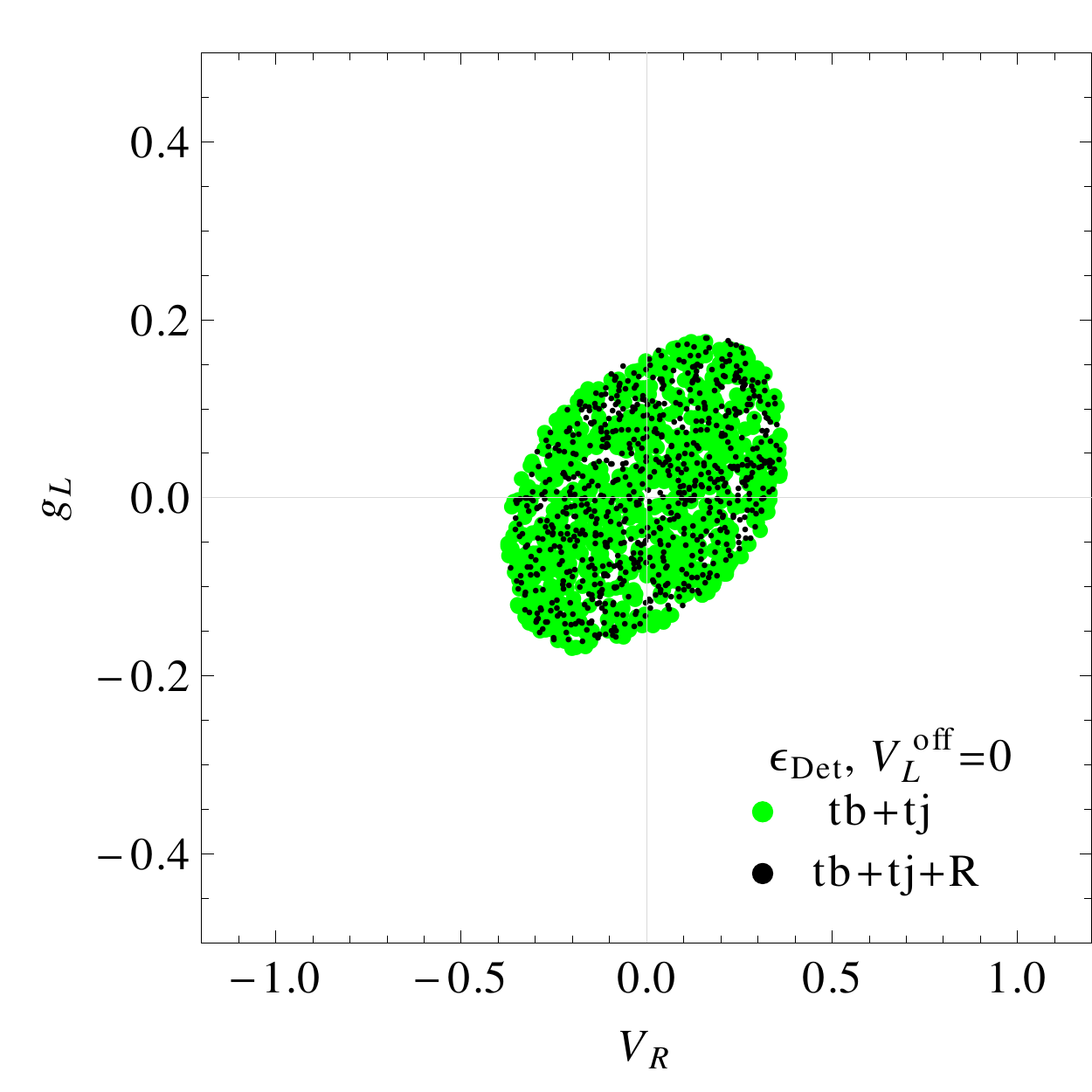}
 \includegraphics[scale=0.55]{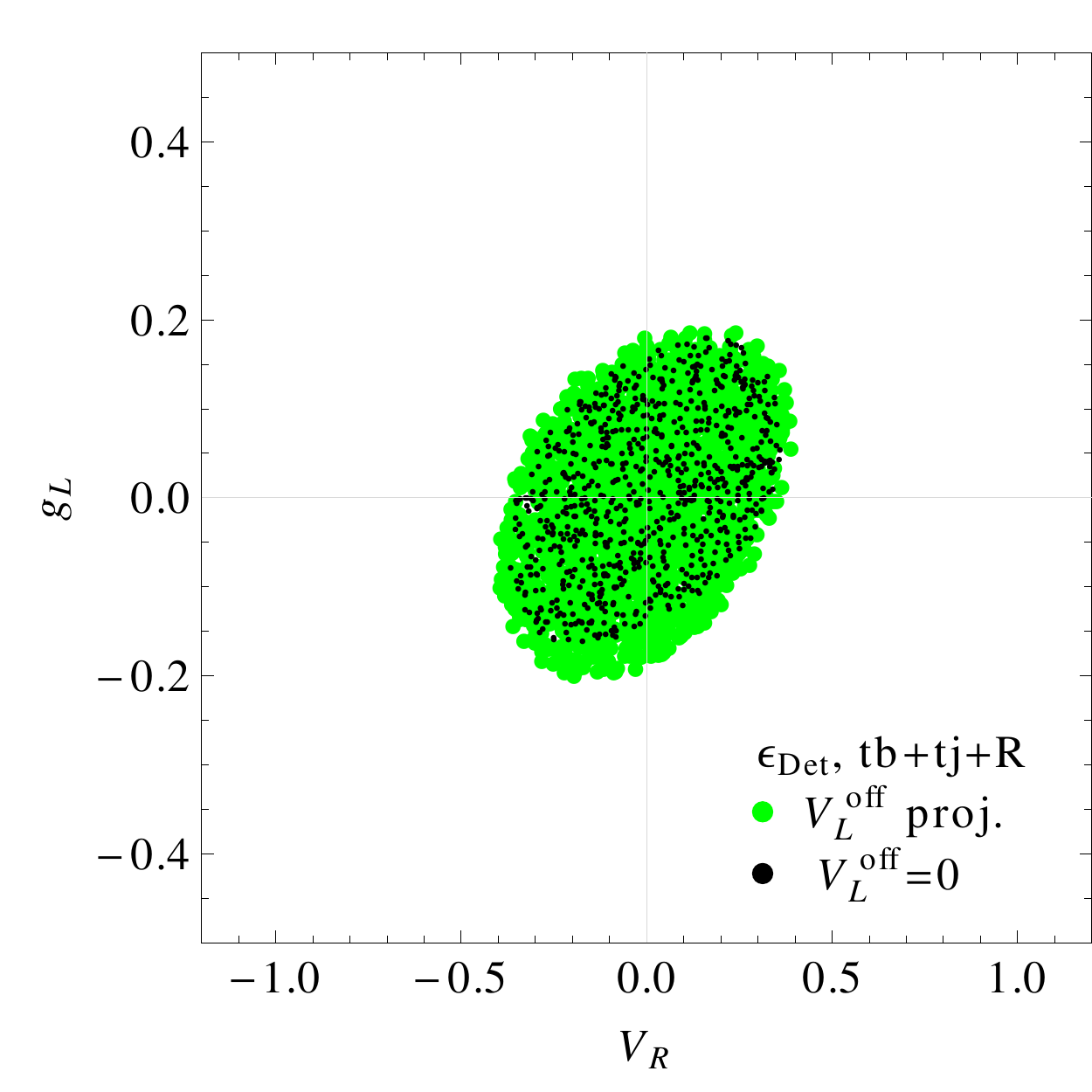}
 \caption{Combined $1\sigma$ contours in the $V_R$--$g_L$ plane,
setting $V_L=1$ and $V_L^\text{off}=0$ (top) or projecting (bottom).
\label{VR_gL}}
\end{figure*}

More explicitly, the top decay observables (mostly related to the charged lepton
distribution) are very sensitive to anomalous $W$ helicity fractions generated by
$V_R$, $g_L$ and $g_R$
(cf.~\cite{AguilarSaavedra:2006fy,AguilarSaavedra:2007rs,AguilarSaavedra:2008gt,Aad2012}).
For example, the limit $\left|g_R\right|\lesssim 0.024$
stated in~\cite{AguilarSaavedra:2008gt} for our LHC reference point
($\unit[10]{fb^{-1}}$ at $\unit[14]{TeV}$) is more than an order of
magnitude below the sensitivity of the cross sections, so we may as well
set $g_R\equiv0$ for our purposes.
On the other hand, the large interference among $V_R$ and $g_L$ leads to rather poor
bounds $\left|V_R\right|\lesssim 0.3$ resp. $\left|g_L\right|\lesssim 0.15$ as
long as they are fine-tuned to $V_R\sim 2g_L$.
Finally, since decay observables basically measure helicity fractions, they are
neither sensitive to the overall vertex normalization nor to the admixture of
$V_L^\text{off}$ to the left-handed vector part. This is where the cross sections
come into play, delivering the most stringent direct constraints.
In Fig.~\ref{VL_VO}, we therefore present combined limits on $V_L$ and $V_L^\text{off}$
from single top cross sections, both setting $V_R=g_L=0$ as well as varying over
$-0.3\leq V_R=2g_L\leq 0.3$. The very different sensitivities of the two
final states greatly help in the combined limit:
the $s$ channel is very sensitive along $V_L^\text{off}$  due to the kinematics,
whereas the $t$ channel basically cuts the substantial interference in the $s$ channel
along $V_L$.
Still, the resulting limit on $V_L$
deteriorates from $0.9<V_L<1.1$ ($V_L^\text{off}=0$) to $0.82<V_L<1.1$
($V_L^\text{off}$ varied). Naturally, projecting over the remaining freedom
in $V_R$ and $g_L$ instead of switching them off further relaxes the combined limits
to $0.68<V_L<1.1$.
Fig.~\ref{VR_gL} displays combined bounds in the
$V_R$--$g_L$ plane, switching off resp. varying over $V_L^\text{off}$.

In the long run, it is perfectly clear that this ambiguity among $V_L$ and
$V_L^\text{off}$ remaining in the single top cross sections can be further resolved,
namely by examining differential distributions, since $V_L^\text{off}$ scales
very differently with the partonic $\sqrt{s}$ than $V_L$.
(In fact, $V_L^\text{off}$ behaves like, or parametrises, a heavy off-resonant
new degree of freedom, cf.~e.g.,~\cite{Boos:2006xe}.)
However, this issue
will have to be tackled in the $s$ channel where the momentum of the $W$ propagator
producing the top becomes timelike. Sensitive observables would obviously be the
total invariant mass $m_{tb}$ of the final state or the pseudorapidity $\eta_b$
of the hard $b$ jet produced along with the single top. However, such a study
is experimentally challenging, since it requires a very good isolation of the
tiny $s$ channel signal from the huge $t$ channel contamination, whereas at present
this signal has not even been established yet individually at the LHC
(the most recent search being~\cite{ATLAS-CONF-2011-118}).
Hence, one should stay careful when stating limits on $V_L$ from measurements
of the overall size of $V_{tb}$ until its kinematic behaviour is further clarified
experimentally.

\section{Conclusions}\label{sum}

In this article, we have revisited the model-independent parametrisation of
anomalous top couplings to SM gauge bosons within the effective operator
approach, paying special attention to the charged-current sector and its
phenomenological implications at current hadron colliders. More explicitly,
addressing the minimal fully general set of anomalous trilinear $tbW$ couplings
coming from dimension six effective operators, there is
a controversy regarding the meaning of ``fully general'', namely whether an
off-shell interaction contained in the original operator basis should be dropped
because it turns out to be related to four-fermion contact interactions after
application of the equations of motion, or not. While dropping it and 
sticking to the usual coupling basis $(V_L,V_R,g_L,g_R)$ naturally
simplifies the analysis, there are good arguments to include it: Since it
emerges from the minimal gauge invariant operator basis that also generates
the trilinear couplings, the respective coupling size $V_L^\text{off}$ is related
to the other ones by the underlying operator coefficients. (For example, an
experimental limit on $\delta V_L$ is ambiguous in the context of
effective operator coefficients, requiring knowledge about either
$\delta V_L^\text{off}$ or the NC sector to be resolved.)
For the same reason, the coupling basis $(V_L,V_R,g_L,g_R,V_L^\text{off})$
parametrises the complete set of anomalous diagrams which interfere with the
SM diagram in a minimal way,
so including it is also consistent at the level of matrix elements.

In the phenomenological part, the dependence of single top cross sections on
anomalous $tbW$ couplings in $s$ and $t$~channel production is examined,
stressing the fact that the couplings do not only affect the total cross sections
but also final state distributions, which determine the selection
efficiencies within the detector acceptance region. While these effects are usually
considered small, working with constant detector efficiencies and modelling the
whole coupling dependence on the basis of on-shell production amplitudes, we use the
\textsc{Whizard} machinery to scan the full off-shell matrix element dependence
on the couplings inside the acceptance window defined by the final state selection cuts.
Comparing to the on-shell approach, one finds considerable deviations in some
regions of the parameter space, especially where the momentum-dependent couplings
$g_{L,R}$ are involved, affecting the sensitivities of the various production
channels to those couplings and therefore also the limits derived from the
experiment. Finally, an adapted polynomial approach of the coupling dependence
is discussed, which is based on quadratic fits to the full off-shell matrix
element response including detector acceptance, and turns out to parametrise
the full scan result rather well while still being fast and efficient.
However, it is also stated as a result of the present study that the theoretical
modelling of the coupling dependence should be adapted as closely as possible to
a given experimental analysis with defined selection criteria to minimize
the systematic uncertainty of the derived limits.

The study concludes with a short discussion of the influence of top decay
observables on combined coupling limits, and the regions of the parameter space
where single top cross sections still provide the crucial input to the bounds,
namely the overall $tbW$ vertex normalization and the interference direction
$V_R\sim2g_L$. In this respect, we address the impact of including
$V_L^\text{off}$ in the coupling basis, and briefly point out the possibilities
to resolve the ambiguity between $V_L$ and $V_L^\text{off}$ experimentally, using
kinematic distributions in the $s$~channel.

\begin{acknowledgments}
FB thanks J.~A.~Aguilar-Saavedra for useful discussions.
FB is supported by
Deutsche Forschungsgemeinschaft through the Research Training Group GRK\,1147
\textit{Theoretical Astrophysics and Particle Physics}.
\textsc{Whizard} development is supported in part by the Helmholtz Alliance
\textit{Physics at the Terascale}.
Parts of this work are supported by the German Ministry of Education and Research
(BMBF) under contract no.~05H09WWE.
\end{acknowledgments}

\providecommand{\href}[2]{#2}
\renewcommand{\eprint}[1]{\href{http://arxiv.org/abs/#1}{{[arXiv:#1]}}}
\bibliography{references}

\end{document}